\documentclass[aps, prd, floats, floatfix, onecolumn, superscriptaddress, nofootinbib, showpacs, letterpaper, 10pt, preprintnumbers, notitlepage]{revtex4-1}

\usepackage{graphicx}
\usepackage[caption=false]{subfig}
\usepackage{amsmath}
\usepackage{amsfonts}
\usepackage{amssymb}
\usepackage{amstext}
\usepackage{amsthm}
\usepackage{tensor}
\usepackage[english]{babel}
\usepackage{helvet}
\usepackage{microtype}
\usepackage{bbm}
\usepackage{fullpage}
\usepackage[pdftex, pdftitle={Statistics of Peaks in Chi-Squared Fields},
pdfauthor={Jolyon K. Bloomfield, Stephen H. P. Face, Alan H. Guth, Saarik Kalia, Zander Moss},
bookmarks,bookmarksopen=false,
pdfstartview={FitH},
linktocpage=true,
psdextra, % Allows for magic like: \texorpdfstring{$\cos(\gamma)$}{\cos \gamma}
pdfencoding=auto
]{hyperref}

\newcommand{\EV}[1]{\left\langle#1\right\rangle}
\newcommand{\ev}[1]{\langle#1\rangle}

\newcommand{\Bev}[1]{\langle#1\rangle_B}
\newcommand{\Cov}{\mathrm{Cov}}
\newcommand{\Var}{\mathrm{Var}}
\newcommand{\Pk}{{\cal P}}

\DeclareMathOperator{\sinc}{sinc}

\begin{document}

% Fix for a bug in revtex4-1 for onecolumn mode footnotes
\count\footins = 1000

\title{Statistics of Peaks in Chi-Squared Fields}

\author{Jolyon K. Bloomfield}
\email{jolyon@mit.edu}
\author{Stephen H. P. Face}
\email{face@mit.edu}
\author{Alan H. Guth}
\email{guth@ctp.mit.edu}
\affiliation{Center for Theoretical Physics, Laboratory for Nuclear Science, and Department of Physics, Massachusetts Institute of Technology, Cambridge, MA 02139, USA}
\author{Saarik Kalia}
\email{saarik@stanford.edu}
\affiliation{Stanford Institute of Theoretical Physics, Stanford University, Stanford, CA 94305, USA}
\author{Zander Moss}
\email{zander@caltech.edu}
\affiliation{Walter Burke Institute for Theoretical Physics, California Institute of Technology, Pasadena, CA 91125, USA}

\date{\today}

\preprint{MIT-CTP/5043}

\begin{abstract}
Chi-squared random fields arise naturally from the study of fluctuations in field theories with $SO(n)$ symmetry. The extrema of chi-squared fields are of particular physical interest. In this paper, we undertake a statistical analysis of the stationary points of chi-squared fields, with particular emphasis on extrema. We begin by describing the neighborhood of a stationary point in terms of a biased chi-squared random field, and then compute the expected profile of this field, as well as a variety of associated statistics. We are interested in understanding how spherically symmetric the neighborhood of a stationary point is, on average. To this end, we decompose the biased field into its spherical harmonic modes about this point, and explore their statistics. Using these mode statistics, we construct a metric to gauge the degree of spherical symmetry of the field in this neighborhood. Finally, we show how to leverage the harmonic decomposition to efficiently sample both Gaussian and chi-squared fields about a stationary point.
\end{abstract}

\pacs{02.50.-r, 05.40.-a, 98.80.Cq}
% Detail PACS numbers here
% 02.50.-r = Probability theory, stochastic processes, and statistics
% 05.40.-a = Fluctuation phenomena, random processes, noise, and Brownian motion
% 98.80.Cq = Particle-theory and field-theory models of the early Universe (including cosmic pancakes, cosmic strings, chaotic phenomena, inflationary universe, etc.)

\maketitle

\section{Introduction} \label{sec:intro}

A $\chi^2$ field is composed of the sum of the squares of $n$ independent and identically-distributed Gaussian random fields. It naturally arises whenever an $SO(n)$ symmetry exists in field space, which often occurs when multiplets of fields are considered. While the $\chi^2$ field itself may not be a fundamental field, it may describe a useful macroscopic quantity such as the energy density associated with the fundamental fields.

We are interested in rare events in a $\chi^2$ field -- those that are far from the mean behavior. These can be either far-above (peaks) or far-below (troughs). Because of the sum-of-squares nature of a $\chi^2$ field, the mean resides at some finite value, and the behavior of troughs and peaks can be quite different. This contrasts to the case of Gaussian fields, where the behavior of peaks and troughs are identical.

In the case of a Gaussian field, Bardeen, Bond, Kaiser and Szalay (BBKS) \cite{Bardeen1986} performed a comprehensive analysis of the statistical behavior of peaks, including the number density, peak-peak correlations, and density profiles of the peaks. This paper gave birth to the approach of ``peaks theory''.

While not as ubiquitous as Gaussian random fields, it is surprising that there is essentially nothing in the literature regarding the statistics of $\chi^2$ fields. In \cite{Bloomfield2016}, we began the treatment of $\chi^2$ fields in an analogous fashion to BBKS by computing the number density of peaks and troughs in a $\chi^2$ field. With the number density of such events now well-understood, we are now interested in what those rare events look like, statistically speaking. The aim of this paper is to continue the investigation of $\chi^2$ fields by understanding the density profiles of such events.

We perform a statistical analysis of the behavior of the $\chi^2$ field near peaks and troughs. In particular, we are interested in the following questions:
\begin{itemize}
\item What is the shape of a peak, on average? What is the envelope on this shape?
\item How spherical do we expect such events to be?
\item How do these properties scale with the field amplitude at the extrema and number of fields?
\item How can we draw random samples that have a peak or trough at a certain amplitude?
\end{itemize}

This paper is roughly divided into two parts. In the first part, we develop the statistical tools to investigate a $\chi^2$ field and its spherical harmonic components. Section \ref{sec:chisquared} introduces our notation and computes basic statistical properties of the field. In Section \ref{sec:ensembles}, we discuss what it means to bias a statistical ensemble, and clarify two equivalent but philosophically distinct approaches to performing calculations. In Section \ref{sec:biasing}, we describe how to bias our fields so that we can enforce the  existence of a peak or trough, and compute some elementary statistical measures surrounding such events. In Section \ref{sec:spherical_decomp}, we perform a spherical decomposition of the $\chi^2$ field, and compute the statistical behavior of the individual mode functions. Having computed all the relevant statistics, we construct peak profiles and envelopes in Section \ref{sec:sphericity}, and develop measures to describe how spherical we expect rare events to be.

In the second part of this paper, we develop the tools to generate samples of $\chi^2$ fields about a peak or trough. The sampling approach we use is critical, as naive approaches can easily require terabytes of storage. While conceptually rather different from the first part, this involves applying many of the tools developed in previous sections to the underlying Gaussian random fields. In Section \ref{sec:sampling}, we construct the spherical harmonic decomposition of the Gaussian fields and understand their statistics, before applying the biasing procedure to them. We pay particular attention to how the statistics of the full $\chi^2$ fields are reconstructed from the spherical harmonic modes of the underlying Gaussian fields. In Section \ref{sec:sampling2}, we discuss how to truncate the spherical harmonic modes in $\ell$, and investigate the effect that this will have on the sampled profiles. Finally, we construct the sampling procedure for both the $\chi^2$ field and its spherical harmonic modes. The paper concludes with a number of technical appendices.

\section{Chi-Squared Fields} \label{sec:chisquared}

We begin by exploring some basic statistical properties of $\chi^2$ fields. Let $\phi^1(\vec r\,),\ldots,\phi^n(\vec r\,)$ be $n$ independent real Gaussian random fields, for $\vec{r} \in\mathbb R^3$.  We define a \emph{generalized $\chi^2$ field} $\Phi$ as the random field given by
\begin{align} \label{eq:Phi_def}
\Phi(\vec r\,) = \sum_{\alpha=1}^n \phi^\alpha(\vec r\,)^2.
\end{align}
If the Gaussian random fields are identically distributed and have vanishing one-point function ($\langle\phi^\alpha(\vec r\,)\rangle=0$), we call $\Phi$ a \emph{$\chi^2$ field}.

The statistics of $\phi^\alpha$ are entirely governed by its one-point function $\langle\phi^\alpha(\vec r\,)\rangle$ (the mean of the field at a point) and two-point function $\langle\phi^\alpha(\vec r_1)\phi^\alpha(\vec{r}_2)\rangle$ (related to the covariance of the field between two points), where $\langle \cdot \rangle$ is the ensemble expectation value\footnote{Note that in this paper, we never use Einstein summation notation on field indices; all field index sums are explicitly written.}. Throughout this paper, we assume that the one- and two-point functions of the underlying Gaussian fields are known.

Working with a generalized $\chi^2$ field, the expectation value of $\Phi$ at a point is given by
\begin{align} \label{eq:Phi_1point}
\langle \Phi(\vec{r}\,) \rangle = \sum_{\alpha=1}^{n} \langle\phi^\alpha(\vec{r}\,)^2\rangle
= \sum_{\alpha=1}^{n} \left(\langle\phi^\alpha(\vec{r}\,)\rangle^2 + \Var(\phi^\alpha(\vec{r}\,))\right).
\end{align}
The two-point function is likewise given by
\begin{align} \label{eq:grf_4point}
\langle \Phi(\vec r_1)\Phi(\vec r_2)\rangle = \sum_{\alpha=1}^{n} \sum_{\beta=1}^{n} \langle \phi_1^\alpha \phi_1^\alpha \phi_2^\beta \phi_2^\beta \rangle,
\end{align}
where $\phi_i^\alpha = \phi^\alpha(\vec r_i)$. This is a four-point function of Gaussian random variables, and can therefore be simplified using a generalization of Wick's theorem (see Appendix~\ref{app:wick}). Substituting $X_1 = X_2 = \phi_1^\alpha$ and $X_3 = X_4 = \phi_2^\beta$ into Eq. \eqref{eq:4point_wick} and simplifying, we obtain
\begin{align} \label{eq:4point_phi}
\langle \phi_1^\alpha \phi_1^\alpha \phi_2^\beta \phi_2^\beta \rangle &= \mathrm{Var}(\phi_1^\alpha)\mathrm{Var}(\phi_2^\beta) + 2 \, \mathrm{Cov}(\phi_1^\alpha,\phi_2^\beta)^2
\nonumber \\
&\qquad + \langle \phi_1^\alpha \rangle^2 \mathrm{Var}(\phi_2^\beta) + 4\langle \phi_1^\alpha \rangle \langle \phi_2^\beta \rangle \mathrm{Cov}(\phi_1^\alpha, \phi_2^\beta)
\nonumber \\
&\qquad + \langle \phi_2^\beta \rangle^2 \mathrm{Var}(\phi_1^\alpha) + \langle \phi_1^\alpha \rangle^2 \langle \phi_2^\beta \rangle^2.
\end{align}
Making use of the independence of the  Gaussian fields, the two-point function of $\Phi$ becomes
\begin{align}
\langle \Phi(\vec r_1)\Phi(\vec r_2)\rangle &= \left(\sum_{\alpha=1}^n\mathrm{Var}(\phi_1^\alpha)\right)\left(\sum_{\beta=1}^n\mathrm{Var}(\phi_2^\beta)\right)
+ 2 \sum_{\alpha=1}^n\mathrm{Cov}(\phi_1^\alpha,\phi_2^\alpha)^2
\nonumber \\
&\qquad +\left(\sum_{\alpha=1}^n\langle \phi_1^\alpha \rangle^2\right) \left(\sum_{\beta=1}^n\mathrm{Var}(\phi_2^\beta)\right) 
+ 4\sum_{\alpha=1}^n \ev{\phi_1^\alpha} \ev{\phi_2^\alpha} \mathrm{Cov}(\phi_1^\alpha, \phi_2^\alpha)
\nonumber \\
&\qquad +\left(\sum_{\beta=1}^n\langle \phi_2^\beta \rangle^2\right) \left(\sum_{\alpha=1}^n\mathrm{Var}(\phi_1^\alpha)\right) 
+ \left(\sum_{\alpha=1}^n\langle \phi_1^\alpha \rangle^2 \right)\left(\sum_{\beta=1}^n\langle \phi_2^\beta \rangle^2 \right)
\nonumber \\
&=\langle\Phi(\vec r_1)\rangle\langle\Phi(\vec r_2)\rangle + 2 \sum_{\alpha=1}^n\mathrm{Cov}(\phi_1^\alpha,\phi_2^\alpha)^2 + 4\sum_{\alpha=1}^n
\ev{\phi_1^\alpha} \ev{\phi_2^\alpha} \mathrm{Cov}(\phi_1^\alpha, \phi_2^\alpha). \label{eq:Phi_2point}
\end{align}
This result assumes only that the fields $\phi^\alpha$ are Gaussian and independent of each other.

In the remainder of this paper, $\Phi$ (when it appears with no subscripts) will denote a $\chi^2$ field constructed from a set of $n$ identically distributed Gaussian random fields which are subject to three additional assumptions. First, we assume they are \textit{centered}, which means that their one-point function vanishes everywhere, $\langle \phi^\alpha(\vec{r}\,) \rangle =0$. This implies that the two-point function is simply the covariance, $\langle\phi^\alpha(\vec{r}\,)\phi^\alpha(\vec{r}\,')\rangle = \mathrm{Cov}(\phi^\alpha(\vec{r}\,),\phi^\alpha(\vec{r}\,'))$. Second, we assume these fields are \emph{homogeneous}, meaning their two-point function depends only on the difference between the points, i.e. $\mathrm{Cov}(\phi^\alpha(\vec{r}\,), \phi^\alpha(\vec{r}\,')) = C(\vec{r}\,' - \vec{r}\,)$, where we have introduced the shorthand $C$ for the homogeneous covariance. Finally, we  assume they are \emph{isotropic}, meaning the covariance depends only on the magnitude of the difference between the points, i.e. $\mathrm{Cov}(\phi^\alpha(\vec{r}\,), \phi^\alpha(\vec{r}\,')) = C(|\vec{r}\,'-\vec{r}\,|)$. For brevity, we will typically omit the absolute value signs.

We now reevaluate the one- and two-point functions of $\Phi$ under these assumptions. Because our Gaussian fields are centered, homogeneous and isotropic, we can express their two-point function (covariance) in terms of a power spectrum $\Pk(k)$ (see Appendix~\ref{app:kspace_integrals} for details and conventions). The covariance of $\phi^\alpha$ is given by\footnote{This is the first result that assumes 3-dimensional space, which is assumed henceforth throughout this paper. While generalizations to other dimensions are typically straightforward, we do not pursue this here.}
\begin{align}
\mathrm{Cov}(\phi^\alpha(\vec{r}\,), \phi^\alpha(\vec{r}\,'))
= C(\vec r\,' - \vec r\,)
= 4\pi \int dk \, k^2 \, \Pk(k) \, \mathrm{sinc}(k|\vec{r}\,'-\vec{r}\,|).
\end{align}
We denote the moments of the power spectrum by
\begin{align} \label{eq:moments}
\sigma^2_n = 4\pi\int dk \, k^{2n+2} \, \Pk(k).
\end{align}
In particular, $\langle\phi^\alpha(\vec r\,)^2\rangle = C(0) = \sigma_0^2$. Substituting this into \eqref{eq:Phi_1point} and \eqref{eq:Phi_2point}, we obtain the background relations
\begin{align}
\langle\Phi(\vec r\,)\rangle &= n\sigma_0^2 
\label{eq:unbiased_1point}
\\
\langle\Phi(\vec r\,)\Phi(\vec r\,')\rangle &= n^2\sigma_0^4+2nC(\vec r\,'-\vec r\,)^2 \label{eq:unbiased_2point}
\\
\Cov(\Phi(\vec r\,), \Phi(\vec r\,')) &= 2nC(\vec r\,'-\vec r\,)^2. 
\label{eq:unbiased_cov}
\end{align}
In the coincidence limit, we find the background variance
\begin{align} \label{eq:backgroundvar}
\Var (\Phi(\vec r\,)) &= 2 n \sigma_0^4.
\end{align}
The $\chi^2$ field $\Phi$ thus has mean value $n \sigma_0^2$ and fluctuations about the mean with standard deviation $\sqrt{2 n} \sigma_0^2$, or as a relative deviation,
\begin{align}
\frac{\sqrt{\Var (\Phi(\vec r\,))}}{\ev{\Phi(\vec r\,)}} = \sqrt{\frac{2}{n}}.
\end{align}
We see that larger $n$ leads to larger expected field values, but smaller relative fluctuations. These results give us a description of the expected background behavior of a $\chi^2$ field, before specializing to the neighborhood of a stationary point.

\section{Ensembles, Expectation Values and Biasing} \label{sec:ensembles}

Throughout the rest of this paper, we will discuss the statistics of a $\chi^2$ field in the vicinity of a stationary point, and understand how they deviate from the homogeneous ``background'' behavior discussed above. We will translate the condition of the existence of a stationary point of fixed amplitude into a set of algebraic constraints on the field. To compute field statistics in this case, we must first understand how to compute expectation values of a random field in the presence of constraints.

Recall that a general random field $\psi$ is defined by an ensemble of scalar functions of $\mathbb{R}^3$, together with a probability density function (PDF) over the ensemble. We will refer to the elements of the ensemble as \textit{realizations} of the field. We can then understand the expectation value of a function of a general random field, $\langle f(\psi) \rangle$ as the result of a process in which all realizations are taken from the field ensemble, acted upon by $f$, weighted by the PDF, and summed point-wise. The expectation is the map from the random field to a scalar function on $\mathbb{R}^3$ defined by this procedure.

If we want to compute an expectation in the presence of constraints, then we need to modify this map so that only the field realizations which satisfy the constraints are included in the sum. For example, expectations of the $\chi^2$ field $\Phi$ conditioned on the existence of a stationary point of fixed amplitude should be computed by summing only over those realizations for which $\Phi(0) = \nu^2$ and $\vec{\nabla} \Phi(0) = 0$, where $\nu^2$ is the field amplitude at the origin and where we have exploited the homogeneity of $\Phi$ to place the stationary point at the origin. These restricted expectation maps are biased expectation maps, which we will denote by $\langle \cdot \rangle_B$. From this point of view, conditions are applied by modifying the expectation map, not the random field $\psi$. The \textit{biased expectation value} is then given by $\langle f(\psi) \rangle_B$.

On the other hand, we could view the set of realizations of $\psi$ which satisfy the given constraints as an ensemble in its own right. We could then compute the PDF of $\psi$ restricted to this set and interpret the new ensemble and PDF as a different random field, called the \textit{biased field}, which we denote by $\psi_B$. Finally, we could apply the standard expectation map to $\psi_B$, yielding $\langle f(\psi_B) \rangle$.

The important thing to note here is that both the biased field and the biased expectation map approaches produce the same result,
\begin{align}
\langle f(\psi_B) \rangle \equiv \langle f(\psi) \rangle_B.
\end{align}
As such, throughout this paper, we will use the two interchangeably. The distinction, however, is not purely philosophical. In cases where the expectation has a simple integral representation, it may be easier to restrict the map explicitly using delta functions (see, for example, Appendix~\ref{app:biasing}). On the other hand, if the PDF of the biased field is available, or if the biased field can be represented in terms of simpler biased fields with known PDFs (or known expectations), then the second approach is preferable. 

In the case of a $\chi^2$ field, we will take the biased field approach, representing $\Phi_B$ in terms of biased Gaussian random fields $\phi_B$. This calculation is done in detail in the next section, but we will pause to make a few observations about $\phi_B$ and $\Phi_B$ before diving in.

It is important to note that while the PDF of a biased field is, in a sense, inherited from the PDF of the unbiased field, it is altered by the biasing constraints. The PDF of a biased field may therefore be qualitatively different from the unbiased PDF, especially in terms of the symmetries it exhibits. Consider a centered, homogeneous, and isotropic Gaussian random field $\phi$ conditioned on the existence of a stationary point of fixed amplitude at the origin. The corresponding biased field, $\phi_B$ turns out to be a Gaussian random field as well, but it is no longer homogeneous and isotropic because we have broken translation symmetry by introducing a privileged point. However, it does retain an $SO(3)$ rotational symmetry about the origin. Further conditions can be applied at the origin while maintaining this rotational symmetry, so long as they are compatible with the spherical symmetry (for example, requiring the field to have zero gradient at the origin).

In the case of our homogeneous, isotropic $\chi^2$ field $\Phi$, conditioning on a stationary, fixed point at the origin yields a biased field $\Phi_B$ which is a \textit{generalized} $\chi^2$ field, but not a $\chi^2$ field. This is a consequence of the fact that $\Phi_B$ can be written as the sum of squares of biased Gaussian fields $\phi_B$, which are no longer zero mean nor identically distributed, though they remain independent. As with the Gaussian fields, isotropy and homogeneity are lost, having been broken down to an $SO(3)$ rotational symmetry about the origin.

When taking the biased field approach, it is important to realize that even though the biased PDF may be qualitatively different from the unbiased PDF, all field realizations from the biased field are also realizations of the unbiased field. We are just selecting those that satisfy our constraints. Hence, while the neighborhoods of stationary points in a $\chi^2$ field may be said to arise from a generalized $\chi^2$ field, they are still realizations of a $\chi^2$ field.

\section{Biasing} \label{sec:biasing}

We will now detail the calculation of biased expectation values of Gaussian and $\chi^2$ random fields. We are interested in biased expectation values of functions of $\Phi$, which we now know to be equivalent to expectation values of functions of $\Phi_B$. Evaluating such expressions will allow us to compute the means and (co)variances of our $\chi^2$ field and its spherical harmonic coefficients. 

To perform these calculations, we will first need to understand how to bias the Gaussian fields underlying $\Phi$ such that all realizations of the sum of their squares will have a stationary point of fixed amplitude at the origin. We will then need to compute the one- and two-point functions of these biased Gaussian fields, $\ev{\phi^\alpha_B}$ and $\ev{\phi^\alpha_B \phi^\beta_B}$. Finally, we will need to express the one- and two-point functions of $\Phi_B$ in terms of the one- and two-point functions of $\phi_B^\alpha$.

\subsection{Biasing Gaussian Random Fields} \label{sec:biasing-procedure}

We begin by biasing our centered, homogeneous and isotropic Gaussian random fields $\phi^\alpha$. We will show that after biasing, they remain Gaussian random fields, but are no longer centered, homogeneous and isotropic.

We wish to bias the $\chi^2$ field $\Phi(\vec{r}\,)$ such that $\Phi(0) = \nu^2$ and $\vec{\nabla} \Phi(0) = 0$. Given that
\begin{align} \label{eq:amplitude_constraint}
\Phi(0) = \nu^2 = \sum_{\alpha=1}^n (\phi^\alpha(0))^2,
\end{align}
we see that biasing the $\chi^2$ field amplitude amounts to a constraint on the sum-of-squares of the fields $\phi^\alpha$. Considering the various $\phi^\alpha$ as a vector in an $n$-dimensional field space, equation \eqref{eq:amplitude_constraint} constrains the vector to lie on a sphere of radius $\nu$ in field space, centered at the origin.

Note that Eq. \eqref{eq:amplitude_constraint} possesses an $SO(n)$ rotational symmetry. Hence, we can perform a global field rotation to choose the coordinate system in field space such that the vector $\phi^\alpha(0)$ lies along the $\hat{\phi}^{(1)}$ axis. Formally, we can choose
\begin{align} \label{eq:gauge_fix}
\Phi(0) &= \nu^2 = (\phi^1(0))^2
\\
\phi^\alpha(0) &= 0 \quad \text{for all} \quad \alpha \in \{2, \ldots, n\}.
\end{align}
The amplitude fixing constraint \eqref{eq:amplitude_constraint} then reduces to the constraint \begin{align} \label{eq:phi_amplitude_constraint}
\phi^1(0) = \pm \nu,
\end{align}
together with the constraint that all other $\phi^\alpha$ vanish at the origin.
The freedom of choice of sign in front of $\nu$ in Eq.~\eqref{eq:phi_amplitude_constraint} can also be thought of as a rotational freedom in field space. In particular, we are free to choose the vector $\phi^\alpha$ to be parallel or antiparallel to $\hat{\phi}^{(1)}$ direction. For simplicity, we will always rotate to the positive root. 

Having treated amplitude fixing, we turn to the stationarity constraint,
\begin{align}
\vec{\nabla}\Phi(0) &= 2\sum_{\alpha=1}^{n} \phi^\alpha(0) \vec{\nabla}\phi^\alpha(0)
= 2\nu \vec{\nabla}\phi^1(0) = 0
\\
\Rightarrow \vec{\nabla}\phi^1(0) &= 0,
\end{align}
where we've substituted $\phi^1(0) = \nu$ and $\phi^{\alpha}(0) = 0$ for\footnote{Note that we assume $\nu \neq 0$. The case $\nu = 0$ is simpler to work with, but much less general. If one is interested in this case, the results can typically be obtained from the results in this paper by setting $D(r) \to 0$.} $\alpha > 1$.

Altogether, the condition of a stationary point of fixed amplitude at the origin corresponds to the following constraints on realizations of $\phi^\alpha$:
\begin{align} \label{eq:constraints}
\phi^1(0) = \nu
, \qquad
\vec{\nabla}\phi^1(0) = 0
, \qquad
\phi^\alpha(0) = 0 \ \ \text{for} \ \ \alpha \in \{2, \ldots, n\}.
\end{align}
It remains to be proven that expectation values of $f(\Phi)$ under the condition that $\Phi(0) = \nu^2$ and $\vec{\nabla} \Phi(0) = 0$ are equivalent to expectation values of $f(\Phi)$ under the conditions given by Eq. \eqref{eq:constraints}. The proof of this equivalence is technical, and we leave the details to Appendix \ref{app:biasing}.

We are now ready to impose the constraints in Eq. \eqref{eq:constraints} onto our fields. By definition, a Gaussian random field has the property that all PDFs $p(\phi(\vec{r}_1), \ldots, \phi(\vec{r}_n))$ for finitely many points are multivariate normal distributions. These multivariate PDFs can be explicitly constructed from the field mean and covariance functions. Furthermore, the derivative of a Gaussian field is itself a Gaussian field, and indeed, vectors containing the field value and its derivatives evaluated at finitely many points are also described by multivariate Gaussian PDFs. Throughout this paper, when we refer to the ``PDF'' of a Gaussian random field, we are implicitly referring to such multi-point PDFs. 

We now exploit the fact that a subset of variables from a multivariate Gaussian distribution, conditioned on fixed values of the complementary subset, is described by another multivariate Gaussian distribution. Hence, by constructing a random vector containing the field and its derivatives at the origin, as well as the field at two arbitrary locations, after conditioning on the values of the field and its derivatives at the origin, we will obtain the means and covariances describing the biased field at two arbitrary locations. This yields the mean and covariance of the biased field, from which all statistical properties can be computed.

We now present this biasing procedure in detail. Consider a multivariate Gaussian random variable $\mathbf{w}$ described by mean $\boldsymbol{\mu}$ and covariance matrix $\boldsymbol{\Sigma}$, where we use bold face to indicate these vectors in order to differentiate from spatial vectors denoted by an arrow. Let us split the vector $\mathbf{w}$ into $\mathbf{w}_1$ and $\mathbf{w}_2$ as
\begin{align}
\mathbf{w} = \left[\begin{array}{c}\mathbf{w}_1 \\ \mathbf{w}_2\end{array}\right]
\end{align}
where $\mathbf{w}$ has dimension $N$, $\mathbf{w}_1$ has dimension $q$, and $\mathbf{w}_2$ has dimension $N - q$. The means of $\mathbf{w}_1$ and $\mathbf{w}_2$ are given by
\begin{align}
\ev{\mathbf{w}} = \boldsymbol{\mu} = \left[\begin{array}{c}\boldsymbol{\mu}_1 \\ \boldsymbol{\mu}_2\end{array}\right]
\end{align}
while the covariance matrix splits into a block form given by
\begin{align}
\boldsymbol{\Sigma} = 
\left[
\begin{array}{cc}
\boldsymbol{\Sigma}_{11} & \boldsymbol{\Sigma}_{12} \\
\boldsymbol{\Sigma}_{21} & \boldsymbol{\Sigma}_{22}
\end{array}
\right].
\end{align}
The sizes of these covariance matrices are
\begin{align}
\left[
\begin{array}{cc}
q \times q & q \times (N - q) \\
(N - q) \times q & (N - q) \times (N - q)
\end{array}
\right].
\end{align}
A well-known result in multivariate normal statistics then gives the expectation values and covariance matrix of $\mathbf{w}_1$ conditioned on $\mathbf{w}_2$. The results are
\begin{subequations} \label{eq:biasing}
\begin{align}
\ev{\mathbf{w}_1}_B &= \boldsymbol{\mu}_1 - \boldsymbol{\Sigma}_{12} \boldsymbol{\Sigma}_{22}^{-1} (\boldsymbol{\mu}_2 - \mathbf{w}_2)
\\
\boldsymbol{\Sigma}_{B,11} &= \boldsymbol{\Sigma}_{11} - \boldsymbol{\Sigma}_{12} \boldsymbol{\Sigma}_{22}^{-1} \boldsymbol{\Sigma}_{21},
\end{align}
\end{subequations}
where $\mathbf{w}_2$ here denotes the constant vector to which the random vector $\mathbf{w}_2$ is fixed, not the random vector itself.

We need to apply this result twice, once for $\phi^1$ and once for $\phi^{\alpha \neq 1}$, as the different fields are subject to different constraints. First though, we introduce some new notation. For the unbiased Gaussian random fields, we have already seen
\begin{align}
\ev{\phi^\alpha(\vec{r}\,) \phi^\beta(\vec{r}\,')} 
= \Cov(\phi^\alpha(\vec{r}\,), \phi^\beta(\vec{r}\,'))
= \delta^{\alpha \beta} C(\vec{r}\,' - \vec{r}\,).
\end{align}
As $C(\vec{r}\,' - \vec{r}\,)$ only depends on the magnitude of its argument, we will write $C(r)$ when a convenient magnitude is available. Note that $C(0) = \sigma_0^2$. We compute in Appendix~\ref{app:kspace_integrals} that
\begin{align}
C(r) = 4 \pi \int dk \, k^2 \, \Pk(k) \sinc(k r).
\end{align}

We will also come across the two-point function $\ev{\phi^\alpha(\vec{r}\,) \vec{\nabla} \phi^\alpha(0)}$. Expanding the gradient in terms of constant unit basis vectors, and observing that these unit vectors can be taken out of the expectation value, we conclude that this expectation value is vector-valued. We define
\begin{align}
\vec{D}(\vec{r}\,) = \ev{\phi^\alpha(\vec{r}\,) \vec{\nabla} \phi^\alpha(0)},
\end{align}
which is also independent of $\alpha$. As $\phi^\alpha$ is homogeneous and isotropic, the only direction that $\vec{D}$ can point in is $\hat{r}$, and $\vec{D}(\vec{r}\,)$ can only depend on $|\vec{r}\,|$. Hence, we can write
\begin{align}
\vec{D}(\vec{r}\,) = D(r) \hat{r}.
\end{align}
We show in Appendix~\ref{app:kspace_integrals} that
\begin{align}
D(r) = 4 \pi \int dk \, k^3 \, \Pk(k) j_1(kr)
\end{align}
where $j_1$ is a spherical Bessel function of the first kind. We will also come across the dot product $\hat{r} \cdot \hat{r}'$ repeatedly, so we define the angle between these two unit vectors as $\gamma$, such that
\begin{align}
\cos \gamma = \hat{r} \cdot \hat{r}'.
\end{align}
Finally, it is convenient to define the dimensionless field amplitude
\begin{align}
\bar{\nu} = \frac{\nu}{\sigma_0}.
\end{align}

Returning to the biasing of the $\phi^\alpha$, we perform the calculation for $\alpha > 1$ first as it involves fewest constraints. Let 
\begin{align}
\mathbf{w}_1 = \begin{pmatrix}
\phi^\alpha(\vec{r}\,) \\ \phi^\alpha(\vec{r}\,')
\end{pmatrix}
, \qquad
\mathbf{w}_2 = (\phi^\alpha(0)),
\end{align}
with $\alpha > 1$. The means of these vectors are
\begin{align}
\ev{\mathbf{w}_1} = \boldsymbol{\mu}_1 = 
\begin{pmatrix}
0 \\ 0
\end{pmatrix}
, \qquad 
\ev{\mathbf{w}_2} = \boldsymbol{\mu}_2 = (0).
\end{align}
The covariance matrices are given by
\begin{align}
\boldsymbol{\Sigma}_{11} = \begin{pmatrix}
\sigma_0^2 & C(\vec{r}\,' - \vec{r}\,) \\
C(\vec{r}\,' - \vec{r}\,) & \sigma_0^2
\end{pmatrix}
, \qquad
\boldsymbol{\Sigma}_{12} =
\boldsymbol{\Sigma}_{21}^T = \begin{pmatrix}
C(r)\\
C(r')\\
\end{pmatrix}
, \qquad
\boldsymbol{\Sigma}_{22} = \begin{pmatrix}
\sigma_0^2
\end{pmatrix}.
\end{align}
Applying Eqs. \eqref{eq:biasing}, we obtain the following from conditioning on $\phi^\alpha(0) = 0$.
\begin{align}
\langle \phi^\alpha_B(\vec{r}\,) \rangle &= 0
\\
\mathrm{Cov}(\phi^\alpha_B(\vec{r}\,), \phi^\alpha_B(\vec{r}\,')) &= 
\langle \phi^\alpha_B(\vec{r}\,) \phi^\alpha_B(\vec{r}\,') \rangle
= C(\vec{r}\,' - \vec{r}\,) - \frac{C(r) C(r')}{\sigma_0^2}
\end{align}
Unsurprisingly, a field with original mean zero, biased to zero amplitude at the origin, does not develop a mean. However, the covariance does shift due to the amplitude fixing at the origin. Note that the covariance given here is the off-diagonal entry of $\boldsymbol{\Sigma}_{B,11}$, which is the same as the diagonal entries when $\vec{r} = \vec{r}\,'$.

We now tackle the more complicated case of $\alpha = 1$. Let 
\begin{align}
\mathbf{w}_1 = \begin{pmatrix}\phi^1(\vec{r}\,) \\ \phi^1(\vec{r}\,')\end{pmatrix},
\qquad
\mathbf{w}_2 = \begin{pmatrix}
\phi^1(0) \\
\partial_1 \phi^1(0) \\
\partial_2 \phi^1(0) \\
\partial_3 \phi^1(0)
\end{pmatrix}.
\end{align}
The means of these vectors are
\begin{align}
\ev{\mathbf{w}_1} = \boldsymbol{\mu}_1 = 
\begin{pmatrix}
0 \\ 0
\end{pmatrix}
, \qquad 
\ev{\mathbf{w}_2} = \boldsymbol{\mu}_2 = \begin{pmatrix}
0 \\
0 \\
0 \\
0
\end{pmatrix}.
\end{align}
The covariance matrices are given by
\begin{align}
\boldsymbol{\Sigma}_{11} &= \begin{pmatrix}
\sigma_0^2 & C(\vec{r}\,' - \vec{r}\,) \\
C(\vec{r}\,' - \vec{r}\,) & \sigma_0^2
\end{pmatrix}
\\
\boldsymbol{\Sigma}_{12} &=
\boldsymbol{\Sigma}_{21}^T = \begin{pmatrix}
C(r) & D_1(\vec{r}\,) & D_2(\vec{r}\,) & D_3(\vec{r}\,) \\
C(r') & D_1(\vec{r}\,') & D_2(\vec{r}\,') & D_3(\vec{r}\,') \\
\end{pmatrix}
\\
\boldsymbol{\Sigma}_{22} &= \begin{pmatrix}
\sigma_0^2 & 0 & 0 & 0 \\
0 & \tfrac{1}{3} \sigma_1^2 & 0 & 0 \\
0 & 0 & \tfrac{1}{3} \sigma_1^2 & 0 \\
0 & 0 & 0 & \tfrac{1}{3} \sigma_1^2
\end{pmatrix}
\end{align}
where $D_i$ denotes the $i^{\textrm{th}}$ component of $\vec{D}$, and we have used the results from Appendix~\ref{app:kspace_integrals} that $\vec{D}(0) = 0$ and $\ev{\partial_i \phi(\vec{r}\,) \partial_j \phi(\vec{r}\,)} = \tfrac{1}{3} \sigma_1^2 \delta_{ij}$. Applying Eqs. \eqref{eq:biasing}, we obtain the following from conditioning on $\phi^1(0) = \nu$ and $\partial_i \phi^1(0) = 0$.
\begin{align}
\langle \phi^1_B(\vec{r}\,) \rangle &= \frac{\bar{\nu}}{\sigma_0} \, C(r)
\\
\mathrm{Cov}(\phi^1_B(\vec{r}\,), \phi^1_B(\vec{r}\,')) &= C(\vec{r}\,' - \vec{r}\,) - \frac{C(r) \, C(r')}{\sigma_0^2} - 3 \frac{D(r) D(r')}{\sigma_1^2} \cos \gamma
\\
\langle \phi^1_B(\vec{r}\,) \phi^1_B(\vec{r}\,') \rangle &= C(\vec{r}\,' - \vec{r}\,) - \frac{C(r) \, C(r')}{\sigma_0^2} - 3 \frac{D(r) D(r')}{\sigma_1^2} \cos \gamma + \bar{\nu}^2 \frac{C(r) \, C(r')}{\sigma_0^2}
\end{align}
Here, the mean does shift, and it is evident that $\langle \phi^1_B(0) \rangle = \nu$ as expected. The covariance matrix is very similar to the $\alpha>1$ case, picking up extra terms from derivatives at the origin. Once again, the covariance is given by the off-diagonal entry of $\boldsymbol{\Sigma}_{B,11}$, which is equivalent to the diagonal entries when $\vec{r} = \vec{r}\,'$.

That all pairs $(\phi^\alpha, \phi^\beta)$ of the unbiased, underlying fields are independent implies that the biased fields $\phi^\alpha_B$ have zero covariance across the field-space index for all pairs of real-space positions. Because these are Gaussian random fields, this zero covariance condition actually implies independence across the field space index. Note that unlike the underlying Gaussian random fields $\phi^\alpha$, the biased fields $\phi_B^\alpha$ are not all centered, identically distributed, homogeneous or isotropic. They are still Gaussian random fields, but their real-space symmetry has been broken down to an $SO(3)$ rotational symmetry about the origin.

This completes the calculation of the biased one- and two-point functions for the Gaussian random fields. We summarize our results here.
\begin{subequations} \label{eq:grf-biasing}
\begin{align}
\langle \phi^\alpha_B(\vec{r}\,) \rangle &= \delta^{\alpha 1} \frac{\bar{\nu}}{\sigma_0} \, C(r)
\\
\mathrm{Cov}(\phi^\alpha_B(\vec{r}\,), \phi^\beta_B(\vec{r}\,')) &= \delta^{\alpha \beta} \left(C(\vec{r}\,' - \vec{r}\,) - \frac{C(r) C(r')}{\sigma_0^2}\right) - \delta^{\alpha 1} \delta^{\beta 1} \frac{3 D(r) D(r')}{\sigma_1^2} \cos(\gamma) \label{eq:biased-cov}
\\
\langle \phi^\alpha_B(\vec{r}\,) \phi^\beta_B(\vec{r}\,') \rangle &= \delta^{\alpha \beta} \left(C(\vec{r}\,' - \vec{r}\,) - \frac{C(r) C(r')}{\sigma_0^2}\right)
+ \delta^{\alpha 1} \delta^{\beta 1} \left(\frac{\bar{\nu}^2}{\sigma_0^2} C(r) C(r') - \frac{3 D(r) D(r')}{\sigma_1^2} \cos(\gamma) \right)
\end{align}
\end{subequations}

At this point, it is worth explaining why we are only biasing for stationarity and field amplitude, but do not constrain the field Hessian $\partial_i \partial_j \Phi$. In principle, one could bias the field Hessian to ensure either a peak or a trough at the stationary point. For example, when a stationary point forms a peak, all eigenvalues of the Hessian are negative. Unfortunately, enforcing this negativity condition on the Hessian is a complicated business. First, the biasing procedure described in this paper allows us to bias field values and derivatives to fixed values. Our biasing procedure is unable to enforce inequalities. Furthermore, the eigenvalues are nontrivial combinations of all entries of the Hessian, which makes constructing the constraint in terms of the Hessian components difficult.

Second, even if one could impose a constraint appropriately on the eigenvalues of the Hessian, this constraint is often rendered meaningless. The behavior of the Hessian at a point is typically dominated by high frequency noise from the high-$k$ tail of the power spectrum. Even if these modes have very little amplitude, the tiny ripples they induce govern the Hessian at a point, which means that biasing the Hessian essentially imposes no constraint on a peak. If one needs a guarantee that a given sample is indeed a peak/trough, we recommend choosing a length scale upon which this requirement must be satisfied, and checking points on a sphere with this radius about the origin to ensure this requirement is met.

\subsection{Constructing Expressions}

Having computed the biased one- and two-point functions for the Gaussian random fields, we will now use them to construct expectation values of functions of the biased $\chi^2$ field, $\Phi_B$. We have seen that the constraints on $\Phi$ which ensure a stationary point of fixed amplitude at the origin can be translated into constraints on the underlying $\phi$ fields. The ensemble of $\phi^\alpha$ realizations satisfying these constraints then defines $\phi_B^\alpha$. It follows that we can expand $\Phi_B$ as the sum of squares of the $\phi_B^\alpha$. The biased expectation value of a function of $\Phi$ can then be written
\begin{align}
\Bev{f(\Phi)} \equiv \ev{f(\Phi_B)} = \EV{f\left(\sum_{\alpha = 1}^n (\phi^\alpha_B)^2\right)}.
\end{align}
We have also seen that $\phi^\alpha_B$ remains a Gaussian random field after biasing. This is critical because it means that Wick's theorem and its generalization (see Appendix~\ref{app:wick}) are available for evaluating expectation values of finite products of the $\phi_B^\alpha$. For given $f$, our strategy will be to find a closed form function $F$ such that
\begin{align}
\left\langle f\left(\sum_{\alpha} (\phi_B^\alpha)^2\right) \right\rangle =  F\left(\langle \phi_B^\alpha \rangle, \langle \phi_B^\alpha \phi_B^\beta \rangle\right),
\end{align}
which allows us to express $\langle f(\Phi) \rangle_B$ in terms of the one- and two- point functions of $\phi_B^\alpha$,
\begin{align} \label{eq:dual_biasing_final}
\Bev{f(\Phi)} = F\left(\langle \phi_B^\alpha \rangle, \langle \phi_B^\alpha \phi_B^\beta \rangle\right).
\end{align}
In the cases we consider, $f$ is always a product or sum of products of the $\phi^\alpha_B$, so the form of $F$ is given explicitly by the generalized Wick theorem. Once we find $F$, we will have reduced the problem of computing biased expectation values of the $\chi^2$ field to the problem of computing a handful of one- and two-point functions of the biased Gaussian fields $\phi^\alpha_B$ and plugging them into $F$.

\subsection{Some Biased Statistics of \texorpdfstring{$\Phi$}{\Phi}}

We are now prepared to compute the one- and two-point functions and the covariance of $\Phi$, conditioned on the existence of a stationary point of fixed amplitude at the origin. In each of these cases, we have already computed the relevant function $F$ in section~\ref{sec:chisquared} for the unbiased case. To compute the biased statistics, we only need to replace $\phi$ with $\phi_B$, and we can then substitute Eqs. \eqref{eq:grf-biasing} into equations~\eqref{eq:Phi_1point} and~\eqref{eq:Phi_2point}. 

Beginning with the one-point function,
\begin{align} \label{eq:1p_biased_Phi}
\Bev{\Phi(\vec{r}\,)} = \sum_{\alpha=1}^{n} \Bev{\phi^\alpha(\vec{r}\,)^2}
= \Bev{\phi^1(\vec{r}\,)^2} + \sum_{\alpha=2}^{n} \Bev{\phi^\alpha(\vec{r}\,)^2}
= n \sigma_0^2 + (\bar{\nu}^2 - n) \frac{C(r)^2}{\sigma_0^2} - 3 \frac{D(r)^2}{\sigma_1^2}.
\end{align}
At $r=0$, $C(0) = \sigma_0^2$ and $D(0) = 0$, so $\Bev{\Phi(\vec{r}\,)} = \nu^2$ as expected. As $r \to \infty$, $C(r) \rightarrow 0$ and $D(r) \rightarrow 0$, leading to the background result of $\Bev{\Phi(\vec{r}\,)} = n \sigma_0^2$. Between these limits, we can integrate $C(r)$ and $D(r)$ numerically to see how a peak or trough behaves on average as a function of radius.

We pause here to point out a subtlety in the effect of the stationarity condition on the mean of $\Phi$.
If we set $\nu^2 = n \sigma_0^2$, the background field value, the biased expectation value becomes
\begin{align}
\Bev{\Phi(\vec{r}\,)} &= n \sigma_0^2 - 3 \frac{D(r)^2}{\sigma_1^2},
\end{align}
which matches the background value at the origin, decreases for intermediate $r$, and then returns asymptotically to the background as $r\rightarrow \infty$ (recall that $D\rightarrow 0$ in this limit). This behavior may come as a surprise, as we have fixed the field to the background at the origin, and demanded stationarity there. Why, then, isn't the expectation constant at the background value for all $r$? 
The $D(r)$ term appears when we constrain $\vec{\nabla}\Phi$ to vanish at the origin. Indeed, if we constrain $\Phi(0)$ to the background value, but leave the gradient unconstrained, we will recover the expected result: a constant expectation.

Recall that the mean of $\Phi$ is given by the sum of variances and means-squared of each of the $\phi^\alpha$ (see Eq. \eqref{eq:Phi_1point}). When we fix the direction of $\phi^\alpha$ in field space and constrain the sum-of-squares to the background value, we shift the mean and variance of the $\phi^\alpha$ in such a way as to cancel the terms dependent on $r$ through $C(r)$ (see the first two terms in Eq. \eqref{eq:1p_biased_Phi}). When we constrain $\vec{\nabla}\Phi(0) = 0$, we constrain $\vec{\nabla} \phi^1(0) = 0$. By forcing this gradient to vanish, we are ``flattening out'' $\phi^1$ in the vicinity of the origin, thereby suppressing its variance. Constraining the variance of $\phi^1$ does not alter its mean, so there is no term to counteract the suppression of the variance. The mean of $\Phi$ therefore decreases from the fixed background value at the origin, and then returns asymptotically, even when we do not bias the field amplitude away from the background value!

Next, we compute the two-point function, $\Bev{\Phi(\vec{r}\,) \Phi(\vec{r}\,')}$. As before, the calculations which led us to Eq. \eqref{eq:Phi_2point} are identical in the biased case. Substituting Eqs. \eqref{eq:grf-biasing} into Eq. \eqref{eq:Phi_2point} yields
\begin{align}
\Bev{\Phi(\vec{r}\,) \Phi(\vec{r}\,')} &= 
\Bev{\Phi(\vec{r}\,)} \Bev{\Phi(\vec{r}\,')} 
+ 2 (n-1) \left(C(\vec{r}\,' - \vec{r}\,) - \frac{C(r) C(r')}{\sigma_0^2}\right)^2
\nonumber\\
& \qquad 
+ 2 \left(C(\vec{r}\,' - \vec{r}\,) - \frac{C(r) C(r')}{\sigma_0^2} - \frac{3 D(r) D(r')}{\sigma_1^2} \cos(\gamma)\right)^2
\nonumber\\
& \qquad 
+ 4 \bar{\nu}^2 \frac{C(r) \, C(r')}{\sigma_0^2} \left(C(\vec{r}\,' - \vec{r}\,) - \frac{C(r) C(r')}{\sigma_0^2} - \frac{3 D(r) D(r')}{\sigma_1^2} \cos(\gamma)\right).
\end{align}
In the limit when $\vec{r} \to 0$, only the first term survives, indicating that the two-point function factorizes. This is expected, because $\Phi(0) = \nu^2$ is a constant. We can straightforwardly compute the covariance $\Cov(\Phi_B(\vec{r}\,), \Phi_B(\vec{r}\,'))$ by subtracting off the product of means, producing
\begin{align} \label{eq:biased_Phi_cov}
\Cov(\Phi_B(\vec{r}\,), \Phi_B(\vec{r}\,')) &= 
2 (n-1) \left(C(\vec{r}\,' - \vec{r}\,) - \frac{C(r) C(r')}{\sigma_0^2}\right)^2
\nonumber\\
& \qquad 
+ 2 \left(C(\vec{r}\,' - \vec{r}\,) - \frac{C(r) C(r')}{\sigma_0^2} - \frac{3 D(r) D(r')}{\sigma_1^2} \cos(\gamma)\right)^2
\nonumber\\
& \qquad 
+ 4 \bar{\nu}^2 \frac{C(r) \, C(r')}{\sigma_0^2} \left(C(\vec{r}\,' - \vec{r}\,) - \frac{C(r) C(r')}{\sigma_0^2} - \frac{3 D(r) D(r')}{\sigma_1^2} \cos(\gamma)\right).
\end{align}
The variance at a point is then given by
\begin{align} \label{eq:Phi-var}
\Var(\Phi_B(\vec{r}\,)) &= 
2 (n-1) \left(\sigma_0^2 - \frac{C(r)^2}{\sigma_0^2}\right)^2
+ 2 \left(\sigma_0^2 - \frac{C(r)^2}{\sigma_0^2} - \frac{3 D(r)^2}{\sigma_1^2}\right) \left(\sigma_0^2 - \frac{C(r)^2}{\sigma_0^2} - \frac{3 D(r)^2}{\sigma_1^2}
+ 2 \bar{\nu}^2 \frac{C(r)^2}{\sigma_0^2} \right).
\end{align}
The variance vanishes at the origin, as expected (we have fixed the field there), while at large radii it approaches the unbiased (background) result. We can use this expression for the variance to compute a 1-$\sigma$ envelope about the average field in the vicinity of a stationary point, as shown in Figure \ref{fig:profiles}.

\section{Spherical Harmonic Decompositions} \label{sec:spherical_decomp}

The point-wise expectation values and covariances computed so far give us only a rough picture of the shape of $\Phi$ in the vicinity of a stationary point. To make more precise statements about the geometry of the field near a stationary point, we will develop the spherical harmonic decomposition of $\Phi$. In particular, knowledge of the one- and two-point functions of the harmonic coefficients of $\Phi$ will allow us to compare contributions from the spherical mode ($\ell = 0$) and aspherical modes ($\ell >0$). In the next section, we will use these comparisons to make quantitative statements about how ``spherical'' peaks and troughs in $\Phi$ are, on average.

In this section and throughout this paper, we will typically write angular variables in terms of the unit vector that they describe. For example, the angular variables $(\theta, \phi)$ are described by the unit vector
\begin{align}
\hat{r} = \sin \theta \cos \phi \, \hat{x} + \sin \theta \sin \phi \, \hat{y} + \cos \theta \, \hat{z}.
\end{align}
This is particularly useful when constructing integrals over the angular variables of a vector, where we can instead write the integral over all possible unit vectors. In particular,
\begin{align}
\int d \hat{r} = \int d\Omega = \int_0^{2 \pi} d\phi \int_0^\pi d\theta \, \sin \theta.
\end{align}
Throughout this paper, we use real spherical harmonics $Y_{\ell m}(\hat{r})$ so that the harmonic coefficients of $\Phi$ become real random fields over the radial coordinate, $r$. The relevant properties of real spherical harmonics are presented in Appendix \ref{app:real_harmonics}.

\subsection{Spherical Harmonic Decomposition of \texorpdfstring{$\Phi$}{\Phi}}

Here, we compute the one- and two-point functions as well as the covariances of the spherical harmonic coefficients of the field $\Phi$. The results in this subsection apply to both biased and unbiased $\chi^2$ fields.

Consider the spherical harmonic expansion of $\Phi$.
\begin{align} \label{eq:decomp_Phi}
\Phi(\vec{r}\,) = \sum_{\ell=0}^{\infty} \sum_{m=-\ell}^{\ell} \Phi_{\ell m}(r) Y_{\ell m}(\hat{r})
\end{align}	
The mode coefficients are random fields given by
\begin{align} \label{eq:def_Phi_mode}
\Phi_{\ell m}(r) = \int d\hat{r} \, \Phi(r \hat{r}) Y_{\ell m}(\hat{r}).
\end{align}
Because we have chosen to expand in the real spherical harmonics, $\Phi_{\ell m}(r)$ is a real random field. The distribution of $\Phi_{\ell m}(r)$ is unknown, as it is complicated by an integral over a $\chi^2$ field with nontrivial correlations. Nevertheless, we can still compute the one- and two-point functions of $\Phi_{\ell m}(r)$.

The one-point functions are given by
\begin{align} \label{eq:1point_Phi_mode}
\langle \Phi_{\ell m}(r) \rangle 
= \int d\hat{r} \, \langle \Phi(r \hat{r}) \rangle Y_{\ell m}(\hat{r})
= \delta_{\ell 0} \delta_{m 0} \sqrt{4\pi} \langle \Phi(\vec{r}\,) \rangle,
\end{align}
where we exploit the fact that $\Phi(\vec{r}\,)$ possesses an $SO(3)$ rotation symmetry about the origin, so $\langle \Phi(\vec{r}\,) \rangle$ only depends on the radial coordinate.

The two-point function of the spherical harmonic coefficients is given in terms of the two-point function of $\Phi$ as
\begin{align} \label{eq:2point_Phi_mode}
\langle \Phi_{\ell m}(r) \Phi_{\ell^\prime m^\prime}(r') \rangle = \int d\hat{r} \, d\hat{r}' \, \langle \Phi(r \hat{r}) \Phi(r \hat{r}') \rangle Y_{\ell m}(\hat{r}) Y_{\ell^\prime m^\prime}(\hat{r}').
\end{align}
We then construct the covariance of the mode coefficients in the usual manner.
\begin{align} 
\mathrm{Cov}(\Phi_{\ell m}(r), \Phi_{\ell^\prime m^\prime} (r')) 
&= \langle \Phi_{\ell m}(r) \Phi_{\ell^\prime m^\prime}(r') \rangle - \langle \Phi_{\ell m}(r) \rangle \langle \Phi_{\ell^\prime m^\prime}(r') \rangle
%\\
%&= \int d\Omega_1 \, d\Omega_2 \, \big(\langle \Phi(\vec{r}) \Phi(\vec{r}') \rangle - \ev{\Phi(\vec{r}} \ev{\Phi(\vec{r}'}\big) Y_{\ell m}(\Omega_1) Y_{\ell^\prime m^\prime}(\Omega_2)
\\
&= \int d\hat{r} \, d\hat{r}' \, \Cov(\Phi(r \hat{r}), \Phi(r \hat{r}')) Y_{\ell m}(\hat{r}) Y_{\ell^\prime m^\prime}(\hat{r}').
\label{eq:cov_Phi_mode}
\end{align}
We prove in Appendix~\ref{app:spherical_covariance} that the covariances and two-point functions of the mode coefficients vanish when the either the modes ($\ell,\ell^\prime$) or the azimuthal numbers ($m,m^\prime$) do not match, and furthermore, that so long as $m = m'$, they are independent of $m$. That is to say,
\begin{gather}
\langle \Phi_{\ell m}(r) \Phi_{\ell^\prime m^\prime}(r') \rangle = \delta_{\ell \ell^\prime} \delta_{m m^\prime} \langle \Phi_{\ell m}(r) \Phi_{\ell m}(r')\rangle
\\
\mathrm{Cov}(\Phi_{\ell m}(r), \Phi_{\ell^\prime,m^\prime}(r')) = \delta_{\ell \ell^\prime} \delta_{m m^\prime} \mathrm{Cov}(\Phi_{\ell m}(r), \Phi_{\ell m}(r')).
\end{gather}

We can also compute the covariance between two points in terms of the individual mode covariances. Consider
\begin{align}
\ev{\Phi(\vec{r}\,) \Phi(\vec{r}\,')} = \sum_{\ell, \ell'=0}^{\infty} \sum_{m=-\ell}^{\ell} \sum_{m'=-\ell'}^{\ell'} Y_{\ell m}(\hat{r}) Y_{\ell' m'}(\hat{r}') \ev{\Phi_{\ell m}(r) \Phi_{\ell' m'}(r')}.
\end{align}
Using $\ev{\Phi_{\ell m}(r) \Phi_{\ell' m'}(r')} \propto \delta_{\ell \ell'} \delta_{m m'}$ with the remainder of the expression independent of $m$, this becomes
\begin{align}
\ev{\Phi(\vec{r}\,) \Phi(\vec{r}\,')} = \sum_{\ell=0}^{\infty} \sum_{m=-\ell}^{\ell} Y_{\ell m}(\hat{r}) Y_{\ell m}(\hat{r}') \ev{\Phi_{\ell 0}(r) \Phi_{\ell 0}(r')}.
\end{align}
We now apply the addition theorem \eqref{eq:addition} to sum over $m$.
\begin{align} \label{eq:twopointPhi}
\ev{\Phi(\vec{r}\,) \Phi(\vec{r}\,')} = \frac{1}{4\pi} \sum_{\ell=0}^{\infty} 
(2\ell+1) P_\ell(\hat{r} \cdot \hat{r}')
\ev{\Phi_{\ell 0}(r) \Phi_{\ell 0}(r')}
\end{align}
Subtracting $\ev{\Phi(\vec{r}\,)} \ev{\Phi(\vec{r}\,')} = \ev{\Phi_{00}(r)} \ev{\Phi_{00}(r')} / (4\pi)$ from both sides, this becomes
\begin{align} \label{eq:cov-full-mode}
\Cov(\Phi(\vec{r}\,), \Phi(\vec{r}\,')) = \frac{1}{4\pi} \sum_{\ell=0}^{\infty} 
(2\ell+1) P_\ell(\hat{r} \cdot \hat{r}')
\Cov(\Phi_{\ell 0}(r), \Phi_{\ell 0}(r')).
\end{align}
In the coincidence limit, the variance is
\begin{align} \label{eq:var_relation}
\Var(\Phi(\vec{r}\,)) = \frac{1}{4\pi} \sum_{\ell=0}^{\infty} 
(2\ell+1) \Var(\Phi_{\ell 0}(r)).
\end{align}
This relation allows us to understand how different modes contribute to the variance at a given point, which we investigate in Section 
\ref{sec:sphericity} (see Figure \ref{fig:stackprofiles}).

\subsection{Unbiased Spherical Harmonics of \texorpdfstring{$\Phi$}{\Phi}}

Now that we have a statistical description of the spherical harmonics of $\Phi$, it's time to examine their behavior about a stationary point. To quickly warm up before we tackle the biased spherical harmonics, we start by looking at the results for the unbiased decomposition of $\Phi$.

We begin with the means by substituting Eq. \eqref{eq:unbiased_1point} into the expression for $\ev{\Phi_{\ell m} (r)}$ \eqref{eq:1point_Phi_mode}.
\begin{align}
\langle \Phi_{\ell m}(r) \rangle = \delta_{\ell 0} \delta_{m 0} \sqrt{4\pi} n\sigma_0^2
\end{align}
As expected, there is only a mean for the spherically symmetric mode, and that is the mean of the field (with an extra factor of $\sqrt{4\pi}$ from $Y_{00}$).

For the covariance, we substitute Eq. \eqref{eq:unbiased_cov} into the expression for $\mathrm{Cov}(\Phi_{\ell m}(r), \Phi_{\ell^\prime m^\prime} (r'))$ \eqref{eq:cov_Phi_mode}.
\begin{align} 
\mathrm{Cov}(\Phi_{\ell m}(r), \Phi_{\ell^\prime m^\prime} (r')) 
= 2 n \int d\hat{r} \, d\hat{r}' \, C(r' \hat r' - r \hat r)^2 Y_{\ell m}(\hat{r}) Y_{\ell^\prime m^\prime}(\hat{r}')
\end{align}
While this integral looks rather unappealing, three of the four angular integrals can be performed analytically. Note that $|r' \hat{r}' - r \hat{r}| = \sqrt{r^2 + r'^2 - 2 r r' \cos \gamma}$, with $\cos \gamma = \hat{r} \cdot \hat{r}'$, so the argument to $C$ depends only on $r$, $r'$ and the single angle $\gamma$. When integrating, the only one of these that will vary is $\gamma$. In Appendix~\ref{app:wigner}, we show how such integrals can be performed. In particular, using Eq. \eqref{eq:foursphere}, we find that
\begin{align} \label{eq:firstuglyintegral}
\mathrm{Cov}(\Phi_{\ell m}(r), \Phi_{\ell^\prime m^\prime} (r')) 
= 4 \pi n \delta_{\ell \ell'} \delta_{m m'} \int_{-1}^1 du \, P_\ell(u) C\left(\sqrt{r^2 + r'^2 - 2 r r' u}\right)^2.
\end{align}
This can alternatively be derived from Eqs. \eqref{eq:unbiased_cov} and \eqref{eq:cov-full-mode} by exploiting the orthogonality of Legendre polynomials.

\subsection{Biased Spherical Harmonics of \texorpdfstring{$\Phi$}{\Phi}}
 
We now turn to the biased spherical harmonics of $\Phi$. The means of the spherical harmonic coefficients, $\langle \Phi_{\ell m}(r)\rangle$, can be straightforwardly computed by substituting the expression for $\Bev{\Phi(\vec{r}\,)}$ \eqref{eq:1p_biased_Phi} into the expression for $\ev{\Phi_{\ell m} (r)}$ \eqref{eq:1point_Phi_mode}, yielding
\begin{align} \label{eq:Phi_biased_1point_mode}
\Bev{\Phi_{\ell m}(r)} = \delta_{\ell 0} \delta_{m 0} \sqrt{4\pi} \left(n \sigma_0^2 + (\bar{\nu}^2 - n) \frac{C(r)^2}{\sigma_0^2} - 3\frac{D(r)^2}{\sigma_1^2}\right).
\end{align}

To compute the full covariance $\Cov(\Phi_{B, \ell m}(r), \Phi_{B, \ell' m'}(r'))$, we start by expanding the biased covariance \eqref{eq:biased_Phi_cov} and inserting the result into the spherical harmonic covariance \eqref{eq:cov_Phi_mode}.
%we start by expanding the biased covariance \eqref{eq:biased_Phi_cov}, collecting terms that have the same angular dependence.
%\begin{align}
%\Cov(\Phi_B(\vec{r}\,), \Phi_B(\vec{r}\,')) &=
%2 n C(\vec{r}\,' - \vec{r}\,)^2 
%\nonumber\\
%& \qquad 
%+ 4 C(\vec{r}\,' - \vec{r}\,) \frac{C(r) \, C(r')}{\sigma_0^2} \left(\bar{\nu}^2 - n\right)
%\nonumber\\
%& \qquad 
%- 12 C(\vec{r}\,' - \vec{r}\,) \frac{D(r) D(r')}{\sigma_1^2} \cos(\gamma) 
%\nonumber\\
%& \qquad 
%+ 2 \frac{C(r)^2 C(r')^2}{\sigma_0^4} \left(n - 2 \bar{\nu}^2\right)
%\nonumber\\
%& \qquad 
%- 12 \, \frac{C(r) C(r')}{\sigma_0^2} \frac{D(r) D(r')}{\sigma_1^2} \left(\bar{\nu}^2 - 1\right) \cos(\gamma)
%\nonumber\\
%& \qquad 
%+ 18 \frac{D(r)^2 D(r')^2}{\sigma_1^4} \cos^2(\gamma)
%\end{align}
%We now insert this expression into the spherical harmonic covariance \eqref{eq:cov_Phi_mode}, where we write the angular integrals as integrals over unit vectors $\hat{r}$ and $\hat{r}'$.
\begin{align}
\mathrm{Cov}(\Phi_{B, \ell m}(r), \Phi_{B, \ell' m'} (r')) 
&= \int d\hat{r} \, d\hat{r}' \, \bigg(
2 n C(r' \hat{r}' - r \hat{r})^2 
\nonumber\\
& \qquad \qquad \qquad \quad 
+ 4 C(r' \hat{r}' - r \hat{r}) \frac{C(r) \, C(r')}{\sigma_0^2} \left(\bar{\nu}^2 - n\right)
\nonumber\\
& \qquad \qquad \qquad \quad 
- 12 C(r' \hat{r}' - r \hat{r}) \frac{D(r) D(r')}{\sigma_1^2} \cos(\gamma) 
\nonumber\\
& \qquad \qquad \qquad \quad 
+ 2 \frac{C(r)^2 C(r')^2}{\sigma_0^4} \left(n - 2 \bar{\nu}^2\right)
\nonumber\\
& \qquad \qquad \qquad \quad 
- 12 \, \frac{C(r) C(r')}{\sigma_0^2} \frac{D(r) D(r')}{\sigma_1^2} \left(\bar{\nu}^2 - 1\right) \cos(\gamma)
\nonumber\\
& \qquad \qquad \qquad \quad 
+ 18 \frac{D(r)^2 D(r')^2}{\sigma_1^4} \cos^2(\gamma)
\bigg) Y_{\ell m}(\hat{r}) Y_{\ell' m'}(\hat{r}')
\end{align}
Note that $\cos \gamma = \hat{r} \cdot \hat{r}'$. Here, we have collected terms that have the same angular dependence. Once again, we can evaluate the angular integrals using the techniques derived in Appendix~\ref{app:wigner}. The first integral is identical to the one performed in Eq. \eqref{eq:firstuglyintegral}, while the others can be evaluated using Eqs. \eqref{eq:foursphere} and \eqref{eq:uglyintegrals} by using $|r' \hat{r}' - r \hat{r}| = \sqrt{r^2 + r'^2 - 2 r r' \cos \gamma}$. The result is
\begin{align} \label{eq:partialcov}
\mathrm{Cov}(\Phi_{B, \ell m}(r), \Phi_{B, \ell' m'} (r')) 
= 4 \pi \delta_{\ell \ell'} \delta_{m m'} \Bigg[ &
n \int_{-1}^1 du \, P_{\ell}(u) C\left(\sqrt{r^2 + r'^2 - 2 r r' u}\right)^2
\\\nonumber
&
+ 2 \frac{C(r) \, C(r')}{\sigma_0^2} \left(\bar{\nu}^2 - n\right) \int_{-1}^1 du \, P_{\ell}(u) C\left(\sqrt{r^2 + r'^2 - 2 r r' u}\right)
\\\nonumber
&
- 6 \frac{D(r) D(r')}{\sigma_1^2} \int_{-1}^1 du \, P_{\ell}(u) C\left(\sqrt{r^2 + r'^2 - 2 r r' u}\right) u
\\\nonumber
&
+ \delta_{\ell 0} 2 \left(\frac{C(r)^2 C(r')^2}{\sigma_0^4} \left(n - 2 \bar{\nu}^2\right) + 3 \frac{D(r)^2 D(r')^2}{\sigma_1^4}\right)
\\\nonumber
&
- \delta_{\ell 1} 4 \, \frac{C(r) C(r')}{\sigma_0^2} \frac{D(r) D(r')}{\sigma_1^2} \left(\bar{\nu}^2 - 1\right)
+ \delta_{\ell 2} \frac{12}{5} \frac{D(r)^2 D(r')^2}{\sigma_1^4} \Bigg].
\end{align}
While we are unable to simplify this expression further, it turns out that the second and third integrals have further significance. In Appendix~\ref{app:kspace_integrals}, we compute the covariance between the spherical harmonic coefficients $\phi^\alpha_{\ell m}$ of the underlying Gaussian fields to be
\begin{align}
\Cov(\phi^\alpha_{\ell m}(r), \phi^\beta_{\ell' m'}(r')) = \delta^{\alpha \beta} \delta_{\ell \ell'} \delta_{m m'} 4 \pi \tilde{C}_\ell (r, r')
\end{align}
where
\begin{align} \label{eq:Ctilde1}
\tilde{C}_\ell(r, r') &= 4 \pi \int dk \, k^2 \Pk(k) j_\ell(kr) j_\ell(kr').
\end{align}
We later show that an alternative representation of $\tilde{C}_\ell$ is given by Eq. \eqref{eq:better_ctilde},
\begin{align}
\tilde{C}_\ell(r, r') = \frac{1}{2} \int_{-1}^1 du \, P_\ell(u) C\left(\sqrt{r^2 + r'^2 - 2 r r' u}\right), \label{eq:Ctilde2}
\end{align}
which is exactly one of the integrals that appears in the covariance. Furthermore, we can use Bonnet's recursion formula for the Legendre polynomials to substitute
\begin{align}
u P_\ell(u) = \frac{\ell + 1}{2\ell + 1} P_{\ell + 1}(u) + \frac{\ell}{2\ell + 1} P_{\ell - 1}(u)
\end{align}
in Eq. \eqref{eq:partialcov} to reduce another integral to the form of $\tilde{C}_\ell$. We thus arrive at our final result for the full covariance of the spherical harmonic modes $\Phi_{\ell m}$.
\begin{align} \label{eq:fullcov}
\mathrm{Cov}(\Phi_{B, \ell m}(r), \Phi_{B, \ell' m'} (r')) 
= \delta_{\ell \ell'} \delta_{m m'} 4 \pi \Bigg[ &
n \int_{-1}^1 du \, P_{\ell}(u) C\left(\sqrt{r^2 + r'^2 - 2 r r' u}\right)^2
\\\nonumber
&
+ 4 \frac{C(r) \, C(r')}{\sigma_0^2} \left(\bar{\nu}^2 - n\right) \tilde{C}_\ell(r, r')
\\\nonumber
&
- \frac{12}{2 \ell+ 1} \frac{D(r) D(r')}{\sigma_1^2} \left((\ell + 1) \tilde{C}_{\ell + 1}(r, r') + \ell \tilde{C}_{\ell - 1}(r, r')\right)
\\\nonumber
&
+ \delta_{\ell 0} 2 \left(\frac{C(r)^2 C(r')^2}{\sigma_0^4} \left(n - 2 \bar{\nu}^2\right) + 3 \frac{D(r)^2 D(r')^2}{\sigma_1^4}\right)
\\\nonumber
&
- \delta_{\ell 1} 4 \, \frac{C(r) C(r')}{\sigma_0^2} \frac{D(r) D(r')}{\sigma_1^2} \left(\bar{\nu}^2 - 1\right)
+ \delta_{\ell 2} \frac{12}{5} \frac{D(r)^2 D(r')^2}{\sigma_1^4} \Bigg]
\end{align}
Note that for $\ell = 0$, the recursion relation gives $\ell \tilde{C}_{\ell - 1} = 0$. This result is proportional to $\delta_{\ell \ell'} \delta_{m m'}$ as expected, with the remainder of the expression independent of $m$. Comparing this to the unbiased case \eqref{eq:firstuglyintegral}, the first term arises from the background covariance, while the remainder are due to the biasing. Specializing to the case where $r = r'$ in order to obtain the variance of a spherical harmonic mode, this becomes
\begin{align} \label{eq:fullvar}
\Var(\Phi_{B, \ell m}(r)) 
= {}&
4 \pi n \int_{-1}^1 du \, P_{\ell}(u) C\left(r\sqrt{2(1-u)}\right)^2
+ 16\pi \frac{C(r)^2}{\sigma_0^2} \left(\bar{\nu}^2 - n\right) \tilde{C}_\ell(r, r)
\nonumber\\
&
- \frac{48 \pi}{2 \ell+ 1} \frac{D(r)^2}{\sigma_1^2} \left((\ell + 1) \tilde{C}_{\ell + 1}(r, r) + \ell \tilde{C}_{\ell - 1}(r, r)\right)
\nonumber\\
&
+ \delta_{\ell 0} 8 \pi \left(\frac{C(r)^4}{\sigma_0^4} \left(n - 2 \bar{\nu}^2\right) + 3 \frac{D(r)^4}{\sigma_1^4}\right)
\nonumber\\
&
- \delta_{\ell 1} 16 \pi \, \frac{C(r)^2}{\sigma_0^2} \frac{D(r)^2}{\sigma_1^2} \left(\bar{\nu}^2 - 1\right)
+ \delta_{\ell 2} \frac{48 \pi}{5} \frac{D(r)^4}{\sigma_1^4}.
\end{align}

\section{The Shape of an Extremum} \label{sec:sphericity}

We now have sufficient information to plot the expected shape of a peak or trough with one-sigma envelopes, and we can develop qualitative understandings about the width and tails of extrema. Unlike Gaussian fields, where troughs are simply negative peaks, a $\chi^2$ field is always positive, and so troughs and peaks must be treated separately. We start by investigating the spherical part of a peak or trough, before looking at the aspherical components.

\subsection{Spherical Part}

We begin by developing a qualitative understanding of the functions $C(r)$ and $D(r)$ and their relative contributions to the mean and variance of $\Phi_B$. By construction, $C(r)$ is the covariance between the origin and a point $\vec{r}$ for an unbiased Gaussian field, while $D(r)$ is the covariance between the gradient at the origin in the direction of a point $\vec{r}$ and the field at that point. To compare the strength of the $C$ and $D$ terms, we need to construct the correlations $\rho_C$ and $\rho_D$. The variance of the Gaussian field at a point is simply $\sigma_0^2$, while the variance of its derivative is $\sigma_1^2/3$ (see Eq. \eqref{eq:deriv-var}). The correlation coefficients are then
\begin{align}
\rho_C(r) = \frac{C(r)}{\sigma_0^2}, \qquad \rho_D(r) = \frac{\sqrt{3} D(r)}{\sigma_0 \sigma_1}.
\end{align}
We can rewrite $\ev{\Phi_B}$ \eqref{eq:1p_biased_Phi} and $\Var(\Phi_B)$ \eqref{eq:Phi-var} very neatly in terms of these correlation coefficients as
\begin{align}\label{eq:correlationmean}
\ev{\Phi_B(r)} &= \ev{\Phi} + \sigma_0^2 \left( (\bar{\nu}^2 - n) \rho_C(r)^2 - \rho_D(r)^2\right)
\\\label{eq:correlationvar}
\Var(\Phi_B(r)) &= 
2 \sigma_0^4 \left[(n-1) \left(1 - \rho_C(r)^2\right)^2
+ \left(1 - \rho_C(r)^2 - \rho_D(r)^2\right) \left(1 - \rho_C(r)^2 - \rho_D(r)^2
+ 2 \bar{\nu}^2 \rho_C(r)^2 \right)\right]
\end{align}
where we write $\ev{\Phi} = n \sigma_0^2$ in the mean to pull out the background value.

At the origin, $\rho_C(0) = 1$ and $\rho_D(0) = 0$, while at large radius, both functions decay to zero. Both $C(r)$ and $D(r)$ obtain their radial dependence from a spherical Bessel function, which decays as $j_\ell(r) \sim \sin(r - \pi \ell/2) / r$ at large radius. Hence, $\rho_C$ and $\rho_D$ decay as $1/r$ at large radii. This implies that the mean of the field decays to the background level as $1/r^2$ at large radii
\begin{align}
\ev{\Phi_B(r)} - \ev{\Phi} = \sigma_0^2 \left((\bar{\nu}^2 - n) \rho_C(r)^2 - \rho_D(r)^2\right) \sim \frac{1}{r^2},
\end{align}
which means that we expect peaks and troughs to have very long tails.

We would like to develop an understanding of the relative strengths of $\rho_C(r)$ and $\rho_D(r)$. Our general intuition is that $\rho_C(r)$ should be much stronger than $\rho_D(r)$. Consider two points, $A$ and $B$, separated by a distance $d$. Knowledge of the field amplitude at $A$ is likely to be a much stronger indicator of the height at $B$ than knowledge of the derivative of the field amplitude (along the line from $A$ to $B$) at $A$. This is borne out in the numerics: for a power spectrum with broad support, $\rho_C(r)$ begins at 1 and decays monotonically with distance, while $\rho_D(r)$ begins at 0, grows slightly, and then decays with distance, with $\rho_D(r) < \rho_C(r)$.

However, our intuition fails us when the power spectrum is sharply peaked (e.g., $\Pk(k) \propto \delta(k - k_0)$). In this situation, there are essentially monochromatic waves with wavelength $\lambda$ filling space, and the strength of the correlation functions depends on the ratio $d/\lambda$. Numerically, the correlation functions $\rho_C(r)$ and $\rho_D(r)$ oscillate between positive (correlated) and negative (anticorrelated) values. For some distances, knowing the field amplitude at $A$ allows one to accurately predict the field amplitude at $B$, while for other distances, the field amplitude at $A$ provides no information about $B$, while the derivative of the field amplitude at $A$ gives sufficient information to predict the field amplitude at $B$.

Hence, we are unable to make any general statements about the ratio $\rho_D(r)/\rho_C(r)$, as it depends strongly on the power spectrum in question. However, for the purpose of obtaining semi-analytic expressions to describe peaks, we can reasonably restrict ourselves to the case of a power spectrum with broad support, in which case we can take $\rho_D$ to be subdominant to $\rho_C$, and we can expect $\rho_C$ to be monotonically decreasing with radius.

For the remainder of this section, we will assume that $\rho_D(r)/\rho_C(r)$ is small enough to be neglected in order to gain intuition about the expected shape and envelope of a peak or trough\footnote{It should be kept in mind that this approximation fails when $\bar{\nu}^2 - n \simeq 0$, which turns off $\rho_C(r)$ in the expression for the mean.}. With this assumption, we can approximate
\begin{align} \label{eq:meanapprox}
\ev{\Phi_B(r)} &\simeq \ev{\Phi} \left[1 + \left(\frac{\bar{\nu}^2}{n} - 1\right) \rho_C(r)^2\right]
\\ \label{eq:varapprox}
\Var(\Phi_B(r)) &\simeq 2 n \sigma_0^4 \left(1 - \rho_C(r)^2\right) \left[1 + \left(2 \frac{\bar{\nu}^2}{n} - 1\right) \rho_C(r)^2 \right]
\\ \label{eq:varapprox2}
&= \frac{2}{n} \left(\ev{\Phi_B(r)}^2 - \nu^4 \rho_C(r)^4\right).
\end{align}
We see that the relevant parameter to describe a peak/trough is $\bar{\nu}^2/n$, while the shape and envelope are entirely controlled by $\rho_C(r)$. Note that the shape of the extrema above or below the background is proportional to $C(r)^2$.

Define the radius $r_\alpha$ to be the expected radius at which the profile for the extrema is a factor of $\alpha$ of the height between the background and the extremum. For example, $r_0$ is at the background level, $r_{1/2}$ is halfway towards the extremum, and $r_1 = 0$ is at the origin. As extrema occur at $\Phi_B(0) = \nu^2$ by construction, we are searching for the radius at which
\begin{align}
\ev{\Phi_B(r_\alpha)} = \alpha (\nu^2 - \ev{\Phi}) + \ev{\Phi}.
\end{align}
Using the approximation for the mean above, we then find
\begin{align}
\rho_C(r_\alpha) = \sqrt{\alpha} \quad \to \quad C(r_\alpha) = \sqrt{\alpha} \sigma_0^2.
\end{align}
This implies the curious property that the profile width is independent of $\bar{\nu}$. While this relation can only be inverted numerically, this gives a very simple picture of the expected shape of a profile. Evidently, $r_1 = 0$ while $r_0 \to \infty$. We define the radius $r_{1/2}$ as a measure of the expected width of a profile. At this radius, $\rho_C(r_{1/2}) = 1/\sqrt{2}$, and the variance is given by
\begin{align}
\Var(\Phi_B(r_{1/2})) = \sigma_0^4 \left(\bar{\nu}^2 + \frac{n}{2} \right).
\end{align}

We define the envelope $\Delta r_{1/2}$ on the peak width to be the 1-$\sigma$ bound on the peak width. Because this will not be symmetric to wider/narrower peaks, separate definitions are required, although we simply refer to them both as $\Delta r_{1/2}$. The envelopes are defined by
\begin{align} \label{eq:envelope_def}
\ev{\Phi_B(r_{1/2} \pm \Delta r_{1/2})} \pm \sqrt{\Var(\Phi_B(r_{1/2} \pm \Delta r_{1/2}))} = \ev{\Phi_B(r_{1/2})}
\end{align}
where the relevant choice of the $\pm$ between the two terms on the left hand side depends on whether one is describing a peak or trough.

\begin{figure}[p]
\subfloat[$\bar{\nu} = 0.1$]{%
  \includegraphics[width=0.56\textwidth]{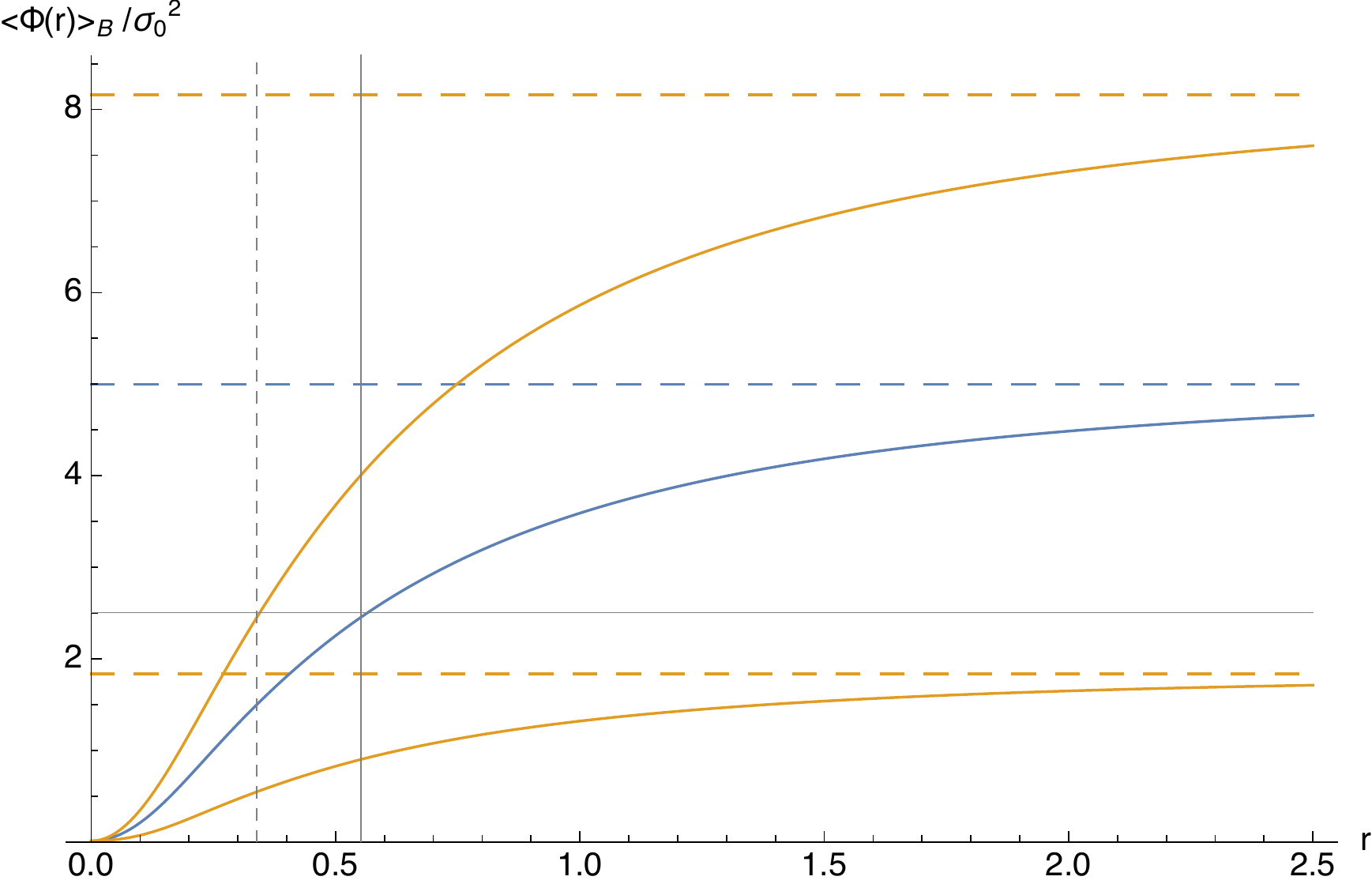}%
  \label{fig:profile-a}
}

\subfloat[$\bar{\nu} = 3$]{%
  \includegraphics[width=0.56\textwidth]{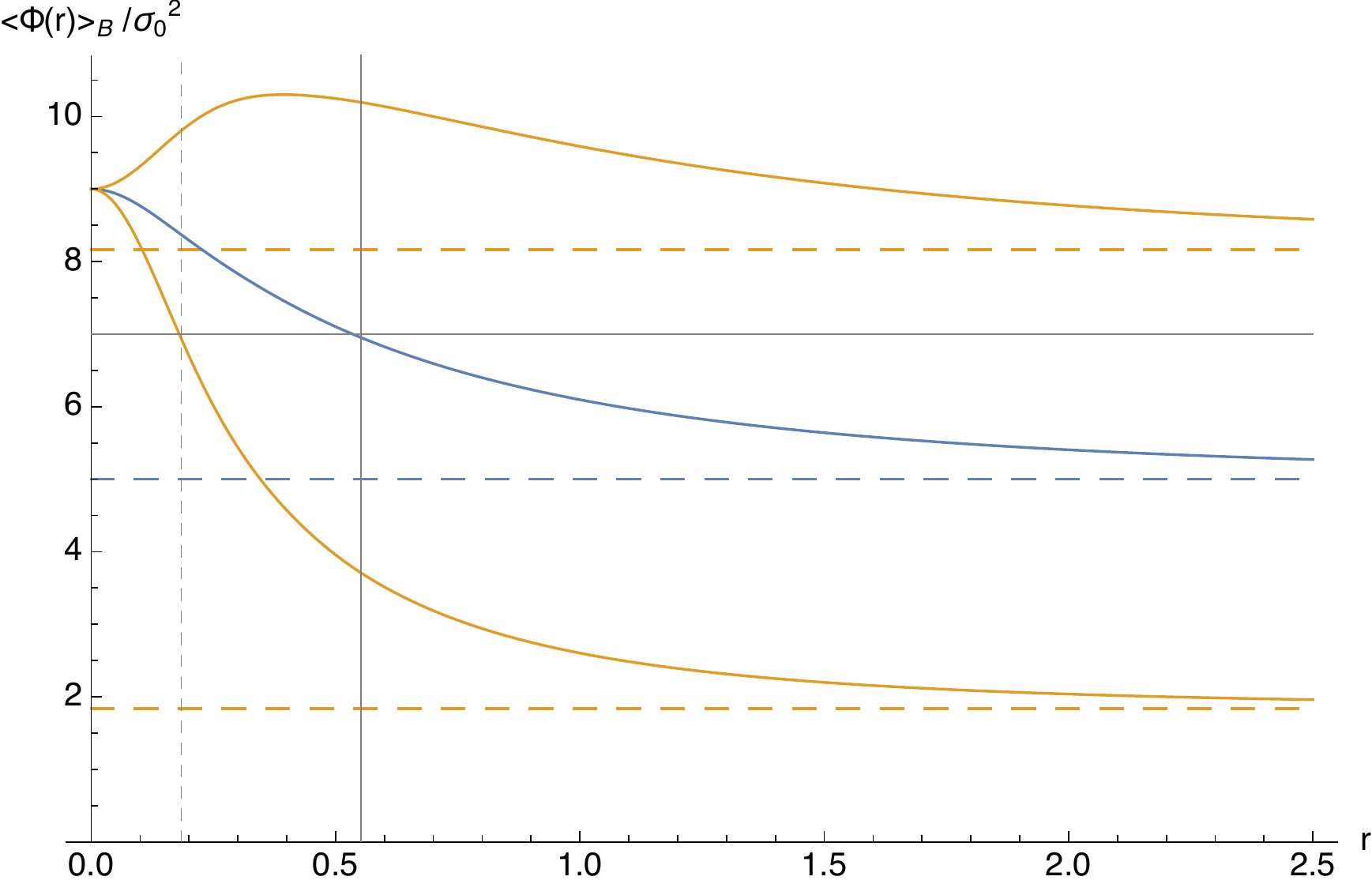}%
  \label{fig:profile-b}
}

\subfloat[$\bar{\nu} = 10$]{%
  \includegraphics[width=0.56\textwidth]{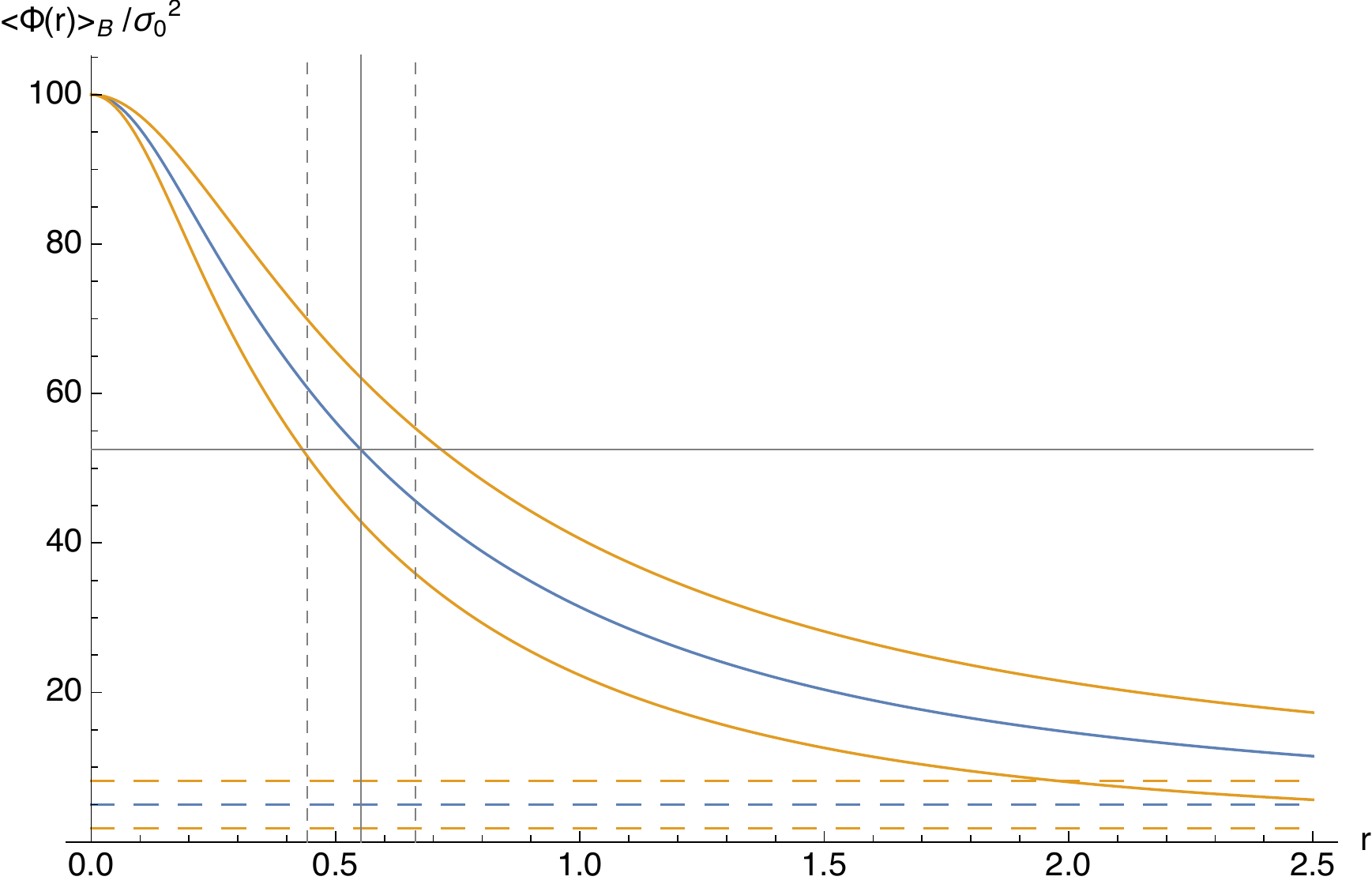}%
  \label{fig:profile-c}
}

\caption{Three plots of expected profile shape (blue) and envelopes (yellow) as functions of radius for a sample power spectrum with different peak/trough heights. Horizontal dashed lines indicate the background values. A horizontal gray line indicates the height half-way between the stationary point and the background, while a vertical gray line indicates the estimated position of $r_{1/2}$. Dashed vertical gray lines indicate the estimated envelope $\Delta r_{1/2}$. All plots in this paper have been generated from the same model power spectrum as this plot, with $n=5$.} \label{fig:profiles}
\end{figure}

We now focus on the vicinity of large peaks, with $\bar{\nu}^2 \gg n$. In this limit, 
\begin{align}
\ev{\Phi_B(r)} &\simeq \nu^2 \rho_C(r)^2
\\
\Var(\Phi_B(r)) &\simeq 4 \nu^2 \sigma_0^2 \rho_C(r)^2 \left(1 - \rho_C(r)^2\right).
\end{align}
While the variance receives a boost from $\nu^2$, it is also suppressed by $1 - \rho_C(r)^2$, which vanishes at the origin. The ratio of the standard deviation to the mean is given by
\begin{align}
\frac{\sqrt{\Var(\Phi_B(r))}}{\ev{\Phi_B(r)}} \simeq \frac{2}{\bar{\nu}} \sqrt{\frac{1}{\rho_C(r)^2} - 1}
\end{align}
which suggests that large peaks very closely follow the expected peak profile. At $r_{1/2}$, the variance is approximately
\begin{align}
\Var(\Phi_B(r_{1/2})) &\approx \nu^2 \sigma_0^2.
\end{align}
Using this, a symmetric envelope with width $\Delta r_{1/2}$ about $r_{1/2}$ can be estimated by assuming a constant variance and a linear slope in $\Phi$ of $-\nu^2/(2 r_{1/2})$ from the peak to the half maximum, yielding
\begin{align}
\frac{\Delta r_{1/2}}{r_{1/2}} = \frac{2}{\bar{\nu}}.
\end{align}
The fact that this is small justifies the assumption of a constant variance. The estimated width $r_{1/2}$ and envelope $\Delta r_{1/2}$ are shown in Figure \ref{fig:profile-c}, where we see that they give an excellent approximation for the true values. However, note that the envelope is lopsided and slightly larger than our estimate on the right. For certain values of $n$ and $\nu$, $\Phi_B(r_{1/2})$ lies inside the background variance envelope, and so the right side of this envelope doesn't exist (see Figure \ref{fig:profile-b}). As $\nu$ increases, the envelope estimation on both sides improves.

We now turn to troughs, where $\bar{\nu}^2 \ll n$. In this limit, the mean cannot be further approximated from Eq. \eqref{eq:meanapprox} without neglecting the $\bar{\nu}$ dependence entirely (which would get the wrong value at the origin), while the variance is well-approximated by
\begin{align}
\Var(\Phi_B(r)) &\approx 2 n \sigma_0^4 \left(1 - \rho_C(r)^2\right)^2.
\end{align}
Using Eq. \eqref{eq:varapprox2} for the variance, the ratio of the standard deviation to the mean is given by
\begin{align}
\frac{\sqrt{\Var(\Phi_B(r))}}{\ev{\Phi_B(r)}} 
=
\sqrt{\frac{2}{n}} \sqrt{1 - \frac{\nu^4 \rho_C(r)^4}{\ev{\Phi_B(r)}^2}}.
\end{align}
Once we get a short distance away from the origin, $\ev{\Phi_B(r)}$ begins to grow and $\rho_C(r)$ begins to fall, so this ratio becomes well-approximated by $\sqrt{2/n}$. This suggests that troughs at higher $n$ more closely follow the expected profile. At $r_{1/2}$, the variance given in the $\bar{\nu}^2 \ll n$ approximation reduces to
\begin{align}
\Var(\Phi_B(r_{1/2})) = \frac{n \sigma_0^4}{2}.
\end{align}
The envelope on the expected width is very lopsided with a long tail to the right, as shown in Figure \ref{fig:profile-a}. Hence, we only attempt to compute a ``left-envelope'' for $\Delta r_{1/2}$. The definition \eqref{eq:envelope_def} yields the relation
\begin{align}
\rho_C(r_{1/2} - \Delta r_{1/2})^2 \approx 1 - \frac{n}{2(n + \sqrt{2n})},
\end{align}
working in the $\nu\rightarrow 0$ limit. For small $r$, we can approximate $\rho_C(r)^2$ as a linear function, using $r = 0$ ($\rho_C(0) = 1$) and $r_{1/2}$ ($\rho_C(r_{1/2}) = 1/\sqrt{2}$) to construct the fit. We obtain
\begin{align}
\rho_C(r)^2 \approx 1 - \frac{r}{2 r_{1/2}}.
\end{align}
Substituting $r = r_{1/2} - \Delta r_{1/2}$ and solving for $\Delta r_{1/2}$, we find
\begin{align}
\frac{\Delta r_{1/2}}{r_{1/2}} = \frac{1}{1 + \sqrt{n/2}}.
\end{align}
This suggests that the approximate minimum trough width shrinks as $n$ increases (recall that $r_{1/2}$ depends only on $C(r)$ and not on $n$). We see that this estimation gives an excellent approximation for the position of the inner envelope in Figure \ref{fig:profile-a}.

The expected profiles and their envelopes are shown for a sample power spectrum in Figure \ref{fig:profiles} for three values of $\bar{\nu}$, where we use $n = 5$. The plot with $\bar{\nu} = 3.5$ has an accurately-predicted left-envelope, but the right-envelope doesn't exist. Note the decay to background values is roughly $1/r^2$ for $r \ge 1.5$ in all plots. While the expected profiles always monotonically approach the background values, the same cannot be said of the envelopes, which often go a little beyond the peak/trough value a short distance away from the origin.

\subsection{Aspherical Part}

In their seminal paper investigating peaks in Gaussian fields, Bardeen, Bond, Kaiser and Szalay \cite{Bardeen1986} used the eigenvalues of the Hessian matrix at the origin to form a ratio to measure how spherical peaks were, on average. We previously discussed how the Hessian is often dominated by high frequency noise, which means that this ratio isn't generally applicable. Furthermore, the Hessian only describes behavior at the origin, and does not provide a holistic picture of the extrema. Hence, in this section, we devise new metrics to measure how ``spherical'', that is, how close to spherically symmetric, our extrema are in expectation.

\begin{figure}[p]
\subfloat[$\bar{\nu} = 0.1$]{%
  \includegraphics[width=0.565\textwidth]{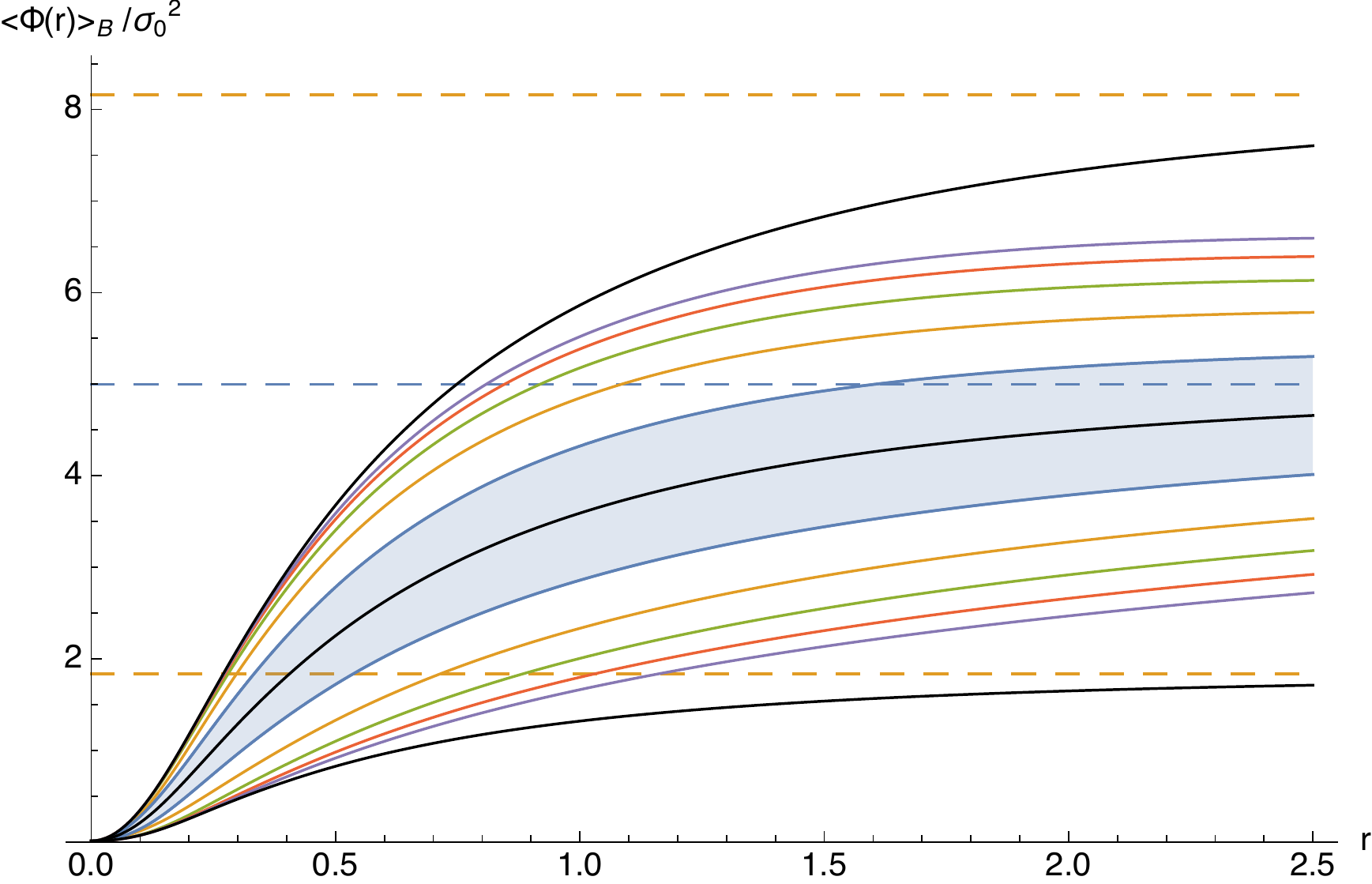}%
  \label{fig:stackprofile-a}
}

\subfloat[$\bar{\nu} = 3$]{%
  \includegraphics[width=0.565\textwidth]{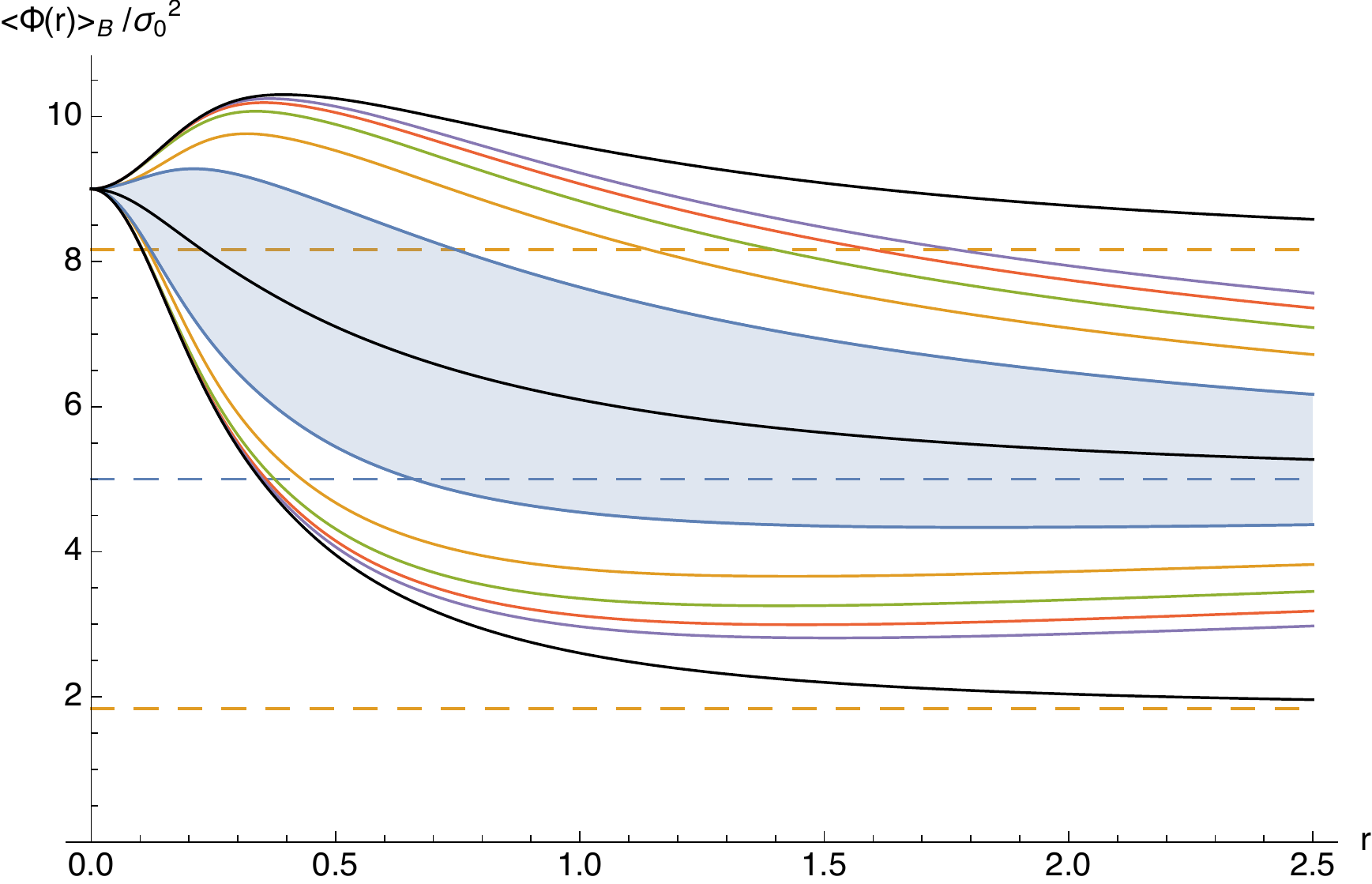}%
  \label{fig:stackprofile-b}
}

\subfloat[$\bar{\nu} = 10$]{%
  \includegraphics[width=0.565\textwidth]{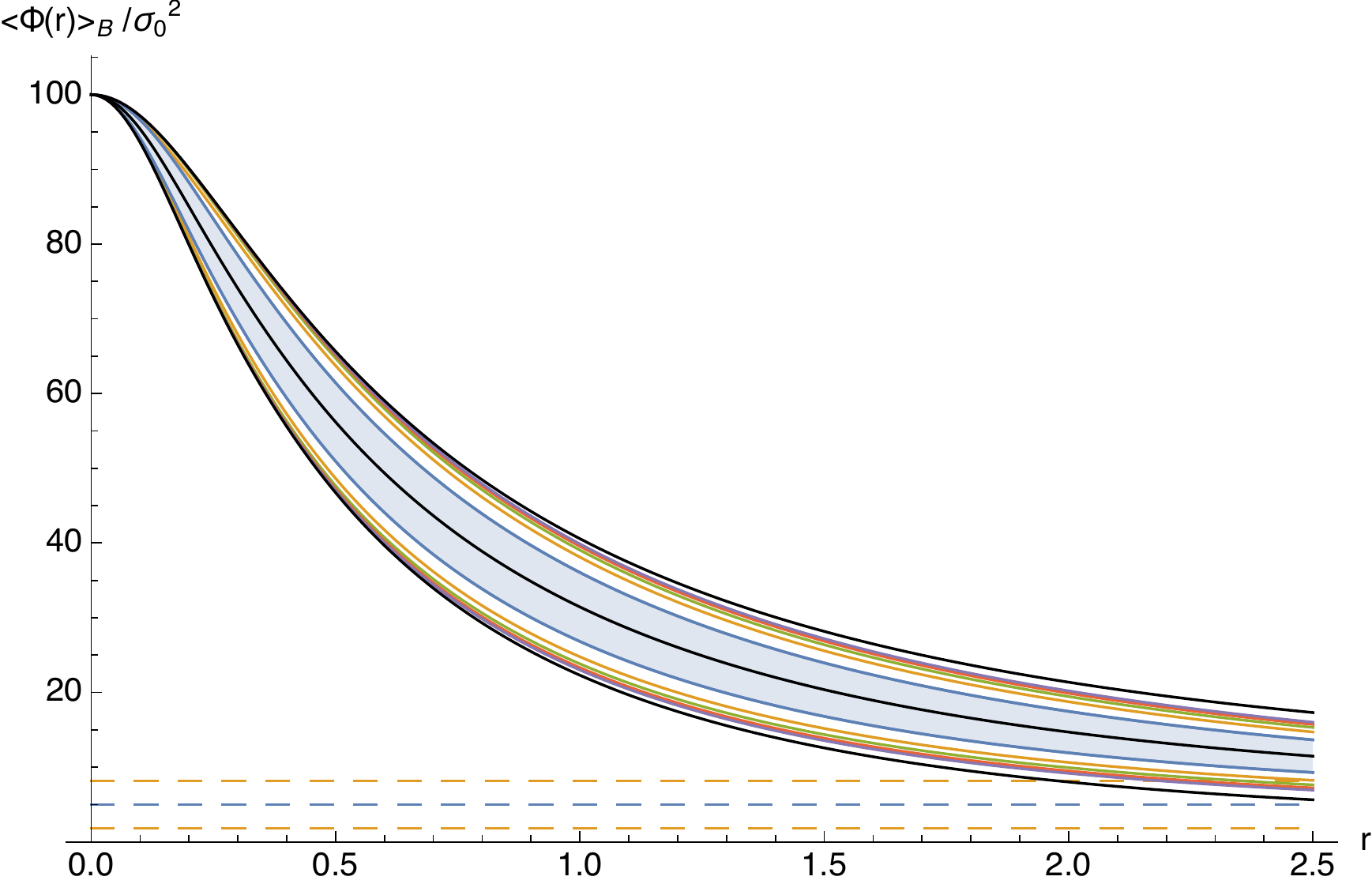}%
  \label{fig:stackprofile-c}
}

\caption{Reproduction of Figure \ref{fig:profiles} showing how the variance envelope is constructed from the variance of individual spherical harmonic modes, as in Eq. \eqref{eq:var_relation}. The mean profile and variance envelopes are plotted in black, with contributions up to $\ell = 4$ included in color. The variance envelope from $\ell = 0$ is shaded. Dashed horizontal lines indicate the background mean and envelope.} \label{fig:stackprofiles}
\end{figure}

The aspherical part of the profile (modes with $\ell > 0$) has zero mean for all radii. The variance of each mode has been computed in Eq. \eqref{eq:fullvar}, while the contribution of all modes to the variance at a point is given by Eq. \eqref{eq:Phi-var}. The relation between the two is given by Eq. \eqref{eq:var_relation}. To visualize these relations, we plot the variance envelope about a mean profile with different $\ell$ truncations in Eq. \eqref{eq:var_relation} in Figure \ref{fig:stackprofiles}. For concreteness, we use the same parameters as in Figure \ref{fig:profiles}. We see that as $\ell_{\rm max}$ increases, the combined variance approaches the expected envelope, with higher and higher $\ell$ modes needed to reconstruct the envelope as radius increases (we later show how to estimate an appropriate cutoff in $\ell$). Note that the proportion of the envelope that is due to variance in the spherical mode (the shaded region) depends on the peak height. As expected, the spherical variance is always the largest individual contributor, with higher $\ell$ modes adding successively less and less to the envelope.

\begin{figure}[p]
\subfloat[Peaks]{%
  \includegraphics[width=0.9\textwidth]{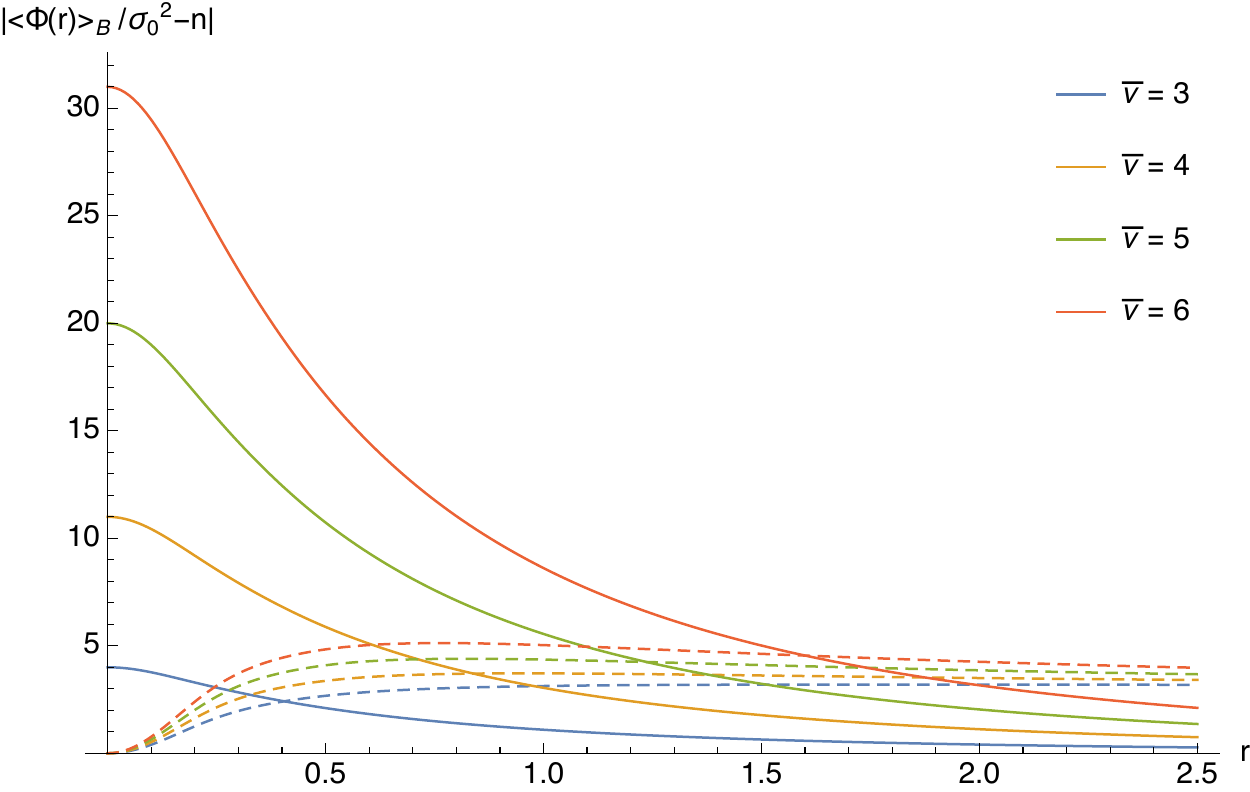}%
  \label{fig:varplot-a}
}

\subfloat[Troughs]{%
  \includegraphics[width=0.9\textwidth]{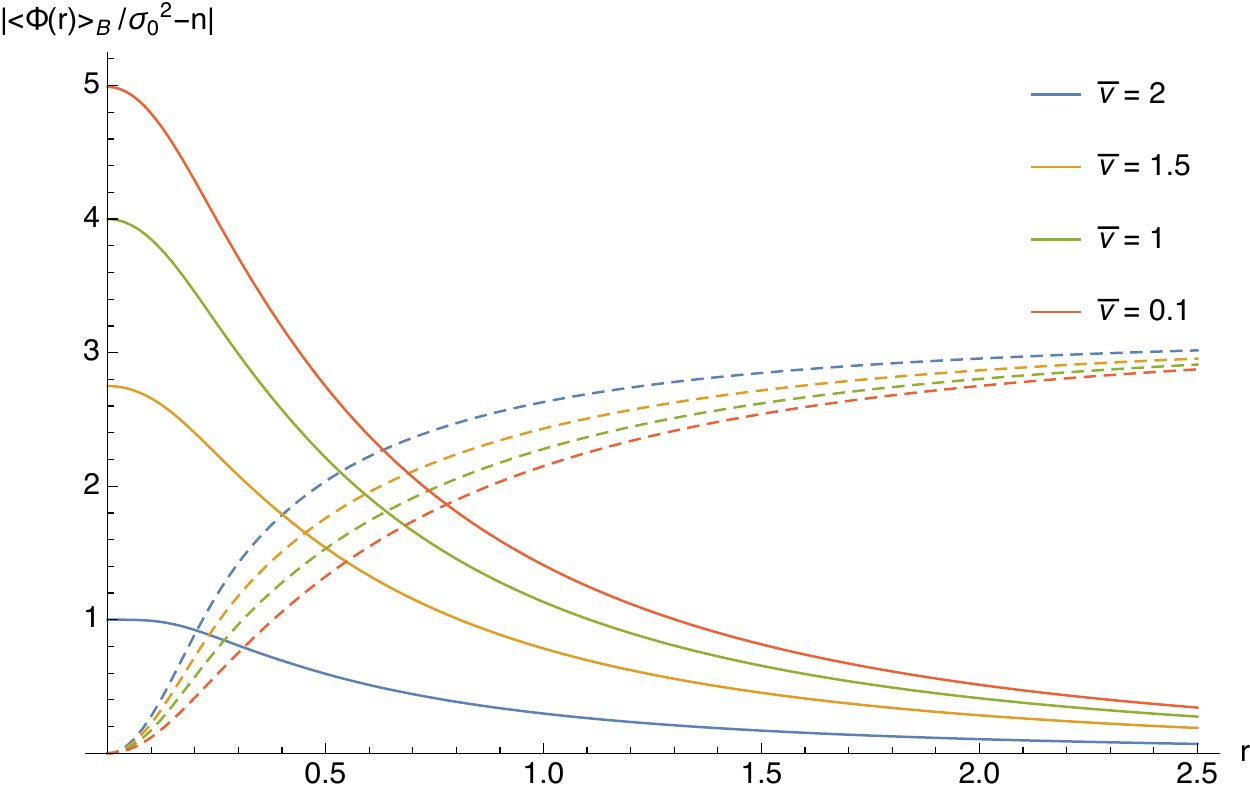}%
  \label{fig:varplot-b}
}

\caption{Plots of expected deviation from background for different values of $\bar{\nu}$ (solid lines) and the corresponding aspherical variance envelope $\sigma_{\rm as}(r)$ (dashed lines). In this plot, $n = 5$, so the background level is $\bar{\nu} = \sqrt{5}$. Note that the peaks and troughs have essentially the same shape; this is a reflection of the dependence on $\bar{\nu}$ in Eq. \eqref{eq:meanapprox}.} \label{fig:varplots}
\end{figure}

A heuristic for how spherical we expect a peak to be emerges from these plots: when does the variance associated with aspherical modes overwhelm the nature of the extrema? To quantify the expected ``aspherical variance'', consider a specific realization from the biased ensemble, $\hat{\Phi}(\vec{r}\,)$. The spatial average $\bar{\Phi}(r)$ of this realization $\hat{\Phi}$ at a radius $r$ is given by
\begin{align}
\bar{\Phi}(r) = \frac{1}{4\pi} \int d\hat{r} \, \hat{\Phi}(r \hat{r}).
\end{align}
The spatial variance of $\hat{\Phi}$ over this sphere is given by
\begin{align}
\Var(\hat{\Phi}(\vec{r}\,))_{\rm sp} = \frac{1}{4\pi} \int d\hat{r} \, \hat{\Phi}(r \hat{r})^2 - \bar{\Phi}(r)^2.
\end{align}
Let us write $\hat{\Phi}(\vec{r}\,)$ using a spherical harmonic decomposition as
\begin{align}
\hat{\Phi}(\vec{r}\,) = \sum_{\ell=0}^{\infty} \sum_{m=-\ell}^{\ell} \hat{\Phi}_{\ell m}(r) Y_{\ell m}(\hat{r}).
\end{align}
The spatial statistics of $\hat{\Phi}$ then become
\begin{align}
\bar{\Phi}(r) &= \frac{\hat{\Phi}_{00}(r)}{\sqrt{4\pi}}
\\
\Var(\hat{\Phi}(\vec{r}\,))_{\rm sp} 
%&= 
%\frac{1}{4\pi} \sum_{\ell=0}^{\infty} \sum_{m=-\ell}^{\ell} \hat{\Phi}_{\ell m}(r)^2 - \frac{\hat{\Phi}_{00}(r)^2}{4\pi}
%\\
&=
\frac{1}{4\pi} \sum_{\ell=1}^{\infty} \sum_{m=-\ell}^{\ell} \hat{\Phi}_{\ell m}(r)^2.
\end{align}
We now compute the statistical expectation value of the spatial variance in $\hat{\Phi}$ to obtain
\begin{align}
\Bev{\Var(\hat{\Phi}(\vec{r}\,))_{\rm sp}} 
%&=
%\frac{1}{4\pi} \sum_{\ell=1}^{\infty} \sum_{m=-\ell}^{\ell} \Bev{\hat{\Phi}_{\ell m}(r)^2}
%\\
&=
\frac{1}{4\pi} \sum_{\ell=1}^{\infty} \sum_{m=-\ell}^{\ell} \Var(\Phi_{B, \ell m}(r))
\equiv \sigma_{\rm as}(r)^2
\end{align}
which we define as the aspherical variance $\sigma_{\rm as}^2$, which represents the (statistically) expected spatial variance over a shell at radius $r$. Intuitively, the smaller this variance, the more spherical $\hat{\Phi}(\vec{r}\,)$ will be at radius $r$. Note that $\sigma_{\rm as}^2$ also has the straightforward interpretation of the expected (statistical) variance from all modes $\ell \ge 1$ at a given radius (c.f. the full variance Eq. \eqref{eq:var_relation}), which is visualized in Figure~\ref{fig:stackprofiles}. Using the full variance \eqref{eq:Phi-var} and Eq. \eqref{eq:var_relation}, $\sigma_{\rm as}^2(r)$ can be rewritten as
\begin{align} \label{eq:sigma_as}
\sigma_{\rm as}^2(r) = \Var(\Phi_B(\vec{r}\,)) - \frac{\Var(\Phi_{B, 00}(r))}{4\pi}.
\end{align}
We plot the deviation from the background $|\Bev{\Phi(r)}/\sigma_0^2 - n|$ along with the envelope from aspherical variance for a variety of values of $\bar{\nu}$ in Figure \ref{fig:varplots}. In these plots, the radius at which the envelope from aspherical modes has the same magnitude as the expected spherical deviation from the background can be seen from the intersection of the curves. Note that the crossover point moves further out as $\bar{\nu}$ deviates further from the background value of $\sqrt{n}$.

This provides a well-defined radius $r_{\rm sph}$ at which we expect aspherical deviations to begin to dominate. The ratio $\beta = r_{\rm sph} / r_{1/2}$ then gives us an indication of how spherical we expect peaks to be. For $\beta \gg 1$, we expect very spherical events, while for $\beta \ll 1$, we expect very aspherical events. Note that $\beta$ will be strongly dependent upon $\bar{\nu}$.

To compare $\sigma_{\rm as}^2$ to the mean deviation from background, we define the \textit{asphericity}
\begin{align}
\mathrm{As}(r) = \frac{\sigma_{\rm as}^2(r)}{\sigma_{\rm as}^2(r) + (\ev{\Phi_B(\vec{r}\,)} - \ev{\Phi(\vec{r}\,)})^2}.
\end{align}
By construction, $0 \le \mathrm{As}(r) \le 1$, with 0 indicating that there is zero variance at radius $r$ (perfectly spherical), and 1 indicating that the spherical component is expected to vanish, so that all aspherical perturbations dominate over the spherical component. We also have $\mathrm{As}(r_{\rm sph}) = 0.5$ by construction. The asphericity then suggests a natural metric for ``how spherical'' extrema are in expectation: given $\mathrm{As}(r_{1/2})$, values between 0 and 0.5 indicate mostly spherical extrema, and values between 0.5 and 1 indicate mostly aspherical extrema.

For the purpose of computing $\mathrm{As}(r)$, we use Eq. \eqref{eq:sigma_as} to write
\begin{align}
\mathrm{As}(r) &= 
\frac{4 \pi \Var(\Phi_B(\vec{r}\,)) - \Var(\Phi_{B, 00}(r))}{4 \pi \Var(\Phi_B(\vec{r}\,)) - \Var(\Phi_{B, 00}(r)) + 4\pi (\ev{\Phi_B(\vec{r}\,)} - n \sigma_0^2)^2}.
\end{align}
We can also understand the contributions to the asphericity from each mode by writing
\begin{align}
\mathrm{As}(r) &= 
\frac{\sum_{\ell=1}^{\infty} \sum_{m=-\ell}^{\ell} \Var(\Phi_{B, \ell m}(r))}{4 \pi \Var(\Phi_B(\vec{r}\,)) - \Var(\Phi_{B, 00}(r)) + 4\pi (\ev{\Phi_B(\vec{r}\,)} - n \sigma_0^2)^2}.
\end{align}

\begin{figure}[p]
\subfloat[Peaks]{%
  \includegraphics[width=0.85\textwidth]{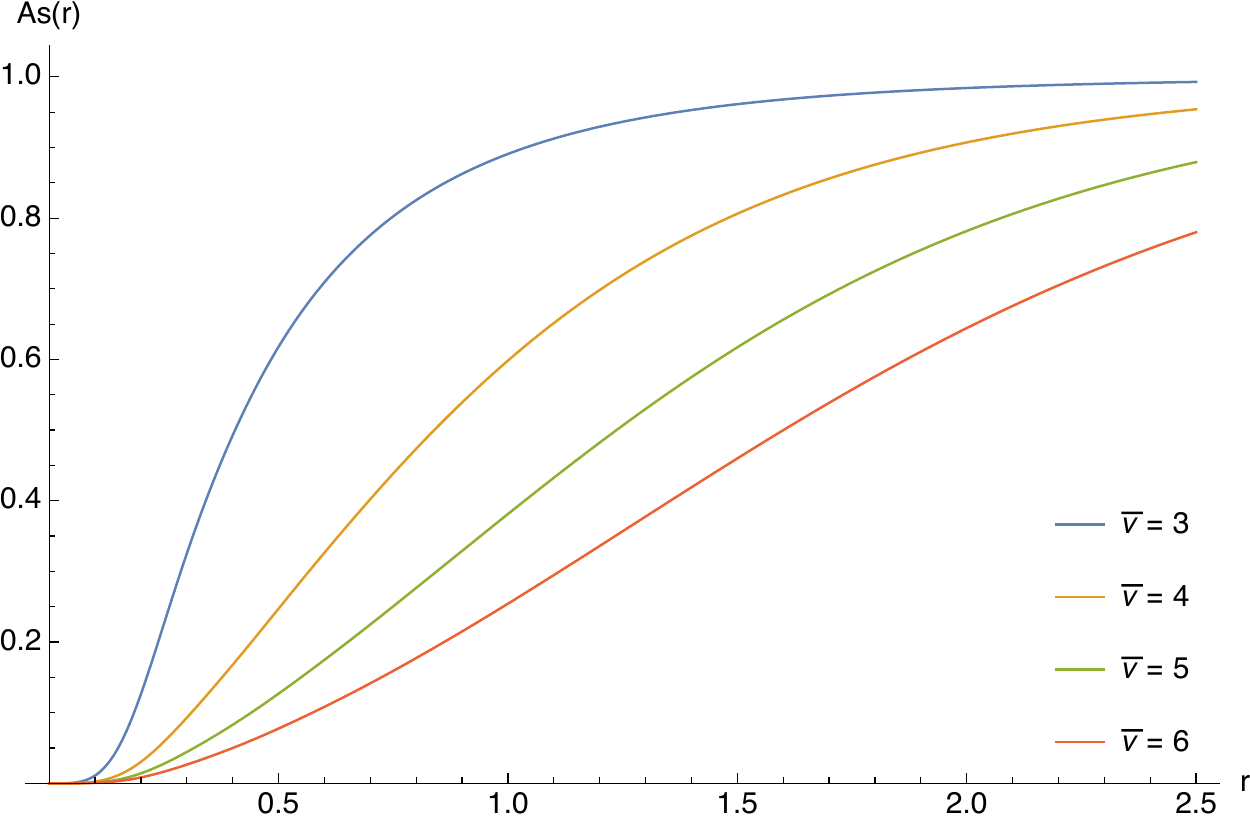}%
  \label{fig:Asplot-a}
}

\subfloat[Troughs]{%
  \includegraphics[width=0.85\textwidth]{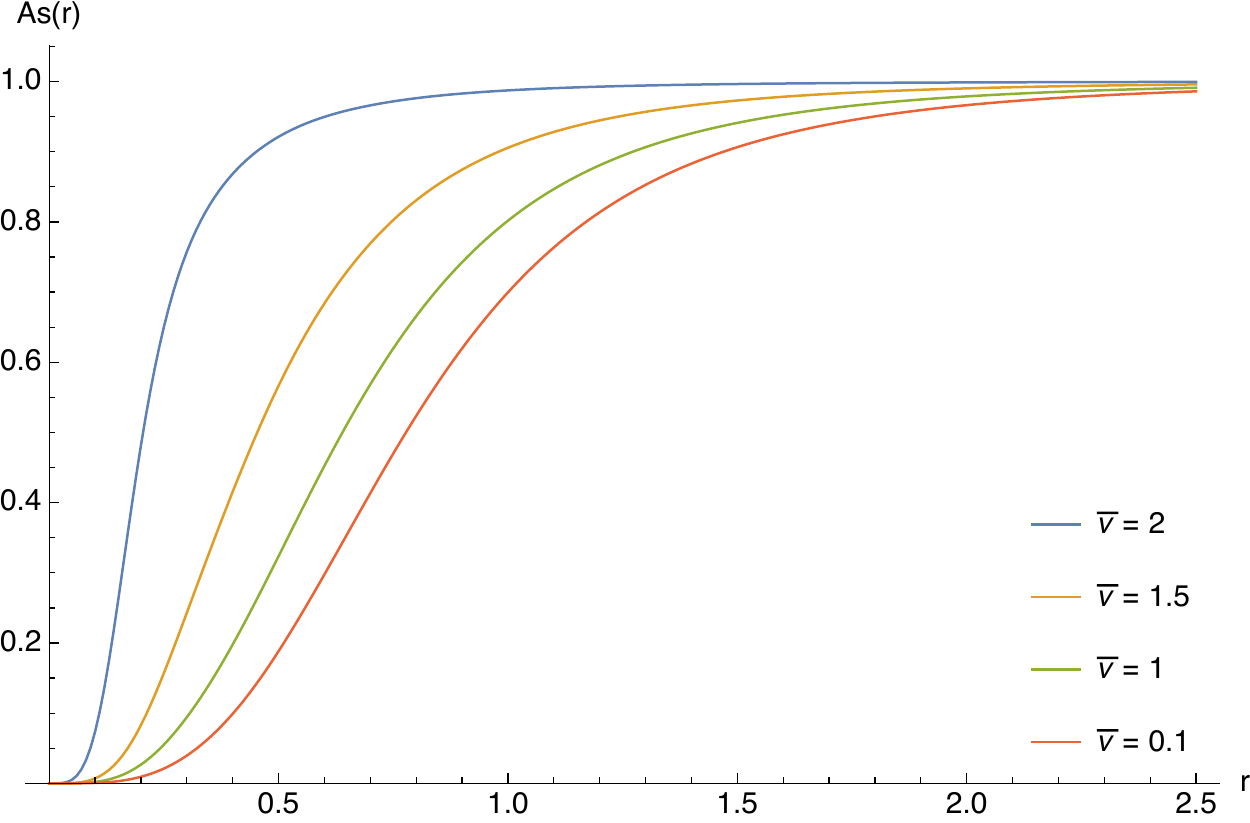}%
  \label{fig:Asplot-b}
}

\caption{Plots of asphericity as a function of radius for different values of $\bar{\nu}$. For values above 0.5, the profiles are mostly aspherical, while for values below 0.5, the profiles are mostly spherical. We see that as the extrema deviate further from the background value of $\bar{\nu} = \sqrt{n} = \sqrt{5}$, the profiles are more spherical for larger radii. Recall that $r_{1/2} \approx 0.55$ for these profiles.}
\label{fig:Asplots}
\end{figure}

To obtain an analytic estimate of $\mathrm{As}(r)$, we approximate
\begin{align}
\ev{\Phi_B(r)} &\approx n \sigma_0^2 \left[1 + \left(\frac{\bar{\nu}^2}{n} - 1\right) \rho_C(r)^2\right]
\\
\Var(\Phi_B(\vec{r}\,)) &\approx 2 n \sigma_0^4 \left(1 - \rho_C(r)^2\right) \left[1 + \left(2 \frac{\bar{\nu}^2}{n} - 1\right) \rho_C(r)^2 \right]
\end{align}
as in Eqs. \eqref{eq:meanapprox} and \eqref{eq:varapprox} above. We further approximate Eq. \eqref{eq:fullvar} to obtain
\begin{align}
\Var(\Phi_{B, 00}(r))
\approx
8 \pi n \sigma_0^4 \left[
\tilde{\rho}_{C2}(r)
+ 2 \rho_C(r)^2 \left(\frac{\bar{\nu}^2}{n} - 1\right) \tilde{\rho}_C(r) 
+ \rho_C(r)^4 \left(1 - 2 \frac{\bar{\nu}^2}{n}\right)\right]
\end{align}
where we let
\begin{align}
\tilde{\rho}_C(r) = \frac{\tilde{C}_0(r, r)}{\sigma_0^2} &= \frac{1}{2} \int_{-1}^1 du \, \rho_C\left(r \sqrt{2(1-u)}\right)
\\
\tilde{\rho}_{C2}(r) &= \frac{1}{2} \int_{-1}^1 du \, \rho_C\left(r\sqrt{2(1-u)}\right)^2.
\end{align}
While $\tilde{\rho}_C(r)$ is technically bounded between -1 and 1, under our assumption of a power spectrum with broad support, we expect $\rho_C(r)$ to be monotonically decreasing with radius, which bounds $\tilde{\rho}_C(r)$ between 0 and 1. $\tilde{\rho}_{C2}(r)$ is always bounded between 0 and 1. Both are equal to 1 for $r = 0$ and decay towards 0 as $r \to \infty$. Putting everything together, we obtain
\begin{align}
\Var(\Phi_B(\vec{r}\,)) - \frac{1}{4\pi} \Var(\Phi_{B, 00}(r))
&\approx
2 n \sigma_0^4 \left(
1 
- \tilde{\rho}_{C2}(r)
+ 2 \rho_C(r)^2 \left(\frac{\bar{\nu}^2}{n} - 1\right) (1 - \tilde{\rho}_C(r))
\right)
\end{align}
and
\begin{align}
\mathrm{As}(r) \approx
\frac{1 - \tilde{\rho}_{C2}(r)
+ 2 \rho_C(r)^2 \left(\frac{\bar{\nu}^2}{n} - 1\right) (1 - \tilde{\rho}_C(r))}
{1 - \tilde{\rho}_{C2}(r)
+ 2 \rho_C(r)^2 \left(\frac{\bar{\nu}^2}{n} - 1\right) (1 - \tilde{\rho}_C(r))
+ \frac{n}{2} \left(\frac{\bar{\nu}^2}{n} - 1\right)^2 \rho_C(r)^4}.
\end{align}
Around peaks with large $\bar{\nu}^2/n$, this is roughly
\begin{align}
\mathrm{As}(r) \approx \frac{4 (1 - \tilde{\rho}_C(r))}{\bar{\nu}^2 \rho_C(r)^2} 
\le \frac{4}{\bar{\nu}^2 \rho_C(r)^2}.
\end{align}
This indicates that the asphericity is very small for large peaks, but initially grows as $\sim r^2$. At $r_{1/2}$, we have
\begin{align}
\mathrm{As}(r_{1/2}) \leq \frac{8}{\bar{\nu}^2}.
\end{align}
Around troughs with small $\bar{\nu}^2/n$, the story is quite different.
\begin{align}
\mathrm{As}(r) \approx
\frac{1 - \tilde{\rho}_{C2}(r)
- 2 \rho_C(r)^2 (1 - \tilde{\rho}_C(r))}
{1 - \tilde{\rho}_{C2}(r)
- 2 \rho_C(r)^2 (1 - \tilde{\rho}_C(r))
+ \frac{n}{2} \rho_C(r)^4}.
\end{align}
Here, the asphericity depends strongly on $n$ and how quickly the $n \rho_C(r)^4$ term decays compared to the other terms in the denominator. At $r_{1/2}$, we have
\begin{align}
\mathrm{As}(r_{1/2}) \sim \left(1 + \frac{n}{8(\tilde{\rho}_C(r_{1/2}) - \tilde{\rho}_{C2}(r_{1/2}))}\right)^{-1}.
\end{align}
While $\tilde{\rho}_C(r_{1/2}) - \tilde{\rho}_{C2}(r_{1/2})$ is complicated and requires numerics to fully characterize, it is bounded between -1 and 1 (0 and 1 for $\rho_C(r)$ monotonically decreasing), so
\begin{align}
\mathrm{As}(r_{1/2}) \lesssim \left(1 + \frac{n}{8}\right)^{-1}.
\end{align}
This suggests that troughs are mostly spherical at $r_{1/2}$ for $n \ge 8$, with numerics required for more accuracy on smaller $n$.

We plot $\mathrm{As}(r)$ for a variety of values of $\bar{\nu}^2$ in Figure \ref{fig:Asplots}.

\section{Spherical Modes of Gaussian Random Fields} \label{sec:sampling}

Thus far, we have studied $\chi^2$ fields from a statistical perspective, identifying useful quantities and computing their expectation values with respect to various ensembles of field realizations. We now begin turning our attention to the field realizations themselves, beginning with a discussion of how one can in principle sample discrete approximations of realizations of a biased $\chi^2$ field.

In Section~\ref{sec:ensembles}, we discussed two equivalent viewpoints on the calculation of biased expectation values. There is an analogous pair describing the process of sampling biased realizations from a random field. On one hand, we could describe a sampling process which picks out only realizations which satisfy a given set of constraints, but which acts on the full, unbiased random field. This can be accomplished by drawing random samples that are then projected onto the constraint hypersurface. On the other, we could construct the same biased random field as before, and then an unbiased sampling procedure is guaranteed to return realizations which satisfy the constraints. In what follows, we will take this second, ``biased field'' approach, which we found to be somewhat simpler than the projection approach.

We will show that the constraints of a stationary point of fixed amplitude in the $\chi^2$ field $\Phi$ can be translated to constraints on the harmonic coefficients $\phi^\alpha_{\ell m}(r)$ of its underlying Gaussian fields, which are themselves Gaussian. It follows that the PDFs of these coefficients are determined by their one- and two-point functions. We will compute these functions for the biased coefficient fields $\phi^\alpha_{B,\ell m}(r)$ using the methods of Section~\ref{sec:biasing}. Finally, we will use the resulting PDFs to generate samples of the biased Gaussian coefficient fields over a discrete radial grid, and then show how to assemble these into samples of the biased $\chi^2$ field $\Phi_{B}$.

The unbiased Gaussian fields underlying the $\chi^2$ field $\Phi$ are centered. Their distributions are therefore governed by their two-point functions. The two-point functions, in turn, are determined by a power spectrum in Fourier space. Constructing discrete approximations to the field in space is straightforward, and Fast Fourier Transforms (FFTs) allow us to convert between Fourier space and real space very efficiently. However, this technique only allows us to draw from the full ensemble of fields, not from a biased ensemble. This means that the projection technique mentioned above must be used in this situation. The alternative is to work directly in position space, where the constraints are imposed at a chosen grid point, and all other grid points are sampled based on the appropriate biased means and covariances.

Both of these approaches require computing the expected means at every grid point, and the covariances between every grid point (in the projection technique, the projection matrices are constructed from the means and covariances). This rapidly becomes infeasible as the grid resolution increases. Let us assume that we will sample a Gaussian field in a box with a resolution of $N$ grid points per edge. This yields $N^3$ grid points in total, which would require a covariance matrix with $N^6$ entries. If we take a reasonably low resolution with $N = 100$, we are left with $10^{12}$ entries in the covariance matrix, which, in double precision, would require approximately 8 terabytes of storage. Worse, in order to draw samples using this covariance matrix, we would need to construct its Cholesky decomposition. In short, we need a better method.

Sampling an entire cubic grid is infeasible because the number of entries in the covariance matrix is given by the number of grid points, $N^3$, squared. It would seem that this is the best we can do when working in a three dimensional space. As it turns out, by expanding the fields in spherical harmonics and sampling their harmonic coefficients over a one-dimensional radial grid, we can construct field realizations much more efficiently.

The spherical approach is advantageous because all of the fields we consider have spherically symmetric PDFs. We once again invoke the result from Appendix~\ref{app:spherical_covariance} that the spherical harmonic coefficients of a field with rotational symmetry have zero covariance across mode and azimuthal number. Taking a radial grid with $N$ grid points, we can sample each mode individually using a covariance matrix containing only $N^2$ entries. The storage requirement then scales as $\ell_{\mathrm{max}} N^2$, as each $\ell$ mode is described by a distinct covariance matrix, and these matrices are independent of $m$. For example, if we let $N = 50$ (working in radius, not diameter) and generously take $\ell_{\mathrm{max}} = 50$, all of the covariance matrices can be stored in approximately a megabyte of memory using double precision\footnote{From an information theory perspective, one can ask what extra information is being stored in the cubic lattice. The answer is that the cubic covariance matrix has a massive degeneracy of information due to not taking advantage of the spherical symmetry of the system. If the system were not spherically symmetric, then one can have nontrivial correlations between each $\ell$ and $m$ mode in a spherical harmonic expansion, leading straight back to the dense covariance matrix of the cubic lattice.}. Evidently, much higher resolution samples are possible using this method.

We approach our development of the sampling procedure in two steps. In this section, we begin by constructing the spherical harmonic decomposition of the underlying Gaussian fields, compute their unbiased and biased statistics, and investigate relations to previous results. In the following section, we discuss truncation and discretization, and construct the actual sampling procedure.

\subsection{Spherical Harmonic Decomposition of \texorpdfstring{$\phi$}{\phi}}

We begin with a statistical analysis of the spherical harmonic decomposition of a Gaussian field $\phi$. The results of this subsection apply equally to biased and unbiased fields. Expanding $\phi(\vec{r}\,)$ in spherical harmonics, we have
\begin{align} \label{eq:decomp_phi}
\phi(\vec{r}\,) = \sum_{\ell=0}^{\infty}\sum_{m=-\ell}^{\ell} \phi_{\ell m}(r) Y_{\ell m}(\hat{r}).
\end{align}
Using the orthonormality relation of the $Y_{\ell m}$, we can invert this equation to find the mode coefficients as functions of radius.
\begin{align} \label{eq:def_phi_mode}
\phi_{\ell m}(r) = \int d\hat{r} \, \phi(r\hat{r}) Y_{\ell m}(\hat{r})
\end{align}
Note that these mode coefficients are real functions due to the use of real spherical harmonics.

We are interested in the statistical properties of $\phi_{\ell m}(r)$. As $\phi(\vec{r}\,)$ is a Gaussian random field, and integrals of Guassian random fields are also Gaussian random fields, $\phi_{\ell m}(r)$ is also a Gaussian random field. Hence, its PDF is completely determined by its mean and covariance (or equivalently, its one- and two-point functions).

The mean is given by
\begin{align} \label{eq:1point_phi_mode}
\ev{\phi_{\ell m}(r)} = \int d\hat{r} \, \ev{\phi(r \hat{r})} Y_{\ell m}(\hat{r}).
\end{align}
For our purposes, $\phi(\vec{r}\,)$ will always have an $SO(3)$ rotation symmetry about the origin, and so $\ev{\phi(\vec{r}\,)}$ will only depend on radius. The angular integral can then be taken over the spherical harmonic, yielding
\begin{align} \label{eq:1point_phi_mode_result}
\ev{\phi_{\ell m}(r)} = \delta_{\ell 0} \delta_{m 0} \sqrt{4 \pi} \ev{\phi(\vec{r}\,)}.
\end{align}
The two-point function is given by
\begin{align}
\ev{\phi_{\ell m}(r) \phi_{\ell' m'}(r')} = \int d\hat{r} \, d\hat{r}' \, Y_{\ell m}(\hat{r}) Y_{\ell' m'}(\hat{r}') \ev{\phi(r \hat{r}) \phi(r' \hat{r}')},
\end{align}
and the covariance by
\begin{align} \label{eq:phi_lm_cov}
\Cov(\phi_{\ell m}(r), \phi_{\ell' m'}(r')) = \int d\hat{r} \, d\hat{r}' \, Y_{\ell m}(\hat{r}) Y_{\ell' m'}(\hat{r}') \Cov(\phi(r \hat{r}), \phi(r' \hat{r}')).
\end{align}
We know from Appendix~\ref{app:spherical_covariance} that the two-point function and the covariance will both be proportional to $\delta_{\ell \ell'} \delta_{m m'}$ and otherwise independent of $m$.

\subsection{Unbiased Statistics of \texorpdfstring{$\phi$}{\phi}}

In the unbiased case, the Gaussian fields $\phi^\alpha_{\ell m}(r)$ are described by
\begin{align} \label{eq:phi_one_point}
\ev{\phi^\alpha_{\ell m}(r)} &= 0
\\ \label{eq:phi_two_point}
\ev{\phi^\alpha_{\ell m}(r), \phi^\beta_{\ell' m'}(r')} &= \delta^{\alpha \beta} \delta_{\ell\ell'} \delta_{mm'} 4 \pi \tilde{C}_{\ell}(r, r')
\end{align}
with $\tilde{C}_\ell$ defined in Eq. \eqref{eq:Ctilde1} and Appendix~\ref{app:kspace_integrals}. Note that when $r' = 0$, $\tilde{C}_\ell(r, 0) = \delta_{\ell 0} C(r)$ \eqref{eq:tildeCr0}. Modes with differing $\ell$ and $m$ are uncorrelated (and hence independent), and the two-point function is otherwise independent of $m$. It is worth stating explicitly that $\phi_{\ell m}(r)$ is an \textit{inhomogeneous} random field, as is evident from the form of \eqref{eq:phi_two_point}. Intuitively, we have broken translation invariance by choosing an origin around which to perform the harmonic decomposition. We also note that because the underlying $\phi^\alpha$ fields are independent, the modes $\phi^\alpha_{\ell m}$ and $\phi^\beta_{\ell m}$ are also independent for $\alpha \neq \beta$.

By construction, $C(\vec{r}\,)$ is the covariance of a Gaussian field between the origin and $\vec{r}$. Similarly, $\tilde{C}_\ell(r, r')$ is the covariance of a spherical harmonic mode of a Gaussian field (up to constant factors). We now construct the explicit relationship between the two.

Begin by writing $C(\vec{r} - \vec{r}\,')$ as
\begin{align}
C(\vec{r} - \vec{r}\,') = \int d^3 k \, e^{i \vec{k} \cdot \vec{r}} e^{- i \vec{k} \cdot \vec{r}\,'} \Pk(k).
\end{align}
We now use the plane wave expansion \eqref{eq:planewaveexp} twice, and write $d^3 k = k^2 \, dk \, d\hat{k}$.
\begin{align}
C(\vec{r} - \vec{r}\,') = 
\sum_{\ell, \ell' = 0}^\infty i^{\ell - \ell'} (2 \ell + 1) (2 \ell' + 1) \int dk \, k^2 \Pk(k) j_{\ell}(kr) j_{\ell'}(kr') 
\int d\hat{k} \, P_\ell(\vec{k} \cdot \vec{r}\,) P_{\ell'}(\vec{k} \cdot \vec{r}\,')
\end{align}
To evaluate the angular integral, we use the addition theorem \eqref{eq:addition} twice, integrate over $d\hat{k}$, and then use the addition theorem once again. The result is
\begin{align}
\int d\hat{k} \, P_\ell(\vec{k} \cdot \vec{r}\,) P_{\ell'}(\vec{k} \cdot \vec{r}\,')
%&=
%\frac{16 \pi^2}{(2 \ell + 1)(2 \ell' + 1)} \sum_{m = -\ell}^\ell \sum_{m' = - \ell'}^{\ell'} Y_{\ell m} (\hat{r}) Y_{\ell' m'}(\hat{r}') \int d\hat{k} \, Y_{\ell m}(\hat{k}) Y_{\ell' m'}(\hat{k})
%\\
&=
\delta_{\ell \ell'} \frac{4 \pi}{2 \ell + 1} P_\ell(\hat{r} \cdot \hat{r}').
\end{align}
We thus obtain our desired result,
\begin{align} \label{eq:CCtilderesult}
C(\vec{r} - \vec{r}\,') = \sum_{\ell = 0}^\infty (2 \ell + 1) P_\ell(\hat{r} \cdot \hat{r}') \tilde{C}_{\ell}(r, r').
\end{align}

Equation \eqref{eq:CCtilderesult} is fundamental to understanding how a Gaussian field is constructed from the sum of its spherical harmonics, and it can be manipulated into a number of useful relations. We can immediately invert the relationship by using the orthogonality of the Legendre polynomials. Writing $|\vec{r} - \vec{r}\,'| = \sqrt{r^2 + r'^2 - 2 r r' \cos \gamma}$, we obtain
\begin{align} \label{eq:better_ctilde}
\tilde{C}_\ell(r, r') = \frac{1}{2} \int_{-1}^1 du \, P_\ell(u) C\left(\sqrt{r^2 + r'^2 - 2 r r' u}\right).
\end{align}
This gives us an alternative prescription for computing $\tilde{C}_\ell$ (c.f. Eq. \eqref{eq:ctildedef}).

The limit $\vec{r}\,' = 0$ in Eq. \eqref{eq:CCtilderesult} offers us no new information, as $\tilde{C}_\ell(r, 0) = \delta_{\ell 0} C(r)$. However, in the coincidence limit $\vec{r} = \vec{r}\,'$, we obtain\footnote{This result can be derived directly from the integral definitions of $\sigma_0^2$ and $\tilde{C}_\ell(r, r)$ by using the sum
\begin{align*}
1 = \sum_{\ell = 0}^{\infty} (2\ell + 1) j_\ell(kr)^2,
\end{align*}
which can itself be derived from the plane wave expansion \eqref{eq:planewaveexp} by taking the mod square of both sides, integrating over $\sin \theta \, d \theta$ from $0$ to $\pi$ and using the orthonormality relation for Legendre polynomials \eqref{eq:legendre-ortho}.}
\begin{align} \label{eq:sigma_sum}
\sigma_0^2 = \sum_{\ell = 0}^\infty (2 \ell + 1) \tilde{C}_{\ell}(r, r).
\end{align}
This is a fairly remarkable result: the LHS is constant, while every term on the RHS depends on $r$, with the sum adding up to be independent of $r$. At the origin, only the $\ell = 0$ mode contributes, as $\tilde{C}_{\ell}(0, 0) = \delta_{\ell 0} \sigma_0^2$. As radius increases, more and more $\ell$ modes become important in the sum.

An interesting relation can be obtained by squaring both sides of Eq. \eqref{eq:CCtilderesult} and writing $\hat{r} \cdot \hat{r}' = \cos \gamma = u$.
\begin{align}
C\left(\sqrt{r^2 + r'^2 - 2 r r' u}\right)^2 = 
\sum_{\ell_1, \ell_2 = 0}^\infty (2 \ell_1 + 1) (2 \ell_2 + 1) \tilde{C}_{\ell_1}(r, r')
\tilde{C}_{\ell_2}(r, r')
P_{\ell_1}(u) 
P_{\ell_2}(u)
\end{align}
Multiply both sides by $P_\ell(u)$ and integrate from $u = -1$ to 1.
\begin{align} \label{eq:crazyformula}
\int_{-1}^{1} du \, P_\ell(u) C\left(\sqrt{r^2 + r'^2 - 2 r r' u}\right)^2 = 
\sum_{\ell_1, \ell_2 = 0}^\infty (2 \ell_1 + 1) (2 \ell_2 + 1) \tilde{C}_{\ell_1}(r, r')
\tilde{C}_{\ell_2}(r, r')
\int_{-1}^{1} du \,
P_\ell(u)
P_{\ell_1}(u) 
P_{\ell_2}(u)
\end{align}
The LHS here appears on the RHS of Eq. \eqref{eq:firstuglyintegral} and again in Eq. \eqref{eq:fullcov}. The integral of a product of three Legendre polynomials is known as a Gaunt integral, and the analytic result can be written in terms of the Wigner 3-$j$ symbol. A special case is $\ell = 0$, which can be computed using the orthogonality of Legendre polynomials to give
\begin{align}\label{eq:funnyrelation}
\int_{-1}^{1} du \, C\left(\sqrt{r^2 + r'^2 - 2 r r' u}\right)^2 = 
2 \sum_{\ell}^\infty (2 \ell + 1) \tilde{C}_{\ell}(r, r')^2.
\end{align}

\subsection{Constructing \texorpdfstring{$\chi^2$}{chi-squared} fields from \texorpdfstring{$\phi$}{\phi} - Unbiased case} \label{sec:unbiasedconstruction}

When constructing $\Phi$ from the Gaussian fields, we simply need to square Eq. \eqref{eq:decomp_phi} and sum.
\begin{align} \label{eq:Phi_from_phi}
\Phi(\vec{r}\,) = \sum_{\alpha = 1}^n \left(\sum_{\ell=0}^{\infty}\sum_{m=-\ell}^{\ell} \phi^\alpha_{\ell m}(r) Y_{\ell m}(\hat{r})\right)^2
\end{align}
We now set about understanding how the statistics of $\Phi$ are recovered from the statistics of $\phi^\alpha_{\ell m}$.

Consider the expectation value
\begin{align} \label{eq:background_exp}
\ev{\Phi(\vec{r}\,)} = \sum_{\alpha = 1}^n \sum_{\ell, \ell' = 0}^\infty \sum_{m = -\ell}^\ell \sum_{m' = -\ell'}^{\ell'} Y_{\ell m}(\hat{r}) Y_{\ell' m'}(\hat{r}) \ev{\phi^\alpha_{\ell m}(r) \phi^\alpha_{\ell' m'}(r)}.
\end{align}
As the fields are all identical, the sum over $\alpha$ yields a factor of $n$. Substituting the two-point function \eqref{eq:phi_two_point} collapses the doubled sums down to
\begin{align}
\ev{\Phi(\vec{r}\,)} = 4 \pi n \sum_{\ell = 0}^\infty \sum_{m = -\ell}^\ell Y_{\ell m}(\hat{r}) Y_{\ell m}(\hat{r}) \tilde{C}_\ell(r, r).
\end{align}
We now apply Uns\"old's theorem \eqref{eq:unsold} to compute the sum over $m$. The result is
\begin{align} \label{eq:unbiased_decomp}
\ev{\Phi(\vec{r}\,)} = n \sum_{\ell = 0}^\infty (2 \ell + 1) \tilde{C}_\ell(r, r) = n \sigma_0^2
\end{align}
where we have used Eq. \eqref{eq:sigma_sum} in the second equality. This agrees with our previous computation of $\ev{\Phi(\vec{r}\,)}$ \eqref{eq:unbiased_1point} as expected.

We can go one step further and compute the covariance $\Cov(\Phi(\vec{r}\,), \Phi(\vec{r}\,))$. The derivation is rather tedious, and details are relegated to Appendix~\ref{app:cov}. The result is
\begin{align}\label{eq:expandedcov}
\Cov(\Phi(\vec{r}\,), \Phi(\vec{r}\,'))
&= 2 n \left(\sum_{\ell = 0}^\infty (2 \ell + 1) \tilde{C}_\ell(r, r)\right)^2 = 2 n C(\vec{r} - \vec{r}\,')^2
\end{align}
where the second equality uses Eq. \eqref{eq:CCtilderesult}, and agrees with the background value \eqref{eq:unbiased_cov}.

\subsection{Biasing \texorpdfstring{$\phi^\alpha_{\ell m}(r)$}{spherical harmonics of \phi}}

Our next task is to determine how to impose the stationarity and field amplitude constraints on the spherical harmonic modes $\phi^\alpha_{\ell m}(r)$. We saw previously that the conditions $\Phi(0) = \nu^2$ and $\vec{\nabla}\Phi(0) = 0$ can be imposed by requiring
\begin{align}
\phi^1(0) = \nu
, \qquad
\vec{\nabla}\phi^1(0) = 0
, \qquad
\phi^\alpha(0) = 0 \ \ \text{for} \ \ \alpha \in \{2, \ldots, n\}.
\end{align}
We need to translate these into constraints on the spherical harmonic coefficients of the $\phi^\alpha$, which we can use to construct biased fields $\phi_{B,\ell m}^\alpha$. Sampling from these fields will automatically return coefficient realizations which we can use to construct biased realizations of $\Phi$. This analysis closely follows the methods of Section \ref{sec:biasing}.

We begin by noting that analyticity of $\phi(\vec{r}\,)$ at the origin requires $\phi_{\ell m}(r) \sim r^p$ at the origin, where $p\geq \ell$. This requirement can be understood by noting that $Y_{\ell m}$ can be written as a linear combination of monomials of degree $\ell$ of $x/r$, $y/r$ and $z/r$; sufficiently many powers of $r$ are required to offset the denominator. Hence, at $r=0$, only the $\ell = 0$ mode contributes, and we obtain the $n$ constraints
\begin{align}
\phi^1_{00}(0) = \sqrt{4\pi} \nu
, \qquad
\phi^\alpha_{00}(0) = 0 \ \ \text{for} \ \ \alpha \in \{2, \ldots, n\}.
\end{align}
Note that a factor of $\sqrt{4\pi}$ has been inserted to offset $Y_{00} = 1/\sqrt{4\pi}$.

The derivative constraints on $\phi^1$ are a little trickier. The $\ell = 0$ mode has even parity, and thus doesn't contribute to the first derivative at the origin. Modes with $\ell \ge 1$ must scale as $r^\ell$ at the origin, and so will have vanishing first derivative unless $\ell = 1$. Hence, the gradient of $\phi^1(\vec{r}\,)$ at the origin is given by
\begin{align}
\vec{\nabla} \phi^1(0) &= \vec{\nabla} \left( \phi^1_{11}(r) Y_{11}(\hat{r}) + \phi^1_{10}(r) Y_{10}(\hat{r}) + \phi^1_{1,-1}(r) Y_{1,-1}(\hat{r}) \right)_{\vec{r} = 0}
\\
&= \sqrt{\frac{3}{4\pi}} \vec{\nabla} \left( x \frac{\phi^1_{11}(r)}{r} + z \frac{\phi^1_{10}(r)}{r} + y \frac{\phi^1_{1,-1}(r)}{r} \right)_{\vec{r} = 0},
\end{align}
where we've inserted the explicit functional forms of the real spherical harmonics. Note that $\phi_{1m}(r)/r$ must be analytic at the origin. Acting on the functions of radius, $\vec{\nabla} = \hat{r} \partial_r$, and all these terms must vanish when evaluated at the origin as they are multiplied by $x$, $y$ and $z$. Hence, the only nonzero contributions come from terms where the gradient acts on $x$, $y$ and $z$.
\begin{align}
\vec{\nabla} \phi^1(0) &= \sqrt{\frac{3}{4\pi}} \left( \hat{x} \frac{\phi^1_{11}(r)}{r} + \hat{z} \frac{\phi^1_{10}(r)}{r} + \hat{y} \frac{\phi^1_{1,-1}(r)}{r} \right)_{r=0}
\end{align}
The constraint conditions require these components to vanish at the origin. We can write this condition as a constraint on $\phi^1_{1m}(r)$ using L'H\^opital's rule, yielding $\phi^{1 \prime}_{1m}(0) = 0$. The full set of constraint conditions is then
\begin{align}
\phi^1_{00}(0) = \sqrt{4\pi} \nu
, \qquad
\phi^{1\prime}_{1m}(0) = 0
, \qquad
\phi^\alpha_{00}(0) = 0 \ \ \text{for} \ \ \alpha \in \{2, \ldots, n\}.
\end{align}

The next step is to bias the modes according to these constraints. This computation follows the procedure described in Section~\ref{sec:biasing-procedure}. We will bias the $\ell = 0$ and $\ell = 1$ modes separately, as they have different constraints. Modes with $\ell > 1$ do not need to be biased, because they are unaffected by the constraints.

We start with $\ell = 0$. Let
\begin{align}
\mathbf{w}_1 = \begin{pmatrix}
\phi^\alpha_{00}(r) \\ \phi^\alpha_{00}(r')
\end{pmatrix}
, \qquad
\mathbf{w}_2 = (\phi^\alpha_{00}(0)).
\end{align}
The means of these vectors are
\begin{align}
\ev{\mathbf{w}_1} = \boldsymbol{\mu}_1 =
\begin{pmatrix}
0 \\ 0
\end{pmatrix}
, \qquad
\ev{\mathbf{w}_2} = \boldsymbol{\mu}_2 = (0).
\end{align}
The covariance matrices are given by
\begin{align}
\boldsymbol{\Sigma}_{11} = \begin{pmatrix}
4 \pi \tilde{C}_0(r, r) & 4 \pi \tilde{C}_0(r, r') \\
4 \pi \tilde{C}_0(r, r') & 4 \pi \tilde{C}_0(r', r')
\end{pmatrix}
, \qquad
\boldsymbol{\Sigma}_{12} =
\boldsymbol{\Sigma}_{21}^T = \begin{pmatrix}
4 \pi C(r) \\
4 \pi C(r')\\
\end{pmatrix}
, \qquad
\boldsymbol{\Sigma}_{22} = (4 \pi \sigma_0^2).
\end{align}
Applying Eqs. \eqref{eq:biasing}, we obtain the following from conditioning on $\phi^\alpha_{00}(0) = \delta^{\alpha 0} \sqrt{4\pi} \nu$,
\begin{align}
\Bev{\phi^\alpha_{00}(r)} &= \delta^{\alpha 1} \sqrt{4 \pi} \frac{\bar{\nu}}{\sigma_0} C(r)
\\
\Cov(\phi^\alpha_{B,00}(r), \phi^\alpha_{B,00}(r')) &= 4 \pi \tilde{C}_0(r, r') - 4 \pi \frac{C(r) C(r')}{\sigma_0^2}.
\end{align}
These results are similar in form to those of Section~\ref{sec:biasing}, with some extra factors of $4 \pi$ arising because $Y_{00} = 1/\sqrt{4\pi}$.

Much the same analysis applies to $\phi^1_{1m}(r)$ (only the $\alpha = 1$ case requires biasing). Let
\begin{align}
\mathbf{w}_1 = \begin{pmatrix}
\phi^1_{1m}(r) \\ \phi^1_{1m}(r')
\end{pmatrix}
, \qquad
\mathbf{w}_2 = (\phi^{1 \prime}_{1m}(0)).
\end{align}
The means of these vectors are
\begin{align}
\ev{\mathbf{w}_1} = \boldsymbol{\mu}_1 =
\begin{pmatrix}
0 \\ 0
\end{pmatrix}
, \qquad
\ev{\mathbf{w}_2} = \boldsymbol{\mu}_2 = (0).
\end{align}
The covariance matrices are given by
\begin{align}
\boldsymbol{\Sigma}_{11} = \begin{pmatrix}
4 \pi \tilde{C}_1(r, r) & 4 \pi \tilde{C}_1(r, r') \\
4 \pi \tilde{C}_1(r, r') & 4 \pi \tilde{C}_1(r', r')
\end{pmatrix}
, \qquad
\boldsymbol{\Sigma}_{12} =
\boldsymbol{\Sigma}_{21}^T = \begin{pmatrix}
\frac{4 \pi D(r)}{3} \\[0.4ex]
\frac{4 \pi D(r')}{3} \\
\end{pmatrix}
, \qquad
\boldsymbol{\Sigma}_{22} = \left(\frac{4 \pi \sigma_1^2}{9}\right),
\end{align}
where we have invoked results from Appendix~\ref{app:kspace_integrals}. Applying Eqs. \eqref{eq:biasing}, we obtain the following from conditioning on $\phi^{1\prime}_{1m}(0) = 0$.
\begin{align}
\ev{\phi^1_{B,1m}(r)} &= 0
\\
\Cov(\phi^1_{B,1m}(r), \phi^1_{B,1m}(r')) &= 4 \pi \tilde{C}_1(r, r') - 4 \pi \frac{D(r) D(r')}{\sigma_1^2}
\end{align}

We can summarize the means and covariances of all biased spherical harmonic modes as follows.
\begin{align} \label{eq:biased_sph_mean}
\ev{\phi^\alpha_{B,\ell m}(r)} &= \delta^{\alpha 1} \delta_{\ell 0} \delta_{m 0} \sqrt{4 \pi} \frac{\bar{\nu}}{\sigma_0} C(r)
\\ \label{eq:biased_sph_cov}
\Cov(\phi^\alpha_{B, \ell m}(r), \phi^\beta_{B, \ell' m'}(r')) &=
\delta^{\alpha \beta} \delta_{\ell \ell'} \delta_{m m'} 4 \pi \left(\tilde{C}_\ell(r, r') - \delta_{\ell 0} \delta_{m 0} \frac{C(r) C(r')}{\sigma_0^2}
- \delta^{\alpha 1} \delta_{\ell 1} \frac{D(r) D(r')}{\sigma_1^2}\right)
\end{align}

\subsection{Constructing \texorpdfstring{$\chi^2$}{chi-squared} fields from \texorpdfstring{$\phi$}{\phi} - Biased case}

Given the biased means \eqref{eq:biased_sph_mean} and covariances \eqref{eq:biased_sph_cov}, we can now repeat the analysis of Subsection \ref{sec:unbiasedconstruction} in the biased case. Starting from Eq. \eqref{eq:background_exp} and using the biased means and covariances, we compute the expectation value $\ev{\Phi_B(\vec{r}\,)}$ following essentially the same derivation as for Eq. \eqref{eq:unbiased_decomp}. The result is
\begin{align} \label{eq:mode_sum_mean}
\ev{\Phi_B(\vec{r}\,)} = 
n \sum_{\ell = 0}^\infty (2 \ell + 1) \tilde{C}_\ell(r, r)
+ (\bar{\nu}^2 - n) \frac{C(r)^2}{\sigma_0^2}
- 3 \frac{D(r)^2}{\sigma_1^2}.
\end{align}
As expected, this agrees with our previous result \eqref{eq:1p_biased_Phi} after the substitution of Eq. \eqref{eq:sigma_sum} for the sum. We can similarly compute the covariance of $\Phi_B$. We omit the straightforward but incredibly tedious derivation that generalizes the approach presented in Appendix~\ref{app:cov}. The result is
\begin{align} \label{eq:full_cov_lm}
\Cov(\Phi_B(\vec{r}\,), \Phi_B(\vec{r}\,')) &= 
2 (n-1) \left(\sum_{\ell = 0}^\infty (2 \ell + 1) P_\ell(\hat{r} \cdot \hat{r}') \tilde{C}_{\ell}(r, r') - \frac{C(r) C(r')}{\sigma_0^2}\right)^2
\nonumber\\
& \qquad 
+ 2 \left(\sum_{\ell = 0}^\infty (2 \ell + 1) P_\ell(\hat{r} \cdot \hat{r}') \tilde{C}_{\ell}(r, r') - \frac{C(r) C(r')}{\sigma_0^2} - \frac{3 D(r) D(r')}{\sigma_1^2} \cos(\gamma)\right)^2
\nonumber\\
& \qquad 
+ 4 \bar{\nu}^2 \frac{C(r) \, C(r')}{\sigma_0^2} \left(\sum_{\ell = 0}^\infty (2 \ell + 1) P_\ell(\hat{r} \cdot \hat{r}') \tilde{C}_{\ell}(r, r') - \frac{C(r) C(r')}{\sigma_0^2} - \frac{3 D(r) D(r')}{\sigma_1^2} \cos(\gamma)\right).
\end{align}
As expected, this matches the previous result \eqref{eq:biased_Phi_cov} with the sums substituted from Eq. \eqref{eq:CCtilderesult}.

We can take this even further: any $N$-point function of $\Phi_B$, because of the non-central Wick's theorem, will always be written in terms of the means \eqref{eq:biased_sph_mean} and covariances \eqref{eq:biased_sph_cov} summed over all $\ell$ and $m$ modes. This means that in such $N$-point functions, $\tilde{C}_{\ell}$ will always appear in the combination
\begin{align}\label{eq:combination}
C(\vec{r} - \vec{r}\,') = \sum_{\ell = 0}^\infty (2 \ell + 1) P_\ell(\hat{r} \cdot \hat{r}') \tilde{C}_{\ell}(r, r')
\end{align}
after employing the addition theorem \eqref{eq:addition}. (Recall that in the coincidence limit, the LHS becomes $\sigma_0^2$, which is how this sum appears in the one-point function.) This feature of the statistics of $\Phi_B$ when constructed as sums of spherical harmonics of Gaussian fields is particularly important when the sums over $\ell$ are truncated for numerical purposes, as we discuss in the following section.

\subsection{Investigating \texorpdfstring{$\Phi_{B,\ell m}$}{the spherical harmonics of \Phi}}

It is worth paying some attention to how the statistics of the spherical harmonics $\Phi_{B,\ell m}$ are constructed from the underlying Gaussian fields. We can leverage Eqs. \eqref{eq:1point_Phi_mode} and \eqref{eq:mode_sum_mean} to construct the means as
\begin{align} \label{eq:biased_full_mode_mean}
\langle \Phi_{B, \ell m}(r) \rangle 
= \delta_{\ell 0} \delta_{m 0} \sqrt{4\pi} \left(n \sum_{\ell = 0}^\infty (2 \ell + 1) \tilde{C}_\ell(r, r)
+ (\bar{\nu}^2 - n) \frac{C(r)^2}{\sigma_0^2}
- 3 \frac{D(r)^2}{\sigma_1^2}\right).
\end{align}
Note that the $C(r)$ terms arise from the $\ell = 0$ modes, while the $D(r)$ terms arise from the $\ell = 1$ modes, as can be seen from Eqs. \eqref{eq:biased_sph_mean} and \eqref{eq:biased_sph_cov}. The sum over $\ell$ modes is equal to $\sigma_0^2$ by Eq. \eqref{eq:sigma_sum}, which brings this result into agreement with Eq. \eqref{eq:Phi_biased_1point_mode}.

Similarly, the covariance can be obtained from Eqs. \eqref{eq:cov_Phi_mode} and \eqref{eq:full_cov_lm} by following the same series of steps used to obtain Eq. \eqref{eq:fullcov}.
\begin{align}
\mathrm{Cov}(\Phi_{B, \ell m}(r), \Phi_{B, \ell' m'} (r')) 
= 4 \pi \delta_{\ell \ell'} \delta_{m m'} \Bigg[ &
n \sum_{\ell_1, \ell_2} (2 \ell_1 + 1) (2 \ell_2 + 1) \tilde{C}_{\ell_1}(r, r')
\tilde{C}_{\ell_2}(r, r')
\int_{-1}^{1} du \,
P_\ell(u)
P_{\ell_1}(u) 
P_{\ell_2}(u)
\nonumber\\
&
+ 4 \frac{C(r) \, C(r')}{\sigma_0^2} \left(\bar{\nu}^2 - n\right) \tilde{C}_{\ell}(r, r')
\\\nonumber
&
- \frac{12}{2 \ell + 1} \frac{D(r) D(r')}{\sigma_1^2}
\left((\ell + 1) \tilde{C}_{\ell+1}(r, r') + \ell \tilde{C}_{\ell-1}(r, r')\right)
\\\nonumber
&
+ \delta_{\ell 0} 2 \left(\frac{C(r)^2 C(r')^2}{\sigma_0^4} \left(n - 2 \bar{\nu}^2\right) + 3 \frac{D(r)^2 D(r')^2}{\sigma_1^4}\right)
\\\nonumber
&
- \delta_{\ell 1} 4 \, \frac{C(r) C(r')}{\sigma_0^2} \frac{D(r) D(r')}{\sigma_1^2} \left(\bar{\nu}^2 - 1\right)
+ \delta_{\ell 2} \frac{12}{5} \frac{D(r)^2 D(r')^2}{\sigma_1^4} \Bigg]
\end{align}
Note that this is equivalent to Eq. \eqref{eq:fullcov} after using Eq. \eqref{eq:crazyformula}. In this expression, the contributions to each term from a given $\ell$ mode of $\phi^\alpha_{\ell m}$ can be straightforwardly seen, recalling that $C(r)$ and $D(r)$ terms arise from the biased $\ell = 0$ and 1 modes respectively. As a special case, we look at $\ell = \ell' = m = m' = 0$.
\begin{align}
\Cov(\Phi_{B, 00}(r), \Phi_{B, 00} (r')) 
= 4 \pi \Bigg[ &
2 n \sum_{\ell = 0}^\infty (2 \ell + 1) \tilde{C}_{\ell}(r, r')^2
+ 4 \frac{C(r) \, C(r')}{\sigma_0^2} \left(\bar{\nu}^2 - n\right) \tilde{C}_0(r, r')
\\\nonumber
&
- 12 \frac{D(r) D(r')}{\sigma_1^2} \tilde{C}_1(r, r')
+ 2 \frac{C(r)^2 C(r')^2}{\sigma_0^4} \left(n - 2 \bar{\nu}^2\right) + 6 \frac{D(r)^2 D(r')^2}{\sigma_1^4} \Bigg]
\end{align}
The sum over $\ell$ shows how various $\ell$ modes contribute to the background covariance (c.f. Eqs. \eqref{eq:firstuglyintegral} and \eqref{eq:funnyrelation}), while the remaining $C$ and $D$ terms arise from the biased $\ell = 0$ and 1 modes respectively.

\section{Sampling Biased Fields} \label{sec:sampling2}

We now possess all of the machinery needed to sample our biased fields. In a nutshell, we wish to numerically sample $\phi^\alpha_{\ell m}(r)$ and evaluate
\begin{align}
\Phi(\vec{r}\,) = \sum_{\alpha = 1}^n \left(\sum_{\ell=0}^{\infty}\sum_{m=-\ell}^{\ell} \phi^\alpha_{\ell m}(r) Y_{\ell m}(\hat{r})\right)^2.
\end{align}
Two obvious difficulties immediately present themselves. Firstly, we cannot sum an infinite number of $\ell$ modes, and so we must truncate at some $\ell_{\rm max}$. Secondly, we cannot sample $\phi^\alpha_{\ell m}(r)$ at every possible radius; instead, we must choose a discretization method.

In this section, we discuss how to construct $\ell_{\rm max}$ and investigate the impact that this truncation has on sample statistics, before developing the discretized sampling method for $\phi^\alpha_{\ell m}(r)$ and presenting the algorithm to construct samples of $\Phi_B$. Finally, we discuss the specialization of the construction to the mode $\Phi_{B,00}$.

\subsection{Truncating in \texorpdfstring{$\ell$}{\ell}}

To construct a sample of our $\chi^2$ field, we will need to sample the spherical harmonics of the underlying Gaussian fields. However, the spherical harmonic decomposition of $\phi^\alpha$ involves an infinite sum over $\ell$. As we can't take $\ell \to \infty$ numerically, we must introduce a cutoff, $\ell_{\mathrm{max}}$. We now discuss a heuristic approach to estimating a cutoff value, before following up with a more precise analysis.

First, note that $\ell_{\mathrm{max}}$ will always depend on the outermost radius under consideration, which provides the longest length scale of interest. As the radius increases, so must $\ell_{\mathrm{max}}$ to maintain the same level of accuracy. In the following, we always assume a known $r_{\mathrm{max}}$. We suggest choosing $r_{\mathrm{max}}$ to be a few times the expected width of the peak or trough, noting that the larger $r_{\mathrm{max}}$ is, the more numerical work is necessary.

We also need to estimate an appropriate ``maximum'' wavenumber $\tilde{k}$ to describe the shortest length scale of interest in the model. The method one uses to estimate this should be chosen based on the power spectrum in question. For spectra that decay exponentially in the tail, the decay scale is an appropriate value for $\tilde{k}$. For tails that decay more slowly, one suggestion is to choose $\tilde{k}$ to capture the majority of the power in $\sigma_0^2$, such as
\begin{align}
\int_0^{\tilde{k}} dk \, k^2 \Pk(k) \sim 0.95 \sigma_0^2.
\end{align}
Given $\tilde{k}$, one determines the wavelength of interest in the model $\lambda$ by $\tilde{k} = 2 \pi/\lambda$. 

Having chosen $r_{\rm max}$ and $\tilde{k}$ or $\lambda$, we can now estimate $\ell_{\rm max}$ heuristically. Consider the mode $\phi^\alpha_{\ell_{\mathrm{max}} 0}$. The dependence on $\cos(\theta)$ is given by $P_{\ell_{\mathrm{max}}}(\cos \theta)$, which between $\theta = 0$ and $\theta = \pi$ crosses zero $\ell_{\mathrm{max}}$ times. At radius $r_{\mathrm{max}}$, the distance between these nodes will be $\sim \pi r_{\mathrm{max}} / \ell_{\mathrm{max}}$. We want the distance between the nodes to be roughly\footnote{One can also measure the distance between nodes of the harmonics as functions of $\phi$ (that is, the equatorial nodes) by looking at a maximal $m$ mode; the resulting estimate of $\ell_{\mathrm{max}}$ is identical to the estimate derived here.} $\lambda/2$. Putting this all together, we obtain
\begin{align}
\ell_{\mathrm{max}} \sim r_{\mathrm{max}} \tilde{k}.
\end{align}
Choosing $\ell_{\mathrm{max}}$ in this way, we ensure that we will have sufficiently small angular resolution to capture the most significant features of the field at our maximum radius.

We now make the above heuristic more explicit. We know from Eq. \eqref{eq:biased_full_mode_mean} that the biased expectation value $\ev{\Phi_B(\vec{r})}$ consists of the sum of biasing terms involving $C(r)$ and $D(r)$ and a sum over $\ell$ modes. The contribution from $\ell > 1$ modes to the spherical mode only goes towards constructing the background expectation value $\sigma_0^2$. We now estimate at what point the $\ell$ modes give a negligible contribution to the sum.

From the definition of $\tilde{C}_\ell(r, r')$, we have
\begin{align} \label{eq:Ctildeagain}
\tilde{C}_\ell(r, r) = 4 \pi \int dk \, k^2 \, \Pk(k) j_\ell(kr)^2.
\end{align}
For large $\ell$, $j_\ell(kr)$ is suppressed at small $kr$ (relative to the first root of $j_\ell(kr)$). If the first root of $j_\ell(kr)$ is larger than $\tilde{k} r_{\mathrm{max}}$, then the contribution to the integral from $k = 0$ to $\tilde{k}$ is suppressed by the spherical Bessel function, while contributions from higher $k$ are suppressed by the power spectrum, leaving a small value for $\tilde{C}_\ell$. We can estimate the mode $\ell_{\rm suppress}$ at which this suppression sets in by finding the first nontrivial root $x_1$ of $j_\ell(x)$. This root can be estimated from linear regression between $\ell = 10$ and $\ell = 100$ as $x_1 \approx 5.2 + 1.05 \ell$. Therefore, we estimate
\begin{align} \label{eq:rule_of_thumb}
\ell_{\rm suppress} \approx \frac{\tilde{k} r_{\mathrm{max}} - 5.2}{1.05}.
\end{align}
Choosing $\ell_{\rm max} = \ell_{\rm suppress}$ gives a somewhat tighter bound than our previous geometric argument. Based on this estimate, modes with $\ell > \ell_{\mathrm{max}}$ should make diminishing contributions to the background expectation value. The exact rate at which the contributions decay depends on the decay rate of the tail of the power spectrum.

Note that the truncation in $\ell$ introduces an error which vanishes at the origin and grows with radius. This means that the immediate vicinity of a stationary point should be well described by $\ell \le \ell_\textrm{max}$, with the largest errors introduced at the outer radius.

We would like to develop an understanding of how the introduction of a cutoff $\ell_{\rm max}$ impacts the statistics of samples of $\Phi_B$. We saw above that all contributions from $\ell > 1$ to $N$-point expectation values of $\Phi_B$ occur through the explicit sum \eqref{eq:combination}, which we repeat for good measure.
\begin{align} \label{eq:again}
C(\vec{r} - \vec{r}\,') = \sum_{\ell = 0}^\infty (2 \ell + 1) P_\ell(\hat{r} \cdot \hat{r}') \tilde{C}_{\ell}(r, r')
\end{align}
Hence, we should understand how introducing $\ell_{\rm max}$ affects this sum.

Recall that $\tilde{C}_\ell(r, 0) = \delta_{\ell 0} C(r)$, and note that the integrand in Eq. \eqref{eq:Ctildeagain} is positive definite when $r = r'$. Generically, for fixed $r$, $\tilde{C}_\ell(r, r')$ will be close to its maximum for $r = r'$. Hence, we expect the greatest deficit to Eq. \eqref{eq:again} to arise in the coincidence limit, when every discarded term is providing its maximum (positive) contribution to the sum. In the coincidence limit, we arrive at Eq. \eqref{eq:sigma_sum}. When truncated, this becomes
\begin{align}
\Sigma_{\ell_{\rm max}}(r) = \frac{1}{\sigma_0^2} \sum_{\ell = 0}^{\ell_{\rm max}} (2 \ell + 1) \tilde{C}_{\ell}(r, r).
\end{align}
Note that $\Sigma_\infty(r) \equiv 1$ by Eq. \eqref{eq:sigma_sum}. We thus expect that if the truncation error in $\Sigma$ is minimized, the truncation error in all $N$-point functions of $\Phi_B$ will be similarly minimized. This convergence will then propagate to the individual $\ell$ modes of $\Phi_{B, \ell m}$. This is excellent news, as a single check is then sufficient to ensure that the chosen value of $\ell_{\rm max}$ is appropriate. Note that as higher point functions may depend on various powers of $\Sigma(r)$, the improvement in accuracy with increasing $\ell_{\rm max}$ will differ between different statistics. The important point is that the accuracy of the reconstruction of $\sigma_0^2$ controls the accuracy of all of the statistics.

\begin{figure}[t]
	\centering
	\includegraphics[width=0.8\textwidth]{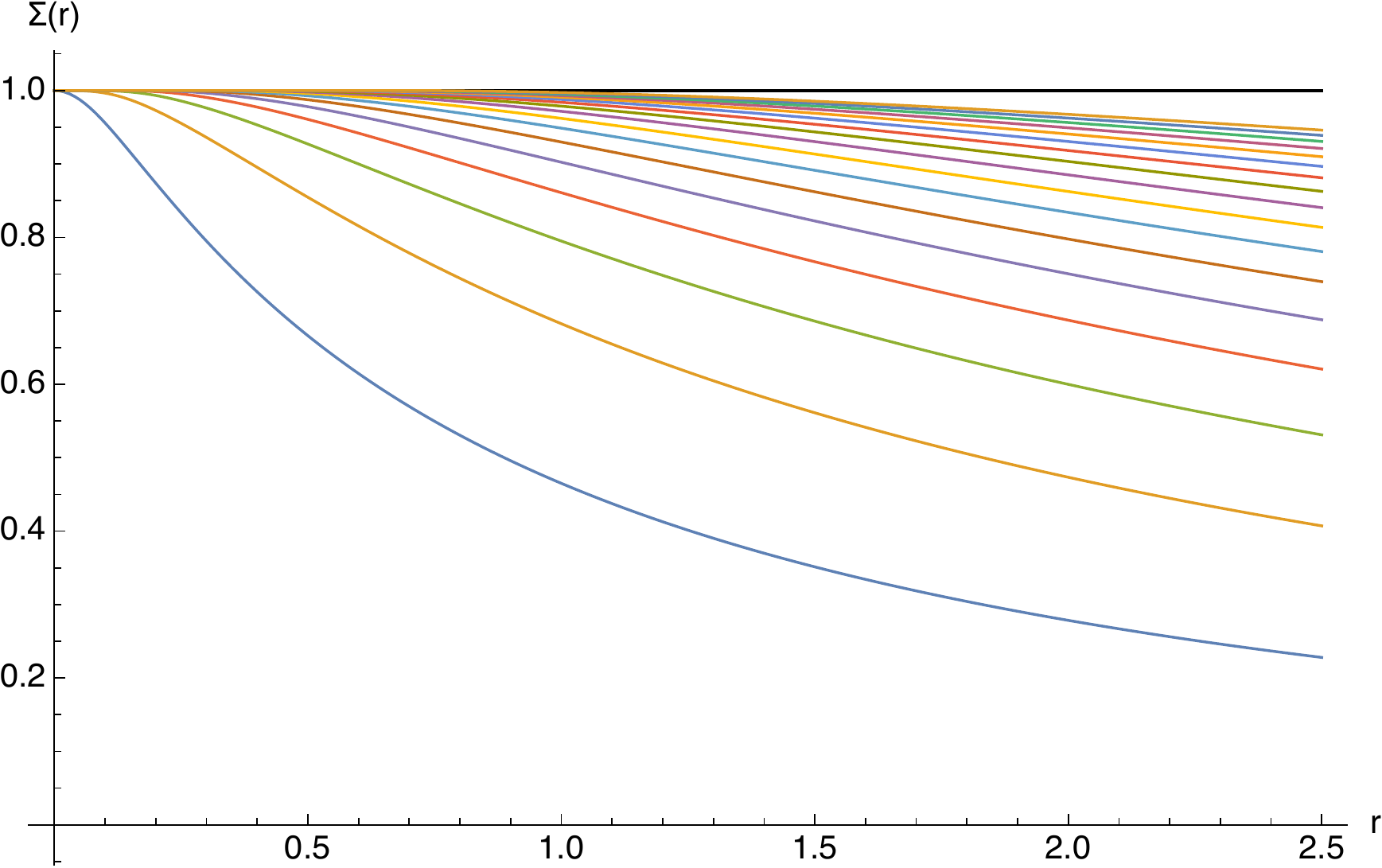}
	\caption{Plot showing $\Sigma_{\ell_{\rm max}} (r)$ with varying $\ell_{\rm max}$. The bottom line has $\ell_{\rm max} = 0$, with each subsequent line increasing $\ell_{\rm max}$ by one. This particular plot has $r_{\rm max} = 2.5$ and a power spectrum that decays like $\exp(-k/\tilde{k})$ with $\tilde{k} = 8$ using dimensionless lengths. The rule of thumb \eqref{eq:rule_of_thumb} suggests $\ell_{\rm max} = 16$, which is the last line we plot. The droop at large $r$ is due to the tail of the power spectrum ($k > \tilde{k}$). We saw in Figure \ref{fig:profiles} that peaks in this spectrum have an expected half-maximum radius around 0.55. We have essentially 100\% of the mode construction of $\sigma_0^2$ out to double this radius, and so this is a reasonable value for $\ell_{\rm max}$.}
	\label{fig:sigma0stack}
\end{figure}

We demonstrate how the different $\ell$ modes sum to the background value in Figure \ref{fig:sigma0stack}. The truncation error at any radius is clearly visible in the gap between the stack and unity. Our task in choosing $\ell_{\rm max}$ is to make sure that we have sufficiently many modes that the truncation error at our maximum radius is acceptable. While Eq. \eqref{eq:rule_of_thumb} can be used as a rule of thumb to estimate $\ell_\mathrm{max}$, the real test of convergence is to generate a plot like Figure \ref{fig:sigma0stack}.

We found from experimentation that when the power spectrum $\Phi(k)$ vanishes beyond $\tilde{k}$, then the rule-of-thumb estimate for $\ell_{\rm max}$ was very good at predicting how many $\ell$ modes were needed to construct the vast majority of $\sigma_0^2$ in Eq. \eqref{eq:sigma_sum} at $r_{\rm max}$. When the power spectrum had a tail beyond $\tilde{k}$, some deviation occurred, as shown in Figure \ref{fig:sigma0stack}. As high $\ell$ modes become increasingly expensive to compute, we suggest choosing an $\ell_{\rm max}$ that yields good coverage at a radius equal to a few times the expected width of the extrema. We thus can have confidence that our truncation in $\ell$ introduces minimal error for our constructed samples.

\subsection{Sampling Profiles}

To perform the actual sampling, we need to choose a radial grid, $\{r_i\}$, at which to sample the spherical harmonic coefficients. Both evenly- and unevenly-spaced grids will work. In view of the Nyquist theorem, the grid spacing should be taken finer than $\lambda/2$, where $\lambda = 2 \pi / \tilde{k}$ from above. We define $\vec{\phi}^\alpha_{B, \ell m} = (\phi^\alpha_{B, \ell m}(r_1), \ldots, \phi^\alpha_{B, \ell m}(r_M))$ where $M$ is the number of radial grid points and $r_M = r_\text{max}$. The elements of this vector are evaluations of a Gaussian random field at discrete grid points, so these vectors belong to a multivariate Gaussian distribution governed by a mean $\vec{\mu}^\alpha_{\ell m}$ and covariance matrix $\boldsymbol{\Sigma}^\alpha_{\ell m}$, the elements of both of which can be computed from Eqs. \eqref{eq:biased_sph_mean} and \eqref{eq:biased_sph_cov}\footnote{There is one caveat concerning the integrals in these equations. Depending on the choice of frequency cutoff and radial grid, the spherical Bessel functions appearing in the integrands of $C$, $D$ and $\tilde{C}_\ell$ may oscillate so rapidly that the integral becomes numerically intractable using standard quadrature methods. In this event, we recommend using the Levin collocation method \cite{Levin1996}. For the interested reader, its application to this integral is discussed in \cite{Bloomfield2017}. We also note that computing these covariance matrices can be quite time-consuming, as each element requires the evaluation of an oscillatory integral, and there are a lot of elements to compute! Finally, we recommend computing $\tilde{C}_\ell(r, r')$ by using the most appropriate of Eqs. \eqref{eq:Ctilde1} and \eqref{eq:Ctilde2}. The former is likely better for small $r$ when there are few oscillations to integrate, while the latter is likely superior at larger $r$ when the number of oscillations in the former become intractable.}.
\begin{align}
\mu^\alpha_{\ell m, i} &= \langle \phi^\alpha_{B, \ell m} (r_i) \rangle,
\\
\boldsymbol{\Sigma}^\alpha_{\ell m, ij} &= \mathrm{Cov}(\phi^\alpha_{B, \ell m}(r_i), \phi^\alpha_{B, \ell m}(r_j)).
\end{align}
Given a cutoff $\ell_{\mathrm{max}}$ and a set of grid points, we need to compute the biased covariance matrices $\boldsymbol{\Sigma}^\alpha_{\ell m}$ containing pairwise covariances between the grid points. Only one matrix is necessary for $\ell = 0$ and $\ell \ge 2$, while two matrices are needed for $\ell = 1$ (one for $\alpha = 1$ and one for $\alpha > 1$). We also need the vector of means $\vec{\mu}^1_0$ (these are the only nonzero means).

With the means and covariance matrices computed, we are ready to sample $\vec{\phi}^\alpha_{B, \ell m}$. We will employ a standard technique to construct a sample of a non-central multivariate Gaussian variable. Let $\vec{Z}$ be a $M$-dimensional random Gaussian vector with covariance matrix equal to the identity (so that each element of $\vec{Z}$ is drawn from the standard unit Gaussian). Given an $M \times M$ matrix $A$ such that\footnote{The matrix $A$ can be found efficiently using a Cholesky decomposition, but one can run into numerical errors this way. A well-defined covariance matrix must be positive definite. However, roundoff errors can lead to violations of this criterion. This problem is made clear by an eigenvalue decomposition, $\boldsymbol{\Sigma}^\alpha_{B, \ell m} = M \Lambda M^T$, where $M$ is orthogonal and $\Lambda$ is diagonal by the real spectral theorem. $\Lambda$ can contain eigenvalues so small that roundoff in numerical diagonalization yields small negative entries in this matrix (similar errors plague the Cholesky method). Once can get around this issue by setting these entries to zero in $\Lambda$, and then defining $A = M \sqrt{\Lambda}$. One must take care to ensure that the magnitudes of these miscalculated eigenvalues are truly small enough to neglect. Eigenvalue decompositions are available through standard linear algebra libraries, such as scipy.linalg.eig in python.} $\boldsymbol{\Sigma}^\alpha_{B, \ell m} = A A^T$, the random variable $A\vec{Z} + \vec{\mu}^\alpha_{B,\ell}$ is distributed according to the multivariate Gaussian distribution described by covariance $\boldsymbol{\Sigma}^\alpha_{B, \ell m}$ and means $\vec{\mu}^\alpha_{B, \ell}$. We sample one such vector $\vec{\phi}^\alpha_{B, \ell m}$ for every combination of $\alpha$, $\ell$, and $m$.

Now that we have samples for $\phi^\alpha_{B, \ell m}(r_i)$, we are ready to construct the $\chi^2$ field $\Phi_B$. Substituting the spherical harmonic decomposition of the $\phi^\alpha$ fields \eqref{eq:decomp_phi} into the definition of $\Phi$ \eqref{eq:Phi_def}, we obtain
\begin{align} \label{eq:Phi-from-philm}
\Phi_B(r_i, \hat{r}) = \sum_{\alpha = 1}^n \left(
\sum_{\ell=0}^{\ell_{\mathrm{max}}} \sum_{m=-\ell}^{\ell} \phi^\alpha_{B, \ell m}(r_i) Y_{\ell m}(\hat{r})
\right)^2.
\end{align}
Essentially, we combine the samples $\phi^\alpha_{B, \ell m}(r_i)$ to construct $\phi^\alpha_B(\vec{r}\,)$ for each $\alpha$, and then square and sum the results. While $\Phi_B$ can only be computed at radii lying on the radial grid, the full angular dependence is available (subject to the limitations of the cutoff $\ell_{\mathrm{max}}$).

If instead of the full $\Phi_B(\vec{r}\,)$ field, we wish to construct particular modes $\Phi_{B, \ell m}(r)$, we can also do that. Observe that we can compute the spherical harmonic modes $\Phi_{B, \ell m}(r_i)$ directly from $\phi^\alpha_{B, \ell m}(r_i)$. Substituting Eq. \eqref{eq:Phi-from-philm} into the expression for $\Phi_{\ell m}(r)$ \eqref{eq:def_Phi_mode}, we obtain
\begin{align} \label{eq:Phi_phi_decomp}
\Phi_{B,\ell m}(r_i) &= \sum_{\alpha=1}^{n} \sum_{\ell_1 \ell_2} \sum_{m_1 m_2} \phi^\alpha_{B, \ell_1 m_1}(r_i) \, \phi^\alpha_{B, \ell_2 m_2}(r_i) \int d\hat{r} \, Y_{\ell m}(\hat{r}) Y_{\ell_1 m_1}(\hat{r}) Y_{\ell_2 m_2}(\hat{r}).
\end{align}
This nasty-looking integral has an analytic result known as Gaunt's formula, which is written in terms of Wigner 3-$j$ symbols. We caution that one will need to convert back to complex spherical harmonics to use this result however. The integral thankfully simplifies for the spherical mode $\ell = m = 0$, as $Y_{00} = 1/\sqrt{4 \pi}$ with the remaining integral following from the orthonormality condition. The result is
\begin{align} \label{eq:Phi_phi_decomp_0}
\Phi_{B,00}(r_i) &= \frac{1}{\sqrt{4 \pi}} \sum_{\alpha=1}^{n} \sum_{\ell = 0}^{\ell_{\rm max}} \sum_{m = - \ell}^\ell (\phi^\alpha_{B, \ell m}(r_i))^2.
\end{align}
This decomposition allows us to efficiently compute samples of the spherical component of our biased $\chi^2$ field from samples of the harmonic coefficients of the underlying Gaussian random fields. Note that with the $\ell$ sum truncated, $\Phi_{B,00}(r)$ becomes a generalized $\chi^2$ field.

It is interesting to compute the mean of $\Phi_{B,00}(r_i)$.
\begin{align}
\langle \Phi_{B, 00}(r_i) \rangle 
&= \sqrt{4\pi} \left(n \sum_{\ell = 0}^{\ell_{\rm max}} (2 \ell + 1) \tilde{C}_\ell(r_i, r_i)
+ (\bar{\nu}^2 - n) \frac{C(r_i)^2}{\sigma_0^2}
- 3 \frac{D(r_i)^2}{\sigma_1^2}\right)
\end{align}
We can compare this to the theoretical value \eqref{eq:biased_full_mode_mean} to compute the truncation error
\begin{align}
E_{\rm trunc}(r_i) = \langle \Phi_{B, 00}(r_i) \rangle_{\rm theory} - \langle \Phi_{B, 00}(r_i) \rangle_{\rm sample}
= \sqrt{4 \pi} n \left(\sigma_0^2 - \sum_{\ell = 0}^{\ell_{\rm max}} (2\ell + 1) \tilde{C}_\ell(r_i, r_i)\right).
\end{align}
The omission of high $\ell$ modes means that the sample mean of modes constructed in this manner will be less than the theoretical expectation ($E_{\rm trunc}(r_i) > 0$). We can correct for this by adding the truncation error back into the sampling as a constant bias,
\begin{align}
\Phi^{\rm corrected}_{B,00}(r_i) &= \frac{1}{\sqrt{4 \pi}} \sum_{\alpha=1}^{n} \sum_{\ell=0}^{\ell_{\rm max}} \sum_{m = - \ell}^\ell (\phi^\alpha_{B, \ell m}(r_i))^2 + E_{\rm trunc}(r_i).
\end{align}
This ensures that $\ev{\Phi^{\rm corrected}_{B,00}}$ matches Eq. \eqref{eq:biased_full_mode_mean}, and is equivalent to manually adding back in the deficit at large $r$ seen in Figure \ref{fig:sigma0stack}. Note that the addition of this constant offset at each radius doesn't impact the (co)variance of $\Phi$ at all. We can further add in this bias to samples of the full field $\Phi_B(\vec{r}\,)$ \eqref{eq:Phi-from-philm} as
\begin{align}
\Phi^{\rm corrected}_B(r_i, \hat{r}) = \sum_{\alpha = 1}^n \left(
\sum_{\ell=0}^{\ell_{\mathrm{max}}} \sum_{m=-\ell}^{\ell} \phi^\alpha_{B, \ell m}(r_i) Y_{\ell m}(\hat{r})
\right)^2 + \frac{E_{\rm trunc}(r_i)}{\sqrt{4\pi}}
\end{align}
to ensure that $\ev{\Phi^{\rm corrected}_B(r_i, \hat{r})} = n \sigma_0^2$ at large radii.

\section{Conclusions} \label{sec:conclusions}

We now have a thorough statistical understanding of how peaks and troughs behave in $\chi^2$ fields. The average shape of a stationary point has been characterized, and statistical measures of departures from this average have been quantified in both a point-wise and a spherical harmonic sense. In addition, the point-wise results have been explicitly decomposed into sums of all spherical harmonic modes.

We investigated the expected shape of profiles, developing scaling relationships for the expected widths, variance on the width, and behavior of the profile tail. We developed a geometric description of how spherical a peak is at a given radius, and used this to define the asphericity of a profile at a given radius, for which we also investigated the scaling behavior.

While the statistical description of the profile about a stationary point at first appears distinct from the sampling of a profile about such a point, we have seen that the two parts are closely intertwined. In particular, the methods used to calculate statistical properties were repeatedly used in the sampling methodology, and the sampling results provided insight into the behavior of the spherical mode of the $\chi^2$ field in terms of the behavior of the underlying Gaussian fields. The sampling method we developed works by efficiently sampling the Gaussian fields before using them to construct the $\chi^2$ field. Particular attention was paid to estimating an appropriate cutoff in $\ell$, and understanding the impact of such a cutoff on the statistical properties of the generated samples.

By working with a numerical implementation of our results, one can learn a lot about the rare events in a model with a $\chi^2$ field. Basic information such as the average width of an event and the variance on this width can be extracted, and an idea of the expected sphericity of the event can be predicted. Furthermore, one can understand the contributions that various spherical harmonic modes make to the event and understand how this changes with the field amplitude. We envisage this information being used to better understand rare events in $\chi^2$ fields, before embarking on expensive simulations based on the repeated sampling of such events.

\acknowledgments

We thank Casey Lam for helpful discussions. This work is supported in part by the U.S. Department of Energy under grant Contract Number DE-SC0012567, and in part by MIT's Undergraduate Research Opportunities Program. 
Zander Moss is supported by the National Science Foundation Graduate Research Fellowship under Grant No. DGE-1745301.
Saarik Kalia is supported by the National Science Foundation Graduate Research Fellowship under Grant No. DGE-1656518.

\bibliographystyle{utphys}
\bibliography{stats}

\appendix

\section{Non-Central Wick's Theorem} \label{app:wick}

Wick's theorem states that for Gaussian random variables $X_1,\ldots,X_N$ with zero mean, we have
\begin{align}
\langle X_1\cdots X_N\rangle=\sum_{P\bowtie\{1,\ldots,N\}}\prod_{(i,j)\in P}\langle X_iX_j\rangle,
\end{align}
where we use $P \bowtie \{1,\ldots,N\}$ to indicate that the sum should run over all possible unordered pairings of the set $\{1,\ldots,N\}$.  (The expectation vanishes if $N$ is odd.) 

If the Gaussian random variables have nonzero mean, Wick's theorem doesn't hold. However, we can define new Gaussian random variables $\tilde X_i = X_i - \langle X_i\rangle$ which do have zero mean, and so we can apply Wick's theorem to them. This yields
\begin{align}
\langle X_1\cdots X_N\rangle
&=\langle(\tilde X_1+\langle X_1\rangle)\cdots(\tilde X_N+\langle X_N\rangle)\rangle
\\
&=\sum_{S\subset\{1,\ldots,N\}}\left\langle\prod_{i\in S}\tilde X_i\right\rangle\prod_{j\notin S}\langle X_j\rangle
\\
&=\sum_{S\subset\{1,\ldots,N\}}\sum_{P\bowtie S}\prod_{(i,j)\in P}\langle\tilde X_i\tilde X_j\rangle\prod_{k\notin S}\langle X_k\rangle
\\
&=\sum_{P\bowtie S\subset\{1,\ldots,N\}}\prod_{(i,j)\in P}\mathrm{Cov}(X_i,X_j)\prod_{k\notin S}\langle X_k\rangle.
\end{align}
Note that $\langle \tilde{X}_i \tilde{X}_j \rangle \equiv \mathrm{Cov}(X_i, X_j)$ by definition. The result is a sum over all pairings of \textit{all subsets} of $\{1,\ldots,N\}$. The indices included in the pairing appear in the product of covariances, while all other indices appear in the product of means. In the case of $N=4$, the result is
\begin{align} \label{eq:4point_wick}
\langle X_1 X_2 X_3 X_4 \rangle &= 
\mathrm{Cov}(X_1,X_2)\mathrm{Cov}(X_3,X_4) 
+ \mathrm{Cov}(X_1,X_3)\mathrm{Cov}(X_2,X_4)
+ \mathrm{Cov}(X_1,X_4)\mathrm{Cov}(X_2,X_3)
\\ \nonumber
&\qquad 
+ \langle X_1 \rangle \langle X_2 \rangle\mathrm{Cov}(X_3,X_4)
+ \langle X_1 \rangle \langle X_3 \rangle \mathrm{Cov}(X_2,X_4)
\\ \nonumber
&\qquad 
+ \langle X_1 \rangle \langle X_4 \rangle \mathrm{Cov}(X_2,X_3)
+ \langle X_2 \rangle \langle X_3 \rangle \mathrm{Cov}(X_1,X_4)
\\ \nonumber
&\qquad 
+ \langle X_2 \rangle \langle X_4 \rangle \mathrm{Cov}(X_1,X_3)
+ \langle X_3 \rangle \langle X_4 \rangle \mathrm{Cov}(X_1,X_2)
\\ \nonumber
&\qquad 
+ \langle X_1 \rangle \langle X_2 \rangle \langle X_3 \rangle \langle X_4 \rangle.
\end{align}

\section{The Chi-Squared PDF and Biasing} \label{app:biasing}

Consider the general expectation value $\Bev{f(\Phi(\vec{r}_1), \ldots, \Phi(\vec{r}_k))}$, where $k$ is the number of spatial coordinates appearing in the expectation value. This expectation value is conditioned on $\Phi(0) = \nu^2$, $\vec{\nabla} \Phi(0) = 0$. We wish to show that this is equivalent to the expectation value $\ev{f(\Phi(\vec{r}_1), \ldots, \Phi(\vec{r}_k))}_\phi$, which is computed under the condition that $\phi^1(0) = \nu$, $\vec{\nabla} \phi^1(0) = 0$ and $\phi^{\alpha > 1}(0) = 0$. For brevity, we write these two expectation values as $\Bev{f}$ and $\ev{f}_\phi$.

Before diving into calculations, let us introduce some compact notation. Let symbols in bold represent vectors in field space. Hence, $\boldsymbol{\phi} = (\phi^1, \ldots, \phi^n)$. We further define the following quantities, all of which are vectors in field space.
\begin{align}
\boldsymbol{\phi}_0 = \boldsymbol{\phi}(0)
, \qquad
\boldsymbol{\phi}_x = \partial_x \boldsymbol{\phi}(0)
, \qquad
\boldsymbol{\phi}_y = \partial_y \boldsymbol{\phi}(0)
, \qquad
\boldsymbol{\phi}_z = \partial_z \boldsymbol{\phi}(0)
, \qquad
\mathbf{y}_i = \boldsymbol{\phi}(\vec{r}_i)
\end{align}
In this section, all dot products are performed in field space. Note that $\Phi(\vec{r}_i) = \mathbf{y}_i \cdot \mathbf{y}_i$. We also employ the shorthand
\begin{align}
\phi_0^i = \phi^i(0), \qquad
\phi_x^i = \partial_x \phi^i(0), \qquad
\phi_y^i = \partial_y \phi^i(0), \qquad
\phi_z^i = \partial_z \phi^i(0).
\end{align}

The quantity $\Bev{f}$ can be calculated from first principles as follows.
\begin{align}
\Bev{f} &= \frac{\int d^n \boldsymbol{\phi}_0 \, d^n \boldsymbol{\phi}_x \, d^n \boldsymbol{\phi}_y \, d^n \boldsymbol{\phi}_z \, d^{nk} \mathbf{y}_i \, 
f(\mathbf{y}_i \cdot \mathbf{y}_i) 
\delta(|\boldsymbol{\phi}_0|^2 - \nu^2) 
\delta(\boldsymbol{\phi}_0 \cdot \boldsymbol{\phi}_x) 
\delta(\boldsymbol{\phi}_0 \cdot \boldsymbol{\phi}_y) 
\delta(\boldsymbol{\phi}_0 \cdot \boldsymbol{\phi}_z) 
p(\boldsymbol{\phi}_0, \boldsymbol{\phi}_x, \boldsymbol{\phi}_y, \boldsymbol{\phi}_z, \mathbf{y}_i)}
{\int d^n \boldsymbol{\phi}_0 \, d^n \boldsymbol{\phi}_x \, d^n \boldsymbol{\phi}_y \, d^n \boldsymbol{\phi}_z \, d^{nk} \mathbf{y}_i \, 
\delta(|\boldsymbol{\phi}_0|^2 - \nu^2) 
\delta(\boldsymbol{\phi}_0 \cdot \boldsymbol{\phi}_x) 
\delta(\boldsymbol{\phi}_0 \cdot \boldsymbol{\phi}_y)
\delta(\boldsymbol{\phi}_0 \cdot \boldsymbol{\phi}_z) 
p(\boldsymbol{\phi}_0, \boldsymbol{\phi}_x, \boldsymbol{\phi}_y, \boldsymbol{\phi}_z, \mathbf{y}_i)}
\end{align}
We use $p(\ldots)$ to indicate the probability density for the given arguments (which may be correlated), and use $\mathbf{y}_i$ to indicate all possible values of $i$. Note that requiring $\vec{\nabla} \Phi(0) = 0$ implies
\begin{align}
\vec{\nabla} \Phi(0) = 2 \sum_{\alpha = 1}^n \phi^\alpha(0) \vec{\nabla} \phi^\alpha(0)
= 2 \boldsymbol{\phi}_0 \cdot (\hat{x} \boldsymbol{\phi}_x + \hat{y} \boldsymbol{\phi}_y + \hat{z} \boldsymbol{\phi}_z) = 0,
\end{align}
where the dot product is in field space. Thus, the delta functions $\delta(\boldsymbol{\phi}_0 \cdot \boldsymbol{\phi}_i)$ enforce the gradient condition in the $i^{\mathrm{th}}$ direction. The corresponding result for $\ev{f}_\phi$ is given by the following.
\begin{align}
\ev{f}_\phi &= \frac{\int d^n \boldsymbol{\phi}_0 \, d^n \boldsymbol{\phi}_x \, d^n \boldsymbol{\phi}_y \, d^n \boldsymbol{\phi}_z \, d^{nk} \mathbf{y}_i \, 
f(\mathbf{y}_i \cdot \mathbf{y}_i) 
\delta(\phi^1_0 - \nu)
\delta(\phi^1_x)
\delta(\phi^1_y)
\delta(\phi^1_z)
\delta^{n-1}(\phi^2_0, \ldots, \phi^n_0)
p(\boldsymbol{\phi}_0, \boldsymbol{\phi}_x, \boldsymbol{\phi}_y, \boldsymbol{\phi}_z, \mathbf{y}_i)}
{\int d^n \boldsymbol{\phi}_0 \, d^n \boldsymbol{\phi}_x \, d^n \boldsymbol{\phi}_y \, d^n \boldsymbol{\phi}_z \, d^{nk} \mathbf{y}_i \, 
\delta(\phi^1_0 - \nu)
\delta(\phi^1_x)
\delta(\phi^1_y)
\delta(\phi^1_z)
\delta^{n-1}(\phi^2_0, \ldots, \phi^n_0)
p(\boldsymbol{\phi}_0, \boldsymbol{\phi}_x, \boldsymbol{\phi}_y, \boldsymbol{\phi}_z, \mathbf{y}_i)}
\end{align}

Our approach to proving that $\Bev{f} \equiv \ev{f}_\phi$ is to use the symmetries of field space to reduce these integrals by integrating over $d^n \boldsymbol{\phi}_0$, and showing that the resulting expressions are identical. 

We begin with $\Bev{f}$. The first step in this process is to rewrite each variable in spherical polar coordinates in field space. Let $r_0 = |\boldsymbol{\phi}_0|$ be the magnitude of $\boldsymbol{\phi}_0$ in field space. We can then write
\begin{align}
d^n \boldsymbol{\phi}_0 = r_0^{n-1} dr_0 d\Omega^{n-2}_0.
\end{align}
For all other variables, we choose to orient their  polar axis in field space along $\boldsymbol{\phi}_0$. The magnitude and measure over these variables becomes
\begin{align}
\begin{aligned}
r_x &= |\boldsymbol{\phi}_x|, \quad& d^n \boldsymbol{\phi}_x &= r_x^{n-1} dr_x d\Omega^{n-2}_x,
\\
r_y &= |\boldsymbol{\phi}_y|, \quad& d^n \boldsymbol{\phi}_y &= r_y^{n-1} dr_y d\Omega^{n-2}_y,
\\
r_z &= |\boldsymbol{\phi}_z|, \quad& d^n \boldsymbol{\phi}_z &= r_z^{n-1} dr_z d\Omega^{n-2}_z,
\\
y_i &= |\mathbf{y}_i|, \quad& d^n \mathbf{y}_i &= y_i^{n-1} dy_i d\Omega^{n-2}_i,
\end{aligned}
\end{align}
although we will not need these explicitly. Having chosen the coordinate system, $\Bev{f}$ becomes
\begin{align}
\Bev{f} &= \frac{\int r_0^{n-1} dr_0 d\Omega^{n-2}_0 \, d^n \boldsymbol{\phi}_x \, d^n \boldsymbol{\phi}_y \, d^n \boldsymbol{\phi}_z \, d^{nk} \mathbf{y}_i \, 
f(y_i^2) 
\delta(r_0^2 - \nu^2) 
\delta(r_0 r_x \cos \theta_x)
\delta(r_0 r_y \cos \theta_y)
\delta(r_0 r_z \cos \theta_z)
p(\ldots)}
{\int r_0^{n-1} dr_0 d\Omega^{n-2}_0 \, d^n \boldsymbol{\phi}_x \, d^n \boldsymbol{\phi}_y \, d^n \boldsymbol{\phi}_z \, d^{nk} \mathbf{y}_i \, 
\delta(r_0^2 - \nu^2) 
\delta(r_0 r_x \cos \theta_x)
\delta(r_0 r_y \cos \theta_y)
\delta(r_0 r_z \cos \theta_z)
p(\ldots)}
\end{align}
where we suppress the arguments to $p(\boldsymbol{\phi}_0, \boldsymbol{\phi}_x, \boldsymbol{\phi}_y, \boldsymbol{\phi}_z, \mathbf{y}_i)$ for brevity. Here, we use $\theta_i$ for the polar angle for the $i^{\mathrm{th}}$ direction. Note that the angles $\Omega_0$ don't explicitly appear in either of these integrands.

The only place that $\Omega_0$ dependence can be hiding in these integrals is in the PDFs $p(\boldsymbol{\phi}_0, \boldsymbol{\phi}_x, \boldsymbol{\phi}_y, \boldsymbol{\phi}_z, \mathbf{y}_i)$. These PDFs are constructed from $n$ independent and identically distributed fields, which means that there is no preferred direction in field space. This implies that the PDFs themselves must be rotationally invariant in field space. The PDFs are multivariate Gaussian distributions in the field space vectors $\boldsymbol{\phi}_0$, $\boldsymbol{\phi}_x$, $\boldsymbol{\phi}_y$, $\boldsymbol{\phi}_z$ and $\mathbf{y}_i$, so the only quantities that the PDFs can depend on must be rotationally invariant combinations of these vectors. Hence, the angles $\Omega_0$ can only appear in the PDFS as part of dot products between $\boldsymbol{\phi}_0$ and $\boldsymbol{\phi}_x$, $\boldsymbol{\phi}_y$, $\boldsymbol{\phi}_z$ and $\mathbf{y}_i$. However, by construction, $\boldsymbol{\phi}_0$ lies along the polar axis of $\boldsymbol{\phi}_x$, $\boldsymbol{\phi}_y$, $\boldsymbol{\phi}_z$ and $\mathbf{y}_i$, and so the angle between $\boldsymbol{\phi}_0$ and each of $\boldsymbol{\phi}_x$, $\boldsymbol{\phi}_y$, $\boldsymbol{\phi}_z$ and $\mathbf{y}_i$ is given by the relevant polar angle of that field space vector, independent of $\Omega_0$. Hence, the angles $\Omega_0$ do not appear in the probability densities, and thus they do not appear anywhere in the integrands.

This means that we can integrate over $d\Omega_0^{n-2}$ in the numerator and denominator, yielding the same volume factor in each, which then cancels. We next integrate over $dr_0$ by using one of the delta functions; again, the contribution to the numerator and denominator cancels. Finally, we can cancel three factors of $\nu$ in the remaining delta functions. The final result is
\begin{align}
\Bev{f} &= \frac{\int d^n \boldsymbol{\phi}_x \, d^n \boldsymbol{\phi}_y \, d^n \boldsymbol{\phi}_z \, d^{nk} \mathbf{y}_i \, 
f(y_i^2) 
\delta(r_x \cos \theta_x)
\delta(r_y \cos \theta_y)
\delta(r_z \cos \theta_z)
p(\nu, \boldsymbol{\phi}_x, \boldsymbol{\phi}_y, \boldsymbol{\phi}_z, \mathbf{y}_i)}
{\int d^n \boldsymbol{\phi}_x \, d^n \boldsymbol{\phi}_y \, d^n \boldsymbol{\phi}_z \, d^{nk} \mathbf{y}_i \, 
\delta(r_x \cos \theta_x)
\delta(r_y \cos \theta_y)
\delta(r_z \cos \theta_z)
p(\nu, \boldsymbol{\phi}_x, \boldsymbol{\phi}_y, \boldsymbol{\phi}_z, \mathbf{y}_i)}.
\end{align}
Note that the probability density will depend on $\nu$ after integrating out $\boldsymbol{\phi}_0$ completely because of correlations with the other variables.

We now treat $\ev{f}_\phi$. Here, we integrate over the $n$ delta functions to fix the value of $\boldsymbol{\phi}_0$. Note that as we are fixing the magnitude to be $\nu$, and the angles $\Omega_0$ are all absent from the integrands, the probability density after integration becomes $p(\nu, \boldsymbol{\phi}_x, \boldsymbol{\phi}_y, \boldsymbol{\phi}_z, \mathbf{y}_i)$ as above. A number of constant factors cancel between the numerator and the denominator, leaving us with
\begin{align}
\ev{f}_\phi &= \frac{\int d^n \boldsymbol{\phi}_x \, d^n \boldsymbol{\phi}_y \, d^n \boldsymbol{\phi}_z \, d^{nk} \mathbf{y} \, 
f(y_i^2) 
\delta(\phi^1_x)
\delta(\phi^1_y)
\delta(\phi^1_z)
p(\nu, \boldsymbol{\phi}_x, \boldsymbol{\phi}_y, \boldsymbol{\phi}_z, \mathbf{y})}
{\int d^n d^n \boldsymbol{\phi}_x \, d^n \boldsymbol{\phi}_y \, d^n \boldsymbol{\phi}_z \, d^{nk} \mathbf{y} \, 
\delta(\phi^1_x)
\delta(\phi^1_y)
\delta(\phi^1_z)
p(\nu, \boldsymbol{\phi}_x, \boldsymbol{\phi}_y, \boldsymbol{\phi}_z, \mathbf{y})}.
\end{align}
Noting that $\phi^1_i = r_i \cos \theta_i$, we finally arrive at the desired result,
\begin{align}
\Bev{f} \equiv \ev{f}_\phi.
\end{align}
Hence the PDF for $\Phi$ conditioned on $\Phi(0) = \nu^2$ and $\vec{\nabla} \Phi(0) = 0$ is identical to the PDF for $\Phi$ conditioned on $\phi^1(0) = \nu$, $\vec{\nabla} \phi^1(0) = 0$ and $\phi^{\alpha > 1}(0) = 0$.

\section{Legendre Polynomials and Real Spherical Harmonics} \label{app:real_harmonics}

Working in spherical polar coordinates, Legendre polynomials and spherical harmonics naturally arise. In this appendix, we list the properties of both that are used throughout this paper.

The plane wave expansion relates Fourier modes to spherical wave modes
\begin{align} \label{eq:planewaveexp}
e^{i \vec{k} \cdot \vec{r}} = \sum_{\ell = 0}^\infty i^\ell (2 \ell + 1) j_{\ell}(kr) P_\ell(\hat{k} \cdot \hat{r}),
\end{align}
where $j_\ell$ is a spherical Bessel function of the first kind and $P_\ell$ is a Legendre polynomial. Legendre polynomials satisfy the orthogonality condition
\begin{align} \label{eq:legendre-ortho}
\int_{-1}^1 du \, P_\ell(u) P_{\ell'}(u) = \frac{2}{2 \ell + 1} \delta_{\ell \ell'}.
\end{align}

Because we are dealing exclusively with real fields, it is useful to work with real spherical harmonics instead of complex ones. This has the benefit of making all of the coefficients in a spherical harmonic expansion real. The real spherical harmonics, written with both indices lowered to differentiate from their complex brethren, are just the appropriate linear combinations of the usual complex spherical harmonics.
\begin{align}
Y_{\ell m}(\hat{r}) = 
\left\{\begin{array}{ll}
\displaystyle \frac{1}{\sqrt{2}} (Y^m_\ell(\hat{r}) + Y^{m\ast}_\ell(\hat{r})) & m > 0 \\
\displaystyle Y^0_\ell(\hat{r}) & m = 0 \\
\displaystyle \frac{1}{\sqrt{2} i} (Y^m_\ell(\hat{r}) - Y^{m\ast}_\ell(\hat{r})) & m < 0
\end{array}
\right.
\end{align}
Real spherical harmonics obey the usual orthonormality relationship.
\begin{align} \label{eq:sph-orthonormality}
\int d\hat{r} \, Y_{\ell m}(\hat{r}) Y_{\ell'm'} (\hat{r}) = \delta_{\ell \ell'} \delta_{mm'}
\end{align}
They have the same parity as the complex spherical harmonics.
\begin{align} \label{eq:parity}
Y_{\ell m}(\pi - \theta, \pi + \phi) = (-1)^\ell Y_{\ell m}(\theta, \phi)
\end{align}
From the plane wave expansion, one can obtain
\begin{align} \label{eq:exp_sph_integral}
\int d\hat{r} \, Y_{\ell m} (\hat{r}) e^{i \vec{k} \cdot \vec{r}} = 4 \pi i^\ell j_\ell(kr) Y_{\ell m}(\hat{k}).
\end{align}
The addition theorem
\begin{align}\label{eq:addition}
P_\ell(\hat{r} \cdot \hat{r}') = \frac{4 \pi}{2 \ell + 1} \sum_{m = - \ell}^\ell Y_{\ell m} (\hat{r}) Y_{\ell m} (\hat{r}')
\end{align}
holds for real spherical harmonics, from which Uns\"old's theorem is readily derived.
\begin{align}\label{eq:unsold}
\sum_{m = - \ell}^\ell Y_{\ell m} (\hat{r}) Y_{\ell m} (\hat{r}) = \frac{2 \ell + 1}{4 \pi}
\end{align}
We will also need to know the following about the explicit forms of the real spherical harmonics.
\begin{align}
Y_{\ell m}(\hat{r}) &\propto \begin{cases}
\cos (m \phi), & m > 0 \\
\sin (m \phi), & m < 0
\end{cases}
\\
Y_{\ell 0}(\hat{r}) &= \sqrt{\frac{2 \ell + 1}{4 \pi}} P_\ell(\cos \theta)
\end{align}
It is also useful to note that $Y_{00}(\hat{r}) = 1/\sqrt{4\pi}$.

\section{Momentum Space Definitions and Integrals} \label{app:kspace_integrals}

In this appendix, we document our conventions for momentum space and the power spectrum of a centered (zero mean) Gaussian random field, and derive some helpful integrals.

We use the following conventions for the Fourier decomposition of a real field $\phi(\vec{r}\,)$,
\begin{align}
\phi(\vec{r}\,) = \int \dfrac{d^3 k}{(2\pi)^{3/2}} e^{i\vec{k}\cdot \vec{r}} \tilde{\phi}(\vec{k})
= \int \dfrac{d^3 k}{(2\pi)^{3/2}} \dfrac{1}{2} \left[ e^{i\vec{k}\cdot \vec{r}}\tilde{\phi}(\vec{k}) + e^{-i\vec{k}\cdot \vec{r}}\tilde{\phi}^*(\vec{k}) \right].
\end{align}
As $\phi$ is real, we have $\tilde{\phi}^*(\vec{k}) = \tilde{\phi}(-\vec{k})$. The inverse Fourier transform is given by
\begin{align}
\tilde{\phi}(\vec{k}) &= \int \dfrac{d^3 r}{(2\pi)^{3/2}} e^{-i\vec{k}\cdot \vec{r}} \phi(\vec{r}\,).
\end{align}

Let $\phi$ be a homogeneous centered random field with two-point function $\ev{\phi(\vec{r}\,) \phi(\vec{r}\,')} = \Cov(\phi(\vec{r}\,), \phi(\vec{r}\,')) = C(\vec{r}\,' - \vec{r}\,)$. The power spectrum $\Pk$ is defined to be the Fourier transform of the two-point function,
\begin{align}
\Pk(\vec{k}) = \frac{1}{(2\pi)^3} \int d^3 r \, e^{-i\vec{k} \cdot \vec{r}} C(\vec{r}\,).
\end{align}
Note that the factor of $(2\pi)^3$ is conventional. If $\phi$ is also isotropic so that $C(\vec{r}\,) = C(r)$, then $\Pk(\vec{k}) = \Pk(|\vec{k}|) = \Pk(k)$. We henceforth assume that $\phi$ is isotropic. The Wiener-Khinchin theorem relates the power spectrum to the two-point function of $\phi$ in Fourier space,
\begin{align} \label{eq:wiener-khinchin}
\ev{\tilde{\phi}^*(\vec{k}) \tilde{\phi}(\vec{k}')} 
= \delta^3(\vec{k} - \vec{k}') \int d^3 r \, e^{-i\vec{k}' \cdot \vec{r}} C(r)
= (2 \pi)^3 \delta^3(\vec{k} - \vec{k}') \Pk(k).
\end{align}

All of the one-point functions $\ev{\phi(\vec{r}\,)}$, $\ev{\partial_i \phi(\vec{r}\,)}$, $\ev{\partial_i \partial_j \phi(\vec{r}\,)}$, etc vanish identically. We thus turn to the two-point function $\ev{\phi(\vec{r}\,) \phi(\vec{r}\,')}$. Without loss of generality, we can take $\vec{r}\,' = 0$ by exploiting homogeneity. Then
\begin{align} \label{eq:twopoint}
\ev{\phi(\vec{r}\,) \phi(0)} 
= \int \dfrac{d^3 k \, d^3 k'}{(2\pi)^{3}} e^{i\vec{k}\cdot \vec{r}} \ev{\tilde{\phi}(\vec{k}') \tilde{\phi}(\vec{k})}
%\\
%= \int \dfrac{d^3 k \, d^3 k'}{(2\pi)^{3}} e^{i\vec{k}\cdot \vec{r}} (2 \pi)^3 \delta^3(\vec{k} - \vec{k}') \Pk(k)
%\\
= \int d^3 k \, e^{i\vec{k} \cdot \vec{r}}\Pk(k).
\end{align}
This can be further evaluated by choosing a coordinate system for $k$-space in which $\hat{z}$ points in the direction of $\vec{r}$. In this coordinate system, $\vec{k} \cdot \vec{r} = k r \cos \theta$, and
\begin{align}
C(\vec{r}\,) = \ev{\phi(\vec{r}\,) \phi(0)} 
= \int k^2 \, dk \int_0^\pi \sin \theta d\theta \int_0^{2\pi} d\phi \,e^{i k r \cos(\theta)} \Pk(k)
= 4 \pi \int dk \, k^2 \, \Pk(k) \sinc(k r).
\end{align}
This expression depends only on the magnitude $r$ as advertised. In the coincidence limit $\vec{r} = \vec{r}\,'$, this becomes
\begin{align}
C(0) = \ev{\phi(\vec{r}\,)^2} = 4 \pi \int dk \, k^2 \, \Pk(k).
\end{align}
This is the first of many moments of the power spectrum that we will come across. For convenience, we define
\begin{align}
\sigma^2_n = 4 \pi \int dk \, k^{2n+2} \, \Pk(k),
\end{align}
so that $C(0) = \sigma_0^2$.

We will also need a variety of two-point functions involving derivatives of $\phi$. Noting that the derivative operation commutes with the expectation value, we obtain
\begin{align}
\ev{\phi(\vec{r}\,) \vec{\nabla} \phi(0)} 
= \int d^3 k \, (- i \vec{k}) e^{i\vec{k} \cdot \vec{r}}\Pk(k)
= - \vec{\nabla} \ev{\phi(\vec{r}\,) \vec{\nabla} \phi(0)}
%= - \vec{\nabla} \int d^3 k \, e^{i\vec{k} \cdot \vec{r}}\Pk(k)
%&= - \vec{\nabla} \ev{\phi(\vec{r}\,) \phi(0)}
%\\
%&= - 4 \pi \int dk \, k^2 \, \Pk(k) \vec{\nabla} \sinc(k r)
%\\
= \hat{r} 4 \pi \int dk \, k^3 \, \Pk(k) j_1(kr).
\end{align}
Here, $j_1$ is a spherical Bessel function of the first kind. This quantity is sufficiently common that we define a vector
\begin{align}
\vec{D}(\vec{r}\,) = \ev{\phi(\vec{r}\,) \vec{\nabla} \phi(0)} = \hat{r} D(r),
\end{align}
with
\begin{align}
D(r) = 4 \pi \int dk \, k^3 \, \Pk(k) j_1(kr).
\end{align}
Note that $\vec{D}$ points in the direction of $\vec{r}$, which indicates that the components of the gradient perpendicular to $\vec{r}$ are uncorrelated with the field at $\vec{r}$. Also note that $D(0) = 0$, implying that in the coincident limit,
\begin{align}
\ev{\phi(\vec{r}\,) \vec{\nabla} \phi(\vec{r}\,)} &= 0.
\end{align}

We need one last spatial two-point function, but this time only in the coincidence limit,
\begin{align}
\ev{\partial_i \phi(0) \partial_j \phi(0)} = \int d^3 k \, (i k_i) (-i k_j) \Pk(k).
\end{align}
To evaluate this, let $\vec{k} = k \hat{n}$, and write the integrand in spherical polar coordinates as
\begin{align} \label{eq:deriv-var}
\ev{\partial_i \phi(0) \partial_j \phi(0)} = \int d\hat{n} \, n_i n_j \int dk \, k^4 \Pk(k)
%\\
%&= \delta_{ij} \frac{4 \pi}{3} \int dk \, k^4 \Pk(k)
%\\
= \delta_{ij} \frac{\sigma_1^2}{3},
\end{align}
where we've used
\begin{align}
\int d\hat{n} \, n_i n_j &= \frac{4\pi}{3} \delta_{ij}.
\end{align}

The final correlation functions we need relate to the spherical harmonic decomposition of $\phi$. Recall that we decompose $\phi(\vec{r}\,)$ as
\begin{align}
\phi(\vec{r}\,) = \sum_{\ell=0}^{\infty}\sum_{m=-\ell}^{\ell} \phi_{\ell m}(r) Y_{\ell m}(\hat{r})
, \quad \text{with} \quad
\phi_{\ell m}(r) = \int d\hat{r} \, \phi(\vec{r}\,) Y_{\ell m}(\hat{r}).
\end{align}
This implies
\begin{align}
\ev{\phi_{\ell m}(r)} = 0,
\end{align}
as $\phi(\vec{r}\,)$ is centered. We now want the two-point function $\ev{\phi_{\ell m}(r) \phi_{\ell' m'}(r')}$. First, insert the Fourier transform of $\phi(\vec{r}\,)$ into the definition of $\phi_{\ell m}(r)$ and make use of Eq. \eqref{eq:exp_sph_integral}.
\begin{align}
\phi_{\ell m}(r)
%&= \int d\Omega \, Y_{\ell m} (\Omega) \int \frac{d^3k}{(2 \pi)^{3/2}} e^{i \vec{k} \cdot \vec{r}} \tilde{\phi} (\vec{k})
%\\
%&= \int \frac{d^3k}{(2 \pi)^{3/2}} \tilde{\phi} (\vec{k}) \int d\Omega \, Y_{\ell m} (\Omega) e^{i \vec{k} \cdot \vec{r}}
%\\
= 4 \pi i^\ell \int \frac{d^3k}{(2 \pi)^{3/2}} \tilde{\phi} (\vec{k}) j_\ell(kr) Y_{\ell m}(\hat{k})
\end{align}
We now compute the two-point function from this expression using the Wiener-Khinchin theorem \eqref{eq:wiener-khinchin}, the parity of the spherical harmonics \eqref{eq:parity}, and the orthonormality relation \eqref{eq:sph-orthonormality}.
\begin{align}
\langle \phi_{\ell m}(r) \phi_{\ell'm'}(r') \rangle
&= 
\frac{16 \pi^2}{(2 \pi)^3}i^{\ell + \ell'}  \int d^3k \, d^3k' \, j_\ell (kr) j_{\ell'}(k'r') Y_{\ell m}(\hat{k}) Y_{\ell 'm'}(\hat{k}') \langle \tilde{\phi} (\vec{k}) \tilde{\phi} (\vec{k}') \rangle
%\\
%&= 
%\frac{16 \pi^2}{(2 \pi)^3} i^{\ell + \ell'} \int d^3k \, d^3k' \, j_\ell(kr) j_{\ell'}(k'r') Y_{\ell m}(\hat{k}) Y_{\ell'm'}(\hat{k}') (2 \pi)^3 \Pk(k) \delta^3(\vec{k} + \vec{k}')
%\\
%&= 
%16 \pi^2 i^{\ell + \ell'} \int d^3k \, j_\ell(kr) j_{\ell'}(kr') Y_{\ell m}(\hat{k}) Y_{\ell'm'}(-\hat{k}) \Pk(k)
%\\
%&= 
%16 \pi^2 i^{\ell + \ell'} \int dk \, k^2 \Pk(k) j_\ell(kr) j_{\ell'}(kr') \int d\hat{k} \, Y_{\ell m}(\hat{k}) Y_{\ell'm'}(-\hat{k})
%\\
%&= 
%16 \pi^2 i^{\ell + \ell'} \int dk \, k^2 \Pk(k) j_\ell(kr) j_{\ell'}(kr') (-1)^\ell \int d\hat{k} \, Y_{\ell m}(\hat{k}) Y_{\ell'm'}(\hat{k})
\\
&= 
\delta_{\ell\ell'} \delta_{mm'} 16 \pi^2 \int dk \, k^2 \Pk(k) j_\ell(kr) j_\ell(kr')
\end{align}
Each spherical harmonic mode coefficient is independent of all others, and the integral is independent of $m$. This is expected from the argument presented in Appendix~\ref{app:spherical_covariance}. For convenience, we define
\begin{align} \label{eq:ctildedef}
\tilde{C}_\ell(r, r') = 4 \pi \int dk \, k^2 \Pk(k) j_\ell(kr) j_\ell(kr'),
\end{align}
so that
\begin{align} \label{eq:ctilde}
\langle \phi_{\ell m}(r) \phi_{\ell'm'}(r') \rangle
= \delta_{\ell\ell'} \delta_{mm'} 4 \pi \tilde{C}_{\ell}(r, r').
\end{align}
Note that when $r = r'$, the integrand is positive semi-definite, while for $r \neq r'$, it can be negative. As $j_\ell(0) = \delta_{\ell 0}$, when $r' = 0$ we have
\begin{align} \label{eq:tildeCr0}
\tilde{C}_\ell(r, 0) = \delta_{\ell 0} 4 \pi \int dk \, k^2 \Pk(k) \sinc(kr) = \delta_{\ell 0} C(r).
\end{align}

We also need $\ev{\phi_{\ell m}(r) \phi'_{\ell'm'}(r')}$ and $\ev{\phi'_{\ell m}(r) \phi'_{\ell'm'}(r')}$. The easiest way to compute these is to exploit the fact that derivatives commute with the expectation value and use the above results. The following derivative is useful.
\begin{align}
\frac{\partial j_\ell(kr)}{\partial r} &= \frac{k}{2\ell + 1}(\ell j_{l-1}(k r) - (\ell + 1) j_{\ell+1}(kr)), \quad \ell \ge 1
\end{align}
For $\ev{\phi_{\ell m}(r) \phi'_{\ell'm'}(r')}$, we find
\begin{align}
\ev{\phi_{\ell m}(r) \phi'_{\ell'm'}(r')} &= \delta_{\ell\ell'} \delta_{mm'} 4 \pi \partial_{r'} \tilde{C}_\ell(r, r')
\\
&= \delta_{\ell\ell'} \delta_{mm'} 16 \pi^2 \int \frac{dk \, k^3 \Pk(k)}{2\ell + 1} j_\ell(kr) (\ell j_{l-1}(k r') - (\ell + 1) j_{\ell+1}(kr')),
\end{align}
which for $r' = 0$ reduces to
\begin{align}
\ev{\phi_{\ell m}(r) \phi'_{\ell'm'}(0)} &= \delta_{\ell 1} \delta_{\ell' 1} \delta_{mm'} \frac{16 \pi^2}{3} \int dk \, k^3 \Pk(k) j_1(kr) = \delta_{\ell 1} \delta_{\ell' 1} \delta_{mm'} \frac{4 \pi}{3} D(r).
\end{align}

For $\ev{\phi'_{\ell m}(r) \phi'_{\ell'm'}(r')}$, we obtain
\begin{align}
\ev{\phi'_{\ell m}(r) \phi'_{\ell'm'}(r')} &= \delta_{\ell\ell'} \delta_{mm'} 4 \pi \partial_r \partial_{r'} \tilde{C}_\ell(r, r')
\\
&= \delta_{\ell\ell'} \delta_{mm'} 16 \pi^2 \int \frac{dk \, k^4 \Pk(k)}{(2\ell + 1)^2}(\ell j_{l-1}(k r) - (\ell + 1) j_{\ell+1}(k r)) (\ell j_{l-1}(k r') - (\ell + 1) j_{\ell+1}(k r')),
\end{align}
which at $r = r' = 0$ reduces to
\begin{align}
\ev{\phi'_{\ell m}(0) \phi'_{\ell'm'}(0)} &= \delta_{\ell 1} \delta_{\ell' 1} \delta_{mm'} \frac{16 \pi^2}{9} \int dk \, k^4 \Pk(k) = \delta_{\ell 1} \delta_{\ell' 1} \delta_{mm'} \frac{4 \pi \sigma_1^2}{9}.
\end{align}

\section{Two-Point Function and Covariance of Spherical Harmonic Components} \label{app:spherical_covariance}

In this appendix, we show that for any field $\psi$, assuming only that its underlying probability distribution is rotationally invariant, the covariance of its spherical harmonic coefficients, defined by
\begin{align} \label{eq:decompdef}
\psi(\vec{r}\,) = \sum_{\ell = 0}^\infty \sum_{m = - \ell}^\ell \psi_{\ell m}(r) Y_{\ell m}(\hat{r})
\end{align}
is given by
\begin{align}
\ev{\psi_{\ell m}(r) \psi_{\ell' m'}(r')} = \delta_{\ell \ell'} \delta_{m m'} C_\ell(r, r'),
\end{align}
where $C_\ell(r, r')$ can depend on $\ell$ but not on $m$.

We start by computing $\psi_{\ell m}(r)$ from $\psi(\vec{r}\,)$ using the orthonormality condition.
\begin{align}
\psi_{\ell m}(r) = \int d\hat{r} \, Y_{\ell m}(\hat{r}) \psi(r \hat{r})
\end{align}
We then construct the two-point function
\begin{align}
\ev{\psi_{\ell m}(r) \psi_{\ell' m'}(r')} = \int d\hat{r} \, d\hat{r}' \, Y_{\ell m}(\hat{r}) Y_{\ell' m'}(\hat{r}') \ev{\psi(r \hat{r}) \psi(r' \hat{r}')}.
\end{align}

The key assumption is that the underlying probability distribution is rotationally invariant. This implies that $\ev{\psi(r \hat{r}) \psi(r' \hat{r}')}$ can depend only on rotationally invariant quantities: $r$, $r'$, and the angle between $\hat{r}$ and $\hat{r}'$, which is completely described by $\hat{r} \cdot \hat{r}'$. We are thus interested in the integral
\begin{align} \label{eq:covariance-integral}
I_{\ell m, \ell' m'}[f] = \int d\hat{r} \, d\hat{r}' \, Y_{\ell m}(\hat{r}) Y_{\ell m}(\hat{r}') f(\hat{r} \cdot \hat{r}'),
\end{align}
noting that the quantities $r$ and $r'$ are constant over the integral.

We now invoke the fact that any spherical harmonic basis can be written as
\begin{align} \label{eq:sphericaltensor}
Y_{\ell m}(\hat{r}) = C^{\ell m}_{i_1 \ldots i_\ell} \hat{r}^{i_1} \ldots \hat{r}^{i_\ell}
\end{align}
where the coefficients $C^{\ell m}_{i_1 \ldots i_\ell}$ are completely symmetric and traceless over all pairs of indices, $\hat{r}^i$ represents the $i^\mathrm{th}$ component of $\hat{r}$ (effectively $x/r$, $y/r$ and $z/r$), and the Einstein summation convention is used to sum over all repeated indices \cite{Applequist1989}. It can be shown that the number of linearly independent symmetric traceless tensors of this form is indeed $2 \ell + 1$, as required. The orthonormality condition imposes the constraint\footnote{This result is for a real basis. A complex basis requires the second factor to be conjugated as $C^{\ell m' *}_{i_1 \ldots i_\ell}$.}
\begin{align} \label{eq:orthonormality}
C^{\ell m}_{i_1 \ldots i_\ell} C^{\ell m'}_{i_1 \ldots i_\ell} = \delta_{mm'}
\end{align}
on the coefficients.

Inserting Eq. \eqref{eq:sphericaltensor} into Eq. \eqref{eq:covariance-integral}, we find
\begin{align}
I_{\ell m, \ell' m'}[f] = 
C^{\ell m}_{i_1 \ldots i_\ell} C^{\ell' m'}_{j_1 \ldots j_{\ell'}}
\int d\hat{r} \, \int d\hat{r}' \, 
\hat{r}^{i_1} \ldots \hat{r}^{i_\ell} \hat{r}^{j_1} \ldots \hat{r}^{j_{\ell'}}
f(\hat{r} \cdot \hat{r}').
\end{align}
Focusing on the integral alone for the moment, note that it is manifestly rotationally invariant, as both $\hat{r}$ and $\hat{r}'$ are integrated over all directions uniformly, and $f(\hat{r} \cdot \hat{r}')$ is rotationally invariant. The only rotationally invariant way of forming an expression with these indices that the integral can evaluate to is to construct the expression entirely from Kronecker delta functions\footnote{One may also try to use the Levi-Civita tensor $\epsilon_{ijk}$, but contractions of this with $C^{\ell m}_{i_1 \ldots i_\ell} C^{\ell' m'}_{j_1 \ldots j_{\ell'}}$ would be vanishing, as at least two indices must contract with one of the symmetric $C^{\ell m}$ coefficients.}. Since each Kronecker delta involves two indices, there can only be a nonzero value if the number of indices is even, which requires $\ell = \ell' + 2n$ with integer $n$. However, note that any Kronecker delta with two $i$ or two $j$ indices will form a trace with the traceless coefficients, and hence yield zero. As all indices are contracted, if $\ell \neq \ell'$, then there must be a trace occurring on at least one of the coefficients. Hence, $I_{\ell m, \ell' m'}[f]$ can only be nonzero if $\ell = \ell'$.

Furthermore, since $C^{\ell m}_{i_1 \ldots i_\ell}$ and $C^{\ell' m'}_{j_1 \ldots j_{\ell'}}$ are fully symmetric, it doesn't matter which $i$ index is contracted with which $j$ index, and we can write
\begin{align}
I_{\ell m, \ell' m'}[f] = \delta_{\ell \ell'}
C^{\ell m}_{i_1 \ldots i_\ell} C^{\ell' m'}_{j_1 \ldots j_\ell}
F_\ell \delta_{i_1 j_1} \ldots \delta_{i_\ell j_\ell}
\end{align}
where $F_\ell$ is the constant of proportionality that results from evaluating the integral. Note that the integral itself has no knowledge of $m$ or $m'$, which has been factored out into the $C^{\ell m}_{i_1 \ldots i_\ell}$ and $C^{\ell' m'}_{j_1 \ldots j_{\ell'}}$ coefficients, and so $F_\ell$ is independent of $m$ and $m'$. Contracting the indices appropriately, we arrive at
\begin{align}
I_{\ell m, \ell' m'}[f] 
= \delta_{\ell \ell'} C^{\ell m}_{i_1 \ldots i_\ell} C^{\ell m'}_{i_1 \ldots i_\ell} F_\ell
= \delta_{\ell \ell'} \delta_{m m'} F_\ell
\end{align}
where we have used the orthonormality condition \eqref{eq:orthonormality}. Hence, $I_{\ell m, \ell' m'}[f]$ is proportional to $\delta_{\ell \ell'} \delta_{m m'}$ and is otherwise independent of $m$ and $m'$.

Bringing this back to our two-point function, this implies that
\begin{align}
\ev{\psi_{\ell m}(r) \psi_{\ell' m'}(r')} = \delta_{\ell \ell} \delta_{m m'} \ev{\psi_{\ell m}(r) \psi_{\ell m}(r')}
\end{align}
where $\ev{\psi_{\ell m}(r) \psi_{\ell m}(r')}$ is independent of $m$ and $m'$.

To construct the covariance of $\psi$, we also need a one-point function. This is given by
\begin{align}
\ev{\psi_{\ell m}(r)} = \int d\hat{r} \, Y_{\ell m}(\hat{r}) \ev{\psi(\vec{r}\,)}.
\end{align}
Again, if the underlying distribution of $\psi(\vec{r}\,)$ is rotationally invariant, $\ev{\psi(\vec{r}\,)}$ can only depend on $r$, and so can be taken outside of the integral. The remaining integral over the spherical harmonic is then only nonzero for $\ell = m = 0$.
\begin{align}
\ev{\psi_{\ell m}(r)} 
= \ev{\psi(\vec{r}\,)} \int d\hat{r} \, Y_{\ell m}(\hat{r})
= \delta_{\ell 0} \delta_{m 0} \sqrt{4\pi} \ev{\psi(\vec{r}\,)}
\end{align}
Putting this together with the two-point function yields the covariance
\begin{align}
\Cov(\psi_{\ell m}(r), \psi_{\ell' m'}(r'))
&= \ev{\psi_{\ell m}(r) \psi_{\ell' m'}(r')} - \ev{\psi_{\ell m}(r)} \ev{\psi_{\ell' m'}(r')}
\\
&= \delta_{\ell \ell'} \delta_{m m'} \Cov(\psi_{\ell m}(r), \psi_{\ell m}(r'))
\end{align}
with $\Cov(\psi_{\ell m}(r), \psi_{\ell m}(r'))$ depending on $r$, $r'$ and $\ell$, but not $m$.

\section{Double Spherical Integrals with \texorpdfstring{$\cos(\gamma)$}{cos \gamma} Dependence} \label{app:wigner}

In the course of computing covariances between spherical harmonic modes of $\Phi$, we came across integrals of the form
\begin{align} \label{eq:appintegrals}
I_{\ell m,\ell'm'}[f(\cos \gamma)] = \int d\hat{r}_1 \, d\hat{r}_2 \, Y_{\ell m}(\hat{r}_1) Y_{\ell' m'}(\hat{r}_2) f(\hat{r}_1 \cdot \hat{r}_2).
\end{align}
Note that $\hat{r}_1 \cdot \hat{r}_2 = \cos \gamma$, and we know from Appendix~\ref{app:spherical_covariance} that the result will be proportional to $\delta_{\ell \ell'} \delta_{m m'}$, and will otherwise be independent of $m$. Here, we explain how to perform these integrals.

Three of the four angular degress of freedom in \eqref{eq:appintegrals} should integrate out analytically, because the function $f$ depends only on $\gamma$, the angle between $\hat{r}_1$ and $\hat{r}_2$. To isolate the $\gamma$ integral, we will need to change coordinates. In particular, we will choose a different (primed) coordinate system for $\hat{r}_2$, such that $\hat{z}' \equiv \hat{r}_1$. In this new coordinate system, $\gamma = \theta_2'$, and the integrals decouple. To proceed, we need to figure out how $Y_{\ell m}(\Omega_2)$ transforms under such a rotation.

Spherical harmonics constitute a complete orthogonal basis of functions. Hence, $Y_{\ell' m'}(\hat{r}_2)$ can be written as a linear combination of spherical harmonics\footnote{Note that while $\hat{r}_2 \equiv \hat{r}_2'$ as a vector, they describe different angles, $(\theta_2, \phi_2)$ and $(\theta_2', \phi_2')$ respectively, which is what the spherical harmonics are written in terms of.} $Y_{\ell'' m''}(\hat{r}_2')$. It turns out that this combination only contains spherical harmonics of degree $\ell'$. This restriction can be understood by viewing spherical harmonics of degree $\ell'$ as linear combinations of monomials with combined degree $\ell$ in $\hat{r}^i$, as was done in Appendix~\ref{app:spherical_covariance}. Note that a rotation of the coordinate system does not change the degree of the monomials because it is a linear transformation. Hence, we can write
\begin{align} \label{eq:wigner-d}
Y_{\ell' m'}(\hat{r}_2) = \sum_{m'' = -\ell'}^{\ell'} D^{\ell'}_{m' m''}(\hat{r}_1) Y_{\ell' m''}(\hat{r}_2').
\end{align}
The coefficients $D^{\ell'}_{m' m''}$ form the Wigner D-matrix (see Chapter 3 of \cite{Sakurai1994} for a detailed discussion of the D-matrix) and depend on $\hat{r}_1$, which provides the angles through which the coordinate system rotates. We will return to the elements of the $D$ matrix shortly.

Substituting Eq. \eqref{eq:wigner-d} into Eq. \eqref{eq:appintegrals} and performing the change of variables to $\hat{r}_2'$, we obtain
\begin{align} \label{eq:transformedintegral}
I_{\ell m,\ell'm'}[f(\cos \gamma)] = \sum_{m'' = -\ell'}^{\ell'} \int d\hat{r}_1 \, Y_{\ell m}(\hat{r}_1) D^{\ell'}_{m' m''}(\hat{r}_1) \int d\hat{r}_2' \, Y_{\ell' m''}(\hat{r}_2') f(\cos \theta_2').
\end{align}
Note that the integrals factorize at this stage. We can now simplify the integral over $d\hat{r}_2'$,
\begin{align}
\int d\hat{r}_2' \, Y_{\ell' m''}(\hat{r}_2') f(\theta_2')
&=
\int_0^{2 \pi} d\phi_2' \int_0^\pi \sin \theta_2' \, d\theta_2' \, Y_{\ell' m''}(\hat{r}_2') f(\theta_2')
\\
&=
2 \pi \sqrt{\frac{2\ell' + 1}{4 \pi}} \delta_{m'' 0} \int_0^\pi \sin \theta_2' \, d\theta_2' \, P_{\ell'}(\cos \theta_2') f(\cos \theta_2')
\\
&=
2 \pi \sqrt{\frac{2\ell' + 1}{4 \pi}} \delta_{m'' 0} \int_{-1}^1 du \, P_{\ell'}(u) f(u).
\end{align}
Evidently, we can go no further here without knowing $f(u)$. Substituting this expression back into Eq. \eqref{eq:transformedintegral}, we obtain
\begin{align} \label{eq:separated}
I_{\ell m,\ell'm'}[f(\cos \gamma)] = 2 \pi \sqrt{\frac{2\ell' + 1}{4 \pi}} \int d\hat{r}_1 \, Y_{\ell m}(\hat{r}_1) D^{\ell'}_{m' 0}(\hat{r}_1) \int_{-1}^1 du \, P_{\ell'}(u) f(u).
\end{align}

At this stage, we need expressions for the matrix elements $D^{\ell'}_{m' 0}(\hat{r}_1)$. Thankfully, we don't need the full machinery of the Wigner D-matrix here. Starting with Eq. \eqref{eq:wigner-d}, multiply both sides by $Y_{\ell' 0}(\hat{r}_2')$, and integrate over $\hat{r}_2$ (note that the angles described by $\hat{r}_2'$ depend on $\hat{r}_2$).
\begin{align} \label{eq:derivingD}
\int d\hat{r}_2 \, Y_{\ell' 0}(\hat{r}_2') Y_{\ell' m'}(\hat{r}_2) = \sum_{m'' = -\ell'}^{\ell'} D^{\ell'}_{m' m''}(\hat{r}_1) \int d\hat{r}_2 \, Y_{\ell' 0}(\hat{r}_2') Y_{\ell' m''}(\hat{r}_2') = D^{\ell'}_{m' 0}(\hat{r}_1)
\end{align}
This picks out the D-matrix component we want. Now, we write $Y_{\ell' 0}(\hat{r}_2')$ as
\begin{align}
Y_{\ell' 0}(\hat{r}_2') &= \sqrt{\frac{2 \ell' + 1}{4 \pi}} P_{\ell'}(\hat{z}' \cdot \hat{r}_2')
\end{align}
where the polar angle associated with $\hat{r}_2'$ is given by $\hat{z}' \cdot \hat{r}_2'$. As $\hat{z}' \equiv \hat{r}_1$ by definition and $\hat{r}_2' = \hat{r}_2$ as a coordinate-free vector equation, this becomes
\begin{align}
Y_{\ell' 0}(\hat{r}_2') &= \sqrt{\frac{2 \ell' + 1}{4 \pi}} P_{\ell'}(\hat{r}_1 \cdot \hat{r}_2).
\end{align}
Substituting this into Eq. \eqref{eq:derivingD} and applying the addition theorem \eqref{eq:addition} yields the surprisingly straightforward result
\begin{align} \label{eq:Dmatrix}
D^{\ell'}_{m' 0}(\hat{r}_1)
&= \sqrt{\frac{4 \pi}{2 \ell' + 1}} \sum_{m = - \ell'}^{\ell'} Y_{\ell' m} (\hat{r}_1) \int d\hat{r}_2 \, Y_{\ell' m} (\hat{r}_2) Y_{\ell' m'}(\hat{r}_2)
= \sqrt{\frac{4 \pi}{2 \ell' + 1}} Y_{\ell' m'} (\hat{r}_1).
\end{align}

Combining this D-matrix result \eqref{eq:Dmatrix} with our previous expression \eqref{eq:separated}, we obtain
\begin{align}
I_{\ell m,\ell'm'}[f(\cos \gamma)] &= 2 \pi \int d\hat{r}_1 \, Y_{\ell m}(\hat{r}_1) Y_{\ell' m'}(\hat{r}_1) \int_{-1}^1 du \, P_{\ell'}(u) f(u)
\\ \label{eq:foursphere}
&= 2 \pi \delta_{\ell \ell'} \delta_{m m'} \int_{-1}^1 du \, P_{\ell}(u) f(u),
\end{align}
where we used the orthonormality condition to evaluate the integral over spherical harmonics. As advertised, this is proportional to $\delta_{\ell \ell'} \delta_{m m'}$, is otherwise independent of $m$, and involves only a single integral over the argument of $f$.

For polynomial $f(u)$, we can evaluate the integral in Eq. \eqref{eq:foursphere} analytically. Any polynomial in $u$ can be decomposed in terms of Legendre polynomials. The three cases of interest to us are
\begin{align}
1 = P_0(u),
\qquad
u = P_1(u)
\quad \text{and} \quad
u^2 = \frac{1}{3} P_0(u) + \frac{2}{3} P_2(u).
\end{align}
We can then use the orthonormality condition on Legendre polynomials \eqref{eq:legendre-ortho} to evaluate
\begin{subequations} \label{eq:uglyintegrals}
\begin{align}
I_{\ell m,\ell'm'}[1] &= \delta_{\ell 0} \delta_{\ell' 0} \delta_{m m'} 4 \pi
\\
I_{\ell m,\ell'm'}[\cos \gamma] &= \delta_{\ell 1} \delta_{\ell' 1} \delta_{m m'} \frac{4 \pi}{3}
\\
I_{\ell m,\ell'm'}[\cos^2 \gamma] &= \delta_{\ell \ell'} \delta_{m m'} \left(\frac{4 \pi}{3} \delta_{\ell 0} + \frac{8 \pi}{15} \delta_{\ell 2}\right).
\end{align}
\end{subequations}

\section{Expansion of the Background Covariance} \label{app:cov}

In this appendix, we sketch the computation of the covariance $\Cov(\Phi(\vec{r}\,), \Phi(\vec{r}\,))$ in the unbiased case in terms of the underlying Gaussian fields. First, consider the two point function
\begin{align}
\ev{\Phi(\vec{r}\,) \Phi(\vec{r}\,')} &= \sum_{\alpha, \beta = 1}^n \left\langle
\left(\sum_{\ell_1 = 0}^\infty \sum_{m_1 = -\ell_1}^{\ell_1} Y_{\ell_1 m_1}(\hat{r}) \phi^\alpha_{\ell_1 m_1}(r)\right)^2
\left(\sum_{\ell_2 = 0}^\infty \sum_{m_2 = -\ell_2}^{\ell_2} Y_{\ell_2 m_2}(\hat{r}') \phi^\beta_{\ell_2 m_2}(r')\right)^2
\right\rangle
\\
&= \sum_{\alpha, \beta} \sum_{\ell_1, \ell_1', \ell_2, \ell_2'} \sum_{m_1, m_1', m_2, m_2'} 
Y_{\ell_1 m_1}(\hat{r}) Y_{\ell_1' m_1'}(\hat{r}) Y_{\ell_2 m_2}(\hat{r}') Y_{\ell_2' m_2'}(\hat{r}')
\nonumber \\
& \qquad \qquad \qquad \qquad \qquad \qquad \qquad \times
\left\langle
\phi^\alpha_{\ell_1 m_1}(r)
\phi^\alpha_{\ell_1' m_1'}(r)
\phi^\beta_{\ell_2 m_2}(r')
\phi^\beta_{\ell_2' m_2'}(r')
\right\rangle
.
\end{align}
This four-point function can be computed by Wick's theorem \eqref{eq:4point_wick} and the two-point function \eqref{eq:phi_two_point}, noting that the unbiased fields are centered.
\begin{align}
\left\langle
\phi^\alpha_{\ell_1 m_1}(r)
\phi^\alpha_{\ell_1' m_1'}(r)
\phi^\beta_{\ell_2 m_2}(r')
\phi^\beta_{\ell_2' m_2'}(r')
\right\rangle
&=
\Cov(\phi^\alpha_{\ell_1 m_1}(r), \phi^\alpha_{\ell_1' m_1'}(r))
\Cov(\phi^\beta_{\ell_2 m_2}(r'), \phi^\beta_{\ell_2' m_2'}(r'))
\nonumber\\
&\quad +
\Cov(\phi^\alpha_{\ell_1 m_1}(r), \phi^\beta_{\ell_2 m_2}(r'))
\Cov(\phi^\alpha_{\ell_1' m_1'}(r), \phi^\beta_{\ell_2' m_2'}(r'))
\nonumber\\
&\quad +
\Cov(\phi^\alpha_{\ell_1 m_1}(r), \phi^\beta_{\ell_2' m_2'}(r'))
\Cov(\phi^\beta_{\ell_2 m_2}(r'), \phi^\alpha_{\ell_1' m_1'}(r))
\\
&=
\delta_{\ell_1 \ell_1'} \delta_{m_1 m_1'} \delta_{\ell_2 \ell_2'} \delta_{m_2 m_2'} 16 \pi^2 \tilde{C}_{\ell_1}(r, r) \tilde{C}_{\ell_2}(r', r')
\nonumber\\
&\quad +
\delta^{\alpha \beta} \delta_{\ell_1 \ell_2} \delta_{m_1 m_2} \delta_{\ell_1' \ell_2'} \delta_{m_1' m_2'} 16 \pi^2 \tilde{C}_{\ell_1}(r, r') \tilde{C}_{\ell_1'}(r, r')
\nonumber\\
&\quad +
\delta^{\alpha \beta} \delta_{\ell_1 \ell_2'} \delta_{m_1 m_2'} \delta_{\ell_1' \ell_2} \delta_{m_1' m_2} 16 \pi^2 \tilde{C}_{\ell_1}(r, r') \tilde{C}_{\ell_2}(r, r')
\end{align}
Combining these results and collapsing as many of the sums as we can yields
\begin{align}
\Cov(\Phi(\vec{r}\,), \Phi(\vec{r}\,'))
&= 32 \pi^2 n \sum_{\ell_1, \ell_1' = 0}^\infty \tilde{C}_{\ell_1}(r, r') \tilde{C}_{\ell_1'}(r, r') 
\sum_{m_1 = -\ell_1}^{\ell_1} 
Y_{\ell_1 m_1}(\hat{r}) 
Y_{\ell_1 m_1}(\hat{r}') 
\sum_{m_1' = - \ell_1'}^{\ell_1'} 
Y_{\ell_1' m_1'}(\hat{r}) 
Y_{\ell_1' m_1'}(\hat{r}')
\end{align}
where we use Eq. \eqref{eq:background_exp} to construct the uncoupled term $\ev{\Phi(\vec{r}\,)} \ev{\Phi(\vec{r}\,')}$. We now apply the addition theorem \eqref{eq:addition} to compute the sums over $m_1$ and $m_1'$, yielding
\begin{align}
\Cov(\Phi(\vec{r}\,), \Phi(\vec{r}\,'))
&= 2 n \left(\sum_{\ell = 0}^\infty (2 \ell + 1) \tilde{C}_{\ell}(r, r') P_{\ell}(\hat{r} \cdot \hat{r}')\right)^2.
\end{align}

\end{document}